\documentclass{article} %
\PassOptionsToPackage{table}{xcolor}
\usepackage{iclr2027_conference,times}
\iclrfinalcopy %

\usepackage{xspace}

\setcitestyle{numbers,square,comma}
\newcommand{\paper}{CompoSE\xspace}

\makeatother

\usepackage{placeins}
\usepackage{float}
\usepackage[most]{tcolorbox}
\usepackage[skip=8pt]{caption}
\usepackage{fvextra}
\usepackage{tikz}
\usetikzlibrary{positioning}
\usepackage{hyperref}
\usepackage{cleveref}
\usepackage{enumitem}
\usepackage{ragged2e}
\usepackage{pifont}
\usepackage{amsmath, amsfonts, amsthm, amsbsy, bbm, latexsym}
\usepackage{xspace}
\usepackage{fontawesome5}
\usepackage{booktabs}
\usepackage{multirow}
\usepackage{colortbl}
\usepackage{graphics}
\usepackage{stmaryrd}
\usepackage{svg}
\usepackage{wrapfig}
\usepackage{svg}

\definecolor{tabcol}{HTML}{a2d2ff}
\usepackage{colortbl}
\definecolor{mygrey}{HTML}{dedede} 

\definecolor{win1}{HTML}{E3F4E1}  %
\definecolor{win2}{HTML}{B9E3B3}  %
\definecolor{win3}{HTML}{8ACF80}  %
\newcommand{\wz}[1]{#1}
\newcommand{\wa}[1]{\cellcolor{win1}#1}
\newcommand{\wb}[1]{\cellcolor{win2}#1}
\newcommand{\wc}[1]{\cellcolor{win3}#1}

\newcommand{\point}[1]{\noindent \textbf{#1}.~~}

\newcommand{\fullfig}[5][-1em]{%
\begin{figure*}[p]
    \centering
    \includegraphics[width=\textwidth]{#2}
    \vspace{#1}
    \caption{\textbf{#3} #4}%
    #5
\end{figure*}
}

\newcommand{\todo}[1][]{\textbf{\textcolor{red}{[TODO\ifx#1\empty\else: #1\fi]}}}


\newsavebox{\maxresizeboxsv}
\newcommand{\maxresizebox}[2]{%
  \sbox{\maxresizeboxsv}{#2}%
  \ifdim\wd\maxresizeboxsv>#1
    \resizebox{#1}{!}{\usebox{\maxresizeboxsv}}%
  \else
    \usebox{\maxresizeboxsv}%
  \fi
}

\newcommand{\iconog}{\textcolor{mygrey}{\faIcon{times-circle}}\xspace}

\newcommand{\cmark}{\textcolor{ForestGreen}{\ding{51}}}

\usepackage{dsfont}

\newif\ifshowedits
\newcommand{\addeditor}[3]{%
  \definecolor{#1color}{rgb}{#3}
  \expandafter\newcommand\csname #1\endcsname[1]{%
  \ifshowedits
    {\color{#1color} ##1}%
  \else
    {##1}%
  \fi
  }%
  \expandafter\newcommand\csname #1rmk\endcsname[1]{%
  \ifshowedits
    {\color{#1color} {\bf [#2: ##1]}}
  \fi
  }%
  \expandafter\newcommand\csname #1rpl\endcsname[2]{%
  \ifshowedits
    {{\color{#1color} ##1} \sout{##2}}
  \else
    {##1}
  \fi
  }%
}

\definecolor{darkgreen}{RGB}{0,110,0}
\definecolor{darkred}{RGB}{170,0,0}

\definecolor{colorA}{HTML}{f58b3c}
\definecolor{colorB}{HTML}{8a55ad}
\definecolor{colorC}{HTML}{e72a0e}
\definecolor{startblue}{HTML}{186ede}
\definecolor{endorange}{HTML}{d97636}

\newcommand{\txtc}[1]{\textcolor{startblue}{#1}}
\newcommand{\layc}[1]{\textcolor{endorange}{#1}}
\definecolor{ForestGreen}{HTML}{228B22}

\newcommand{\filledsquare}[2]{%
  \hspace{0.1em}%
  \begin{tikzpicture}[baseline=0.3ex]
    \fill[#1, rounded corners=1.5pt] (0,0) rectangle (.3,.3) 
      node[pos=.5, white] {\footnotesize #2};
  \end{tikzpicture}%
  \hspace{0.1em}%
}

\newcommand{\pointa}{\filledsquare{colorA}{A}}
\newcommand{\pointb}{\filledsquare{colorB}{B}}
\newcommand{\pointc}{\filledsquare{colorC}{C}}

\newcommand{\toclinesep}{0.55em}

\newcommand{\tocline}[4]{%
\normalsize{
\noindent \hyperref[sec:#1]{#2 \hspace{1em} #3 \leaders\hbox to 5pt{\hss.\hss}\hfill #4} \\[\toclinesep]
}
}

\newcommand{\tocsubline}[4]{%
\normalsize{
\noindent \hyperref[sec:#1]{\hspace{2em} #2 \hspace{1em} #3 \leaders\hbox to 5pt{\hss.\hss}\hfill #4} \\[\toclinesep]
}
}

\usepackage[most]{tcolorbox}
\usepackage{xcolor}
\definecolor{promptbordercolor}{RGB}{30, 90, 160}
\definecolor{promptbgcolor}{RGB}{245, 248, 253}
\definecolor{prompttitlecolor}{RGB}{30, 90, 160}
 
\newtcolorbox{promptbox}[1][]{%
  enhanced,
  breakable,
  colback=promptbgcolor,
  colframe=promptbordercolor,
  colbacktitle=prompttitlecolor,
  coltitle=white,
  fonttitle=\bfseries\sffamily,
  title=#1,
  boxrule=0.8pt,
  arc=3pt,
  left=8pt,
  right=8pt,
  top=10pt,
  bottom=10pt,
  attach boxed title to top left={yshift=-2mm, xshift=4mm},
  boxed title style={
    colframe=promptbordercolor,
    boxrule=0.6pt,
    arc=2pt,
  },
}

\usepackage{hyperref}
\usepackage{url}

\title{CompoSE: Compositional Synthesis and \\ Editing of 3D Shapes via Part-Aware Control}

\author{Habib Slim \\ KAUST \\ \texttt{habib.slim@kaust.edu.sa}
\And
Shariq Farooq Bhat \\ Adobe Research
\And
Mohamed Elhoseiny \\ KAUST
\AND
Yifan Wang \\ Adobe Research
\And
Mike Roberts \\ Adobe Research
}

\begin{document}

\maketitle
\lhead{} %

\begin{abstract}
Creating and editing high-quality 3D content remains a central challenge in computer graphics.
We address this challenge by introducing \textbf{\paper}, a novel method for \textbf{Compo}sitional \textbf{S}ynthesis and \textbf{E}diting of 3D shapes via part-aware control.
Our method takes as input a set of coarse geometric primitives (e.g., bounding boxes) that represent distinct object parts arranged in a particular spatial configuration, and synthesizes as output part-separated 3D objects that support localized granular (i.e., compositional) editing of individual parts.
The key insight that enables our method is our use of a diffusion transformer architecture that alternates between processing each part locally and aggregating contextual information across parts globally, and features a novel conditioning technique that ensures strong adherence to the user's input.
Importantly, our method learns to infer part semantics and symmetries directly from the user's coarse layout guidance, and does not require part-level text prompts.
We demonstrate that our method enables powerful part-level editing capabilities, including context-aware substitution, addition, and style-preserving resizing operations.
We show through extensive experiments that our method significantly outperforms existing approaches on guided synthesis, as measured by objective metrics and LLM-based evaluations.

\end{abstract}

\FloatBarrier
\begin{figure}[t]
  \vspace{-1em}
  \centering
  \includegraphics[width=0.96\linewidth]{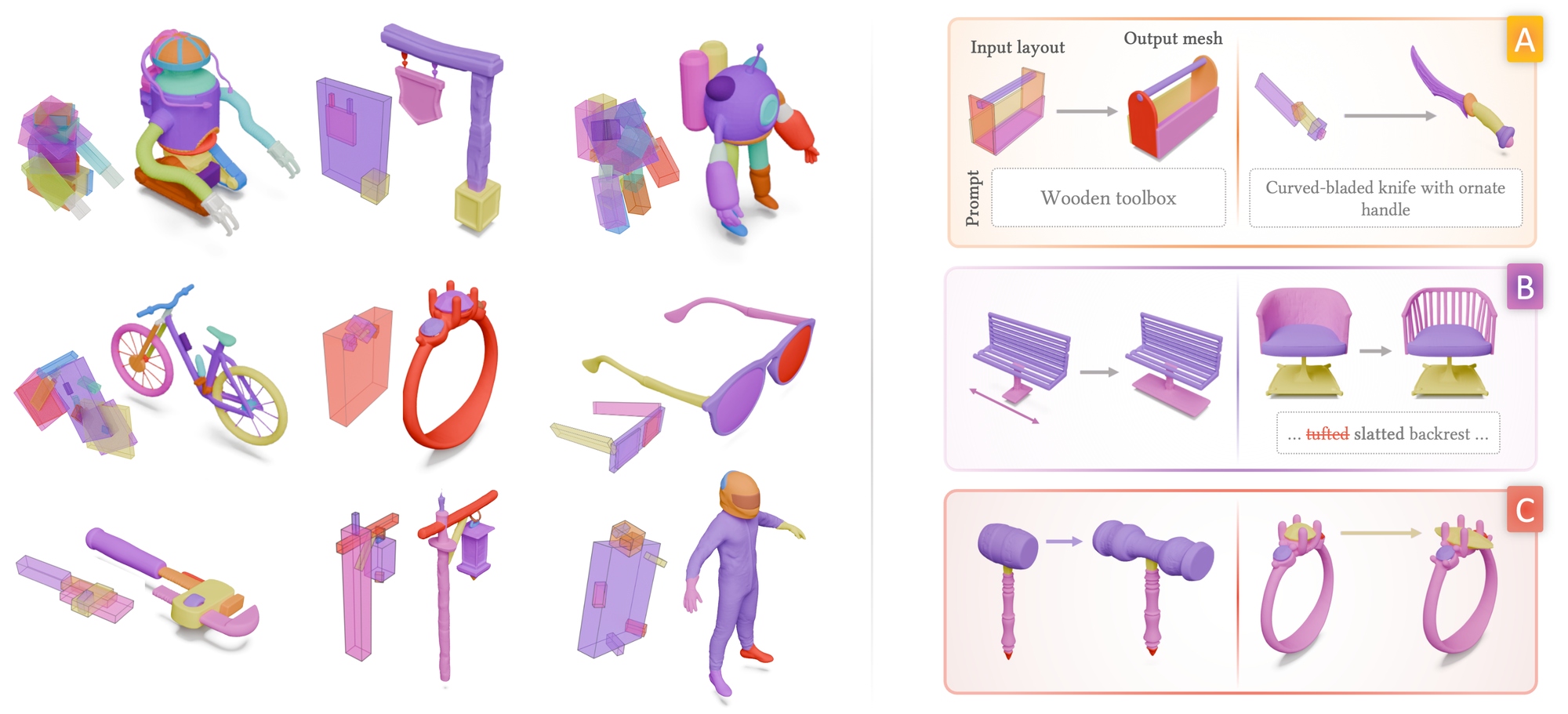}
  \caption{Our method for synthesizing 3D shapes takes as input a text prompt and a set of coarse geometric primitives (e.g., bounding boxes) that represent distinct object parts that a user would like to manipulate individually. From this input, our method synthesizes part-separated 3D objects that align with the user's geometric primitives and text prompt (\pointa~and additional results on the left). Our method enables \textit{compositional synthesis and editing}, i.e., we synthesize 3D shapes in a way that allows for localized granular editing of individual parts. For example, our method enables style-preserving resizing (\pointb, left) and text-based editing of individual parts (\pointb, right), as well as context-aware part substitution (\pointc).}
  \label{fig:teaser}
  \vspace{-1em}
\end{figure}

\section{Introduction}
\label{sec:intro}
Creating and editing high-quality 3D content remains a central challenge in many creative industries, including film, design, simulation, and manufacturing. In response, the computer graphics community has developed a variety of generative AI methods for synthesizing 3D shapes from high-level user input~\cite{3DS2VS_zhang_2023, 3dilg_zhang_2022, autosdf_mittal_2022, chen_meshanything_2024, CLIPSculptor_sanghi_2023, deepsdf_park_2019, difffacto_nakayama_2023, dreamfusion_poole_2023, fastpointcloudgeneration_wu_2023, geomimages_Gu_2002, geometryid_elizarov_2025, giraffe_niemeyer_2021, holodiffusion_karnewar_2023, LION_zeng_2022, nash_polygen_2020, object_yan_2025, overviewgeom_Hoppe_2004, p2pbridge_vogel_2024, pointe_nichol_2022, shapewalk_slim_2024, siddiqui_meshgpt_2024, sparseflex_he_2025, ultra3d_chen_2025, yang2024hunyuan3d,hunyuan3d22025tencent,lai2025hunyuan3d25highfidelity3d}.
However, many of these methods rely on language descriptions to guide their output, and the resulting user interfaces can be notoriously unpredictable~\cite{unpredictable_agrawala_2023}, especially when a user is attempting to specify the precise spatial configuration of individual object parts using text prompts.\\[-0.5em]

In an effort to provide more precise user controls, a variety of methods have been proposed that accept richer forms of guidance, such as 2D images~\cite{craftsman3d_li_2025, triposg_li_2025, xcube_ren_2024, trellis_xiang_2024, CLAY_zhang_2024,hunyuan3d22025tencent,lai2025hunyuan3d25highfidelity3d,yang2024hunyuan3d, Meshy2024, Rodin2024, HexaGen2024}, a global bounding box~\cite{CLAY_zhang_2024, Cube_roblox_2025}, or proxy shapes~\cite{fantasia3d_chen_2023, spice_sella_2024, CLAY_zhang_2024, DECOLLAGE_chen_2025, ARTDECO_chen_2025, dong2024coin3d}.
Additionally, methods have been proposed to generate part-separated outputs~\cite{Holopart_Yang_2025,PartPacker_Tang_2025, PartCrafter_Lin_2025, OneToMore_Dong_2025, OmniPart_Yang_2025, BANG_Zhang_2025} to facilitate downstream applications.
However, although some of these methods generate part-separated \textit{outputs}, none of them accepts part-separated \textit{input layouts}.
The choice made by existing methods to model the user's intent through a single monolithic object (i.e., without individual parts) also limits predictability, because small changes in the user's input can lead to significant changes in the synthesized output, potentially altering the entire generated 3D shape.
More generally, no existing method supports \textit{compositional synthesis and editing} (i.e., synthesizing part-separated objects in a way that allows for localized granular editing of individual parts), which is essential in professional content creation scenarios.\\[-0.5em]

In this work, we introduce \textbf{\paper}, a novel method for \textbf{Compo}sitional \textbf{S}ynthesis and \textbf{E}diting of 3D shapes via part-aware control.
Our method synthesizes part-separated 3D shapes from a set of coarse geometric primitives (e.g., bounding boxes) in a particular spatial configuration, and a global text prompt.
In our method, each geometric primitive represents a distinct object part that the user can manipulate individually.
The key insight that enables compositional synthesis and editing in our method is our use of a diffusion transformer (DiT) architecture~\cite{attention_vaswani_2017, denoisingdiffusion_ho_2020, scalablediffusion_peebles_2023} that processes each input geometric primitive in parallel, and strategically alternates a single DiT block between two modes.
In \emph{local stages}, the block processes spatial layout information for individual primitives, and provides high-fidelity control over the geometry of each synthesized object part; while in \emph{global stages}, it aggregates context cues and text information across object parts to ensure stylistic coherence and structural validity.

Our architecture design enables compositional editing by allowing the user to spatially manipulate a specific geometric primitive that is given as input to a local stage, while optionally freezing the latent variables associated with the other parts at inference time.
The 3D shapes resulting from this inference procedure adhere to user edits, while also preserving the identity of existing object parts exactly. To summarize, our main contributions are as follows:

\vspace{1em}
\begin{enumerate}[nosep]
    \item We present the first method to synthesize part-separated 3D shapes directly from coarse part layouts, featuring a novel conditioning technique that ensures strong adherence to the user's input layout, and a high degree of control over individual object parts.
    \item We enable a versatile set of compositional editing capabilities, including substitution, addition, and style-preserving resizing of individual object parts.
    \item We provide a fully automated data processing pipeline that enables training on large-scale 3D shape datasets, and does not require manual part-level segmentations or text annotations.
\end{enumerate}

\vspace{1em}
We demonstrate through extensive experiments that our method significantly outperforms existing approaches on guided synthesis, as measured by objective metrics and LLM-based evaluations.

\section{Related Work}
\label{sec:related_work}

\point{Controllable Shape Synthesis}
The advent of diffusion models~\cite{denoisingdiffusion_ho_2020, scalablediffusion_peebles_2023} has significantly advanced the field of 3D shape synthesis. Due to the expansive literature, we refer readers to a recent survey by Wang et al.~\cite{wang2025diffusion}. We review the most representative methods in \cref{sec:intro}, where we categorize them according to each method's choice of user guidance.
A common thread among these methods is that they lack the ability to accept part-separated user guidance, which is a gap we address here.\\[-0.5em]

\begin{wrapfigure}{l}{0.46\textwidth}
    \vspace{-1.1em}
    \centering
    \includegraphics[width=0.44\textwidth]{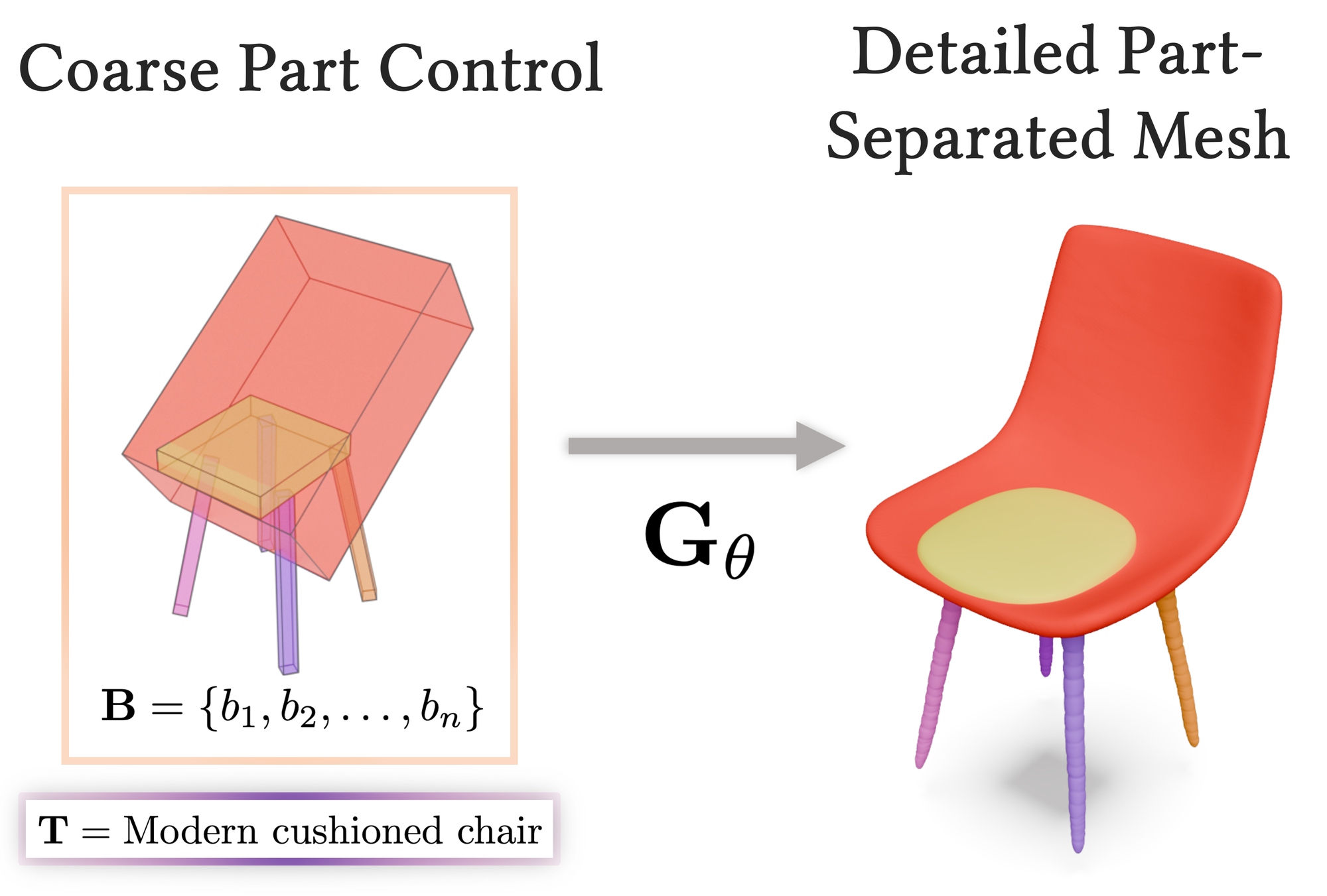}
    \caption{
    \textbf{Our problem setting} for shape synthesis via part-aware control.
    We propose to synthesize a 3D shape from a set of coarse geometric primitives (e.g., oriented bounding boxes) arranged in a particular spatial configuration, and a global text prompt. In our problem setting, we do not rely on image inputs or part-level text prompts.}%
    \label{fig:method_setting}
    \vspace{-1.2em}
\end{wrapfigure}

\point{Part-Aware Shape Synthesis}
Early efforts in part-aware synthesis introduced part-disentangled representations for language- or sketch-guided generation~\cite{spaghetti_hertz_2022, salad_koo_2023, sketchandtextguided_wu_2023}. However, these pioneering methods were often limited by their reliance on category-specific models and typically produced low-quality geometry.
Recent approaches leverage DiT-based diffusion models~\cite{scalablediffusion_peebles_2023} to generate part-separated output shapes, starting either from an input 3D mesh~\cite{Holopart_Yang_2025} or a 2D image~\cite{PartPacker_Tang_2025, PartCrafter_Lin_2025, OneToMore_Dong_2025, OmniPart_Yang_2025, BANG_Zhang_2025}.

While these image-conditioned methods have shown impressive results, their design imposes fundamental limitations on editability and controllability. First, conditioning on 2D image features --- reinjected at every stage of the pipeline --- globally couples all parts together, so resizing, substituting, or iterating on individual parts requires re-conditioning on a new image and regenerating the entire shape. Second, the conditioning image is itself typically produced by a 2D diffusion model, which offers no part-level control and provides no information about occluded or internal geometry, restricting the user to parts visible from a single camera viewpoint.
Although some of these methods (e.g., OmniPart~\cite{OmniPart_Yang_2025}) include an intermediate bounding-box stage, the boxes are axis-aligned, not user-controllable, and cannot be used for part-level editing, since image features are still reinjected downstream. In contrast, our method accepts user-controlled oriented bounding boxes paired with a global text prompt as its primary guidance, enabling specification of parts invisible from any camera angle and supporting compositional editing --- adding, resizing, and replacing parts --- while keeping all other parts frozen.

Distinct from methods that synthesize geometry directly, methods based on \emph{3D shape programs}~\cite{shapeassembly_jones_2020, plankassembly_hu_2023, improvingunsupervised_ganeshan_2023} generate executable code capable of enforcing structural constraints like symmetry, but are often limited to specific categories and produce lower-fidelity geometry than diffusion-based methods.
Our work combines the fine-grained control associated with 3D shape programs with the high geometric fidelity and cross-category generality of diffusion-based methods.

\point{3D Shape Editing}
A common approach for editing 3D shapes involves leveraging 2D diffusion models, either through per-scene optimization via Score Distillation Sampling~\cite{chen2024shap, park2023ed, zhang2023text, zhuang2023dreameditor, kamata2023instruct, decatur20243d, dong2024interactive3d, dinh2025geometry, liu2024make, zhuang2024tip, voleti2024sv3d, xu2024tiger, sella2023vox}, or by using multi-view diffusion models to generate edits that are then lifted to 3D~\cite{barda2025instant3dit, li2025cmd, chi2025disco3d, chen20243d, bar2025editp23, erkocc2025preditor3d, qi2024tailor3d, gaussctrl2024, chen2024gaussianeditor, wen2025intergsedit, haque2023instruct, igs2gs, chen2024dge}. %
Another common approach is to directly leverage a 3D diffusion prior and use an existing proxy shape for guidance~\cite{CLAY_zhang_2024, ARTDECO_chen_2025, dong2024coin3d, spacecontrol_fedele_2026, proxe_sella_2026} while synthesizing geometric details across the entire shape. More recent work has achieved localized editing by leveraging grid-based latent representations~\cite{li2025voxhammer,trellis_xiang_2024,Rodin2024, maruani2025shapeshifter,potamias2024shapefusion}. However, these methods either achieve editing through text or image instructions, or operate on monolithic shapes whose parts cannot be edited independently. In contrast, our work provides an intuitive interface where users can add, resize, and replace individual object parts by manipulating bounding boxes.

\begin{figure*}[t]
    \centering
    \includegraphics[width=0.95\textwidth]{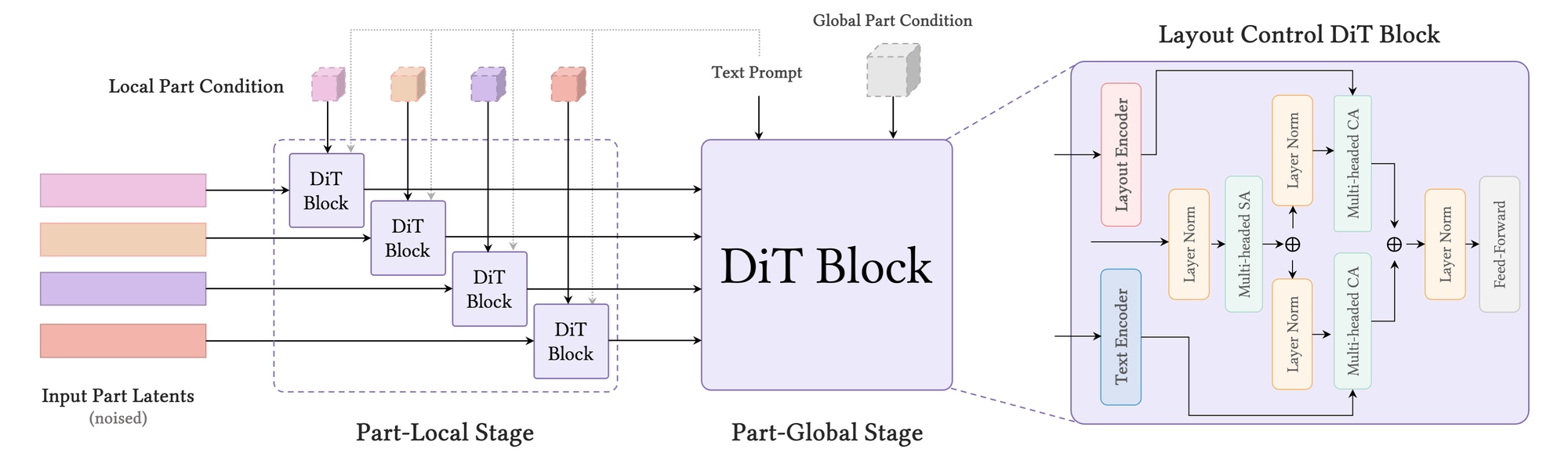}
    \caption{\textbf{Our diffusion transformer (DiT) architecture} for compositional synthesis and editing with part-aware control.
    Our architecture alternates between local and global stages.
    Local stages spatially ground each individual part to its location, and global stages aggregate context cues and text information across primitives.
    }
    {\label{fig:method_overview}}
    \vspace{-1em}
\end{figure*}

\section{Part-Aware Control, Synthesis, and Editing}
\label{sec:prim_guided_synthesis}
Our goal in this section is to design a latent diffusion transformer model that enables compositional synthesis and editing of 3D shapes, where the user's intent is communicated through a global text prompt describing an object, and a spatial layout of coarse geometric primitives (e.g., oriented bounding boxes) describing the arrangement of the object's individual parts.

\point{Technical Requirements} Enabling compositionality in our model imposes three requirements on its design. First, the model must accept coarse geometric primitives (e.g., oriented bounding boxes) that specify the intended structure and part layout. Second, the model must fuse part-level geometric conditions with a global text prompt so that each synthesized part respects its local geometric constraints while remaining semantically coherent with the overall description. Third, the model must support localized editing of individual parts through manipulation of individual geometric primitives at inference time. We discuss our editing procedure in \cref{sec:method-editing}.\\

\point{Problem Setting} Motivated by the technical requirements in the previous paragraph, we consider the following formal problem setting, which we illustrate in~\cref{fig:method_setting}. Given a conditioning set $\layc{\mathbf{B}} = \{\layc{b_1}, \layc{b_2}, \ldots, \layc{b_n}\}$ of coarse oriented boxes and a shape-level text prompt $\txtc{T}$, we want to synthesize \textit{separate parts} $p_i$ for each box in $\layc{\mathbf{B}}$, such that the synthesized parts $\mathbf{P} = \{p_1, p_2, \ldots, p_n\}$ together form a coherent 3D shape aligning with $\txtc{T}$. Formally, we want:
\begin{equation}
    \mathbf{P} = \mathbf{G}_\theta\left( \mathbf{\mathcal{E}_B}( \layc{\mathbf{B}}), \mathbf{\mathcal{E}_T}(\txtc{T}) \right)
\end{equation}

\noindent where $\mathbf{G}_\theta$ is a generative model parameterized by $\theta$, $\mathbf{\mathcal{E}_B}$ is a learned box encoder that maps each box $\layc{b_i}$ to an appropriate latent representation, and $\mathbf{\mathcal{E}_T}$ is a pre-trained text encoder~\cite{clip_radford_2021}.

\subsection{Layout-Controlled Diffusion Transformer}
\label{sec:layout_control_dit}

\point{Shared Latent Space} We encode each part $p_i$ with a pre-trained shape VAE into a latent $\mathbf{z}_i \in \mathbb{R}^{D \times d}$ of $D$ tokens of width $d$, which the decoder turns back into an SDF. For the box encoder $\mathbf{\mathcal{E}_B}$, we use that \emph{same} VAE encoder, so that a box is encoded exactly as a part would be:
\begin{equation}
    \layc{\mathbf{c}_i} = \mathbf{\mathcal{E}_B}(\layc{b_i}) = \textsc{Vae}_{\textsc{enc}}(\textsc{Sample}(\layc{b_i})) \in \mathbb{R}^{D \times d}.
\end{equation}

\noindent The conditions and the denoising targets then share one space, so the model does not have to learn how they relate from scratch. We encode the minimal oriented box $\layc{b_\star}$ enclosing all primitives in the same way, as $\layc{\mathbf{c}_\star} = \mathbf{\mathcal{E}_B}(\layc{b_\star})$.\\[-0.5em]

\point{Layout Control DiT Block} The user gives a single global prompt, but the model must fill several primitives, each with a different spatial role. We therefore inject layout and text conditioning through two \emph{parallel} cross-attention branches (see~\cref{fig:method_overview}). For a block input $\mathbf{u}$, layout condition $\layc{\mathbf{c}}$, and text tokens $\txtc{\mathbf{y}} = \mathbf{\mathcal{E}_T}(\txtc{T})$, the block computes
\begin{equation}
\begin{aligned}
\mathbf{h} &= \mathbf{u} + \mathrm{SA}\!\left(\mathrm{LN}(\mathbf{u}),\, \mathbf{M}\right), \\
\mathbf{h}' &= \mathbf{h} + \layc{\alpha_c}\,\mathrm{CA}\!\left(\mathrm{LN}(\mathbf{h}),\, \layc{\mathbf{c}}\right) + \mathrm{CA}\!\left(\mathrm{LN}(\mathbf{h}),\, \txtc{\mathbf{y}}\right),
\end{aligned}
\end{equation}

\noindent where $\mathrm{LN}$ is layer normalization and $\mathbf{M}$ a self-attention mask that depends on the stage (described next). The scalar $\layc{\alpha_c}$ sets the layout control strength in each layer: we keep $\layc{\alpha_c} = 1$ during training, and anneal it for the negative samples at inference time (\cref{sec:cfgca} of the supplementary).\\[-1em]

\point{Part-Local and Part-Global Stages} The same block runs in two modes, differing only in the attention mask and in the layout condition we inject (\cref{fig:method_overview}). Local stages process each part on its own, conditioned on its primitive $\layc{\mathbf{c}_i}$, which grounds the part spatially: its geometry is synthesized at the position, scale, and orientation of its target box. Global stages process all parts jointly over $\mathbf{Z} = [\mathbf{z}_1; \ldots; \mathbf{z}_n]$, conditioned on the enclosing primitive $\layc{\mathbf{c}_\star}$, so that the parts fit together into one coherent object. These stages also carry the semantics of the injected text prompt $\txtc{\mathbf{y}}$ across parts, and let the model pick up relations between them, such as symmetries, without any explicit supervision. We interleave the two: local stages keep control fine-grained, while global stages keep the parts mutually consistent.\\[-0.5em]

\point{Training} Our generative model $\mathbf{G}_\theta$ is a conditional flow model trained end-to-end. Let $\layc{\mathbf{C}} = \{\layc{\mathbf{c}_1}, \ldots, \layc{\mathbf{c}_n}, \layc{\mathbf{c}_\star}\}$ denote the layout conditions read by the layout branches, $\txtc{\mathbf{y}}$ the text prompt, and $\mathbf{Z}_t = (1-t)\mathbf{Z}_0 + t\boldsymbol{\epsilon}$ the part latents interpolated toward noise $\boldsymbol{\epsilon} \sim \mathcal{N}(0, \mathbf{I})$. Following rectified flow~\cite{rectifiedflow_liu_2022}, we train the velocity network $\mathbf{v}_\theta$ with
\begin{equation}
\mathcal{L}(\theta) = \mathbb{E}_{(\mathbf{Z}_0,\, \layc{\mathbf{C}},\, \txtc{\mathbf{y}}),\, \boldsymbol{\epsilon},\, t}\left[\left\| \mathbf{v}_\theta\!\left(\mathbf{Z}_t,\, \layc{\mathbf{C}},\, \txtc{\mathbf{y}}\,;\, t\right) - (\boldsymbol{\epsilon} - \mathbf{Z}_0) \right\|_2^2\right],
\end{equation}

\noindent so that each denoising step is conditioned on both the part layout (via $\layc{\mathbf{C}}$) and the prompt (via $\txtc{\mathbf{y}}$). Full training details are in~\cref{sec:training}.\\[-0.5em]

\point{Part-Aware Editing} Our model supports part-level edits at inference time by freezing the latents of untouched parts, which controls \textit{where} an edit applies, and reinjecting cached attention keys and values, which controls \textit{how much} of the edited part's original identity is preserved. We provide full details in~\cref{sec:method-editing}.

\fullfig[0.5em]
{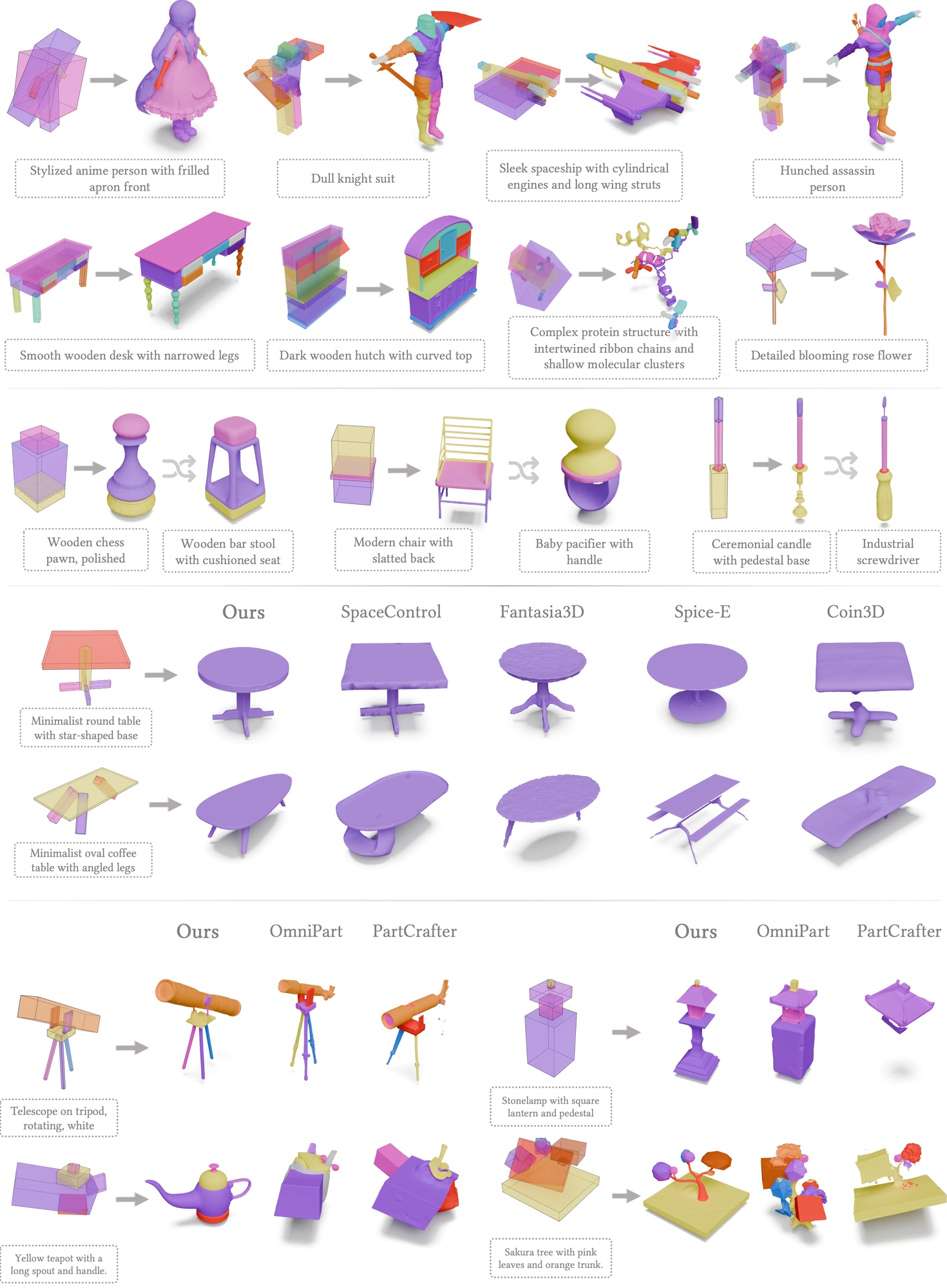}
{Part-Controlled Synthesis.}{
We demonstrate our method's ability to generate diverse shapes that satisfy part layout constraints while matching input text prompts.
We showcase multiple generations for various prompt-layout pairs (first two rows) using our highest-resolution model, followed by variations in text prompts for a fixed layout (third row).
We then compare against proxy-guided methods (rows four and five), which receive the same coarse guidance shape and text prompt as our method, and against image-conditioned part synthesis methods (rows six and seven), which we condition on an image of the target layout synthesized with Nano Banana Pro~\cite{nanobananapro_2025}.
}
{\label{fig:prim-guided-synth}}

\begin{figure*}[p]
    \vspace{-2em}
    \centering
    \includegraphics[width=0.96\textwidth]{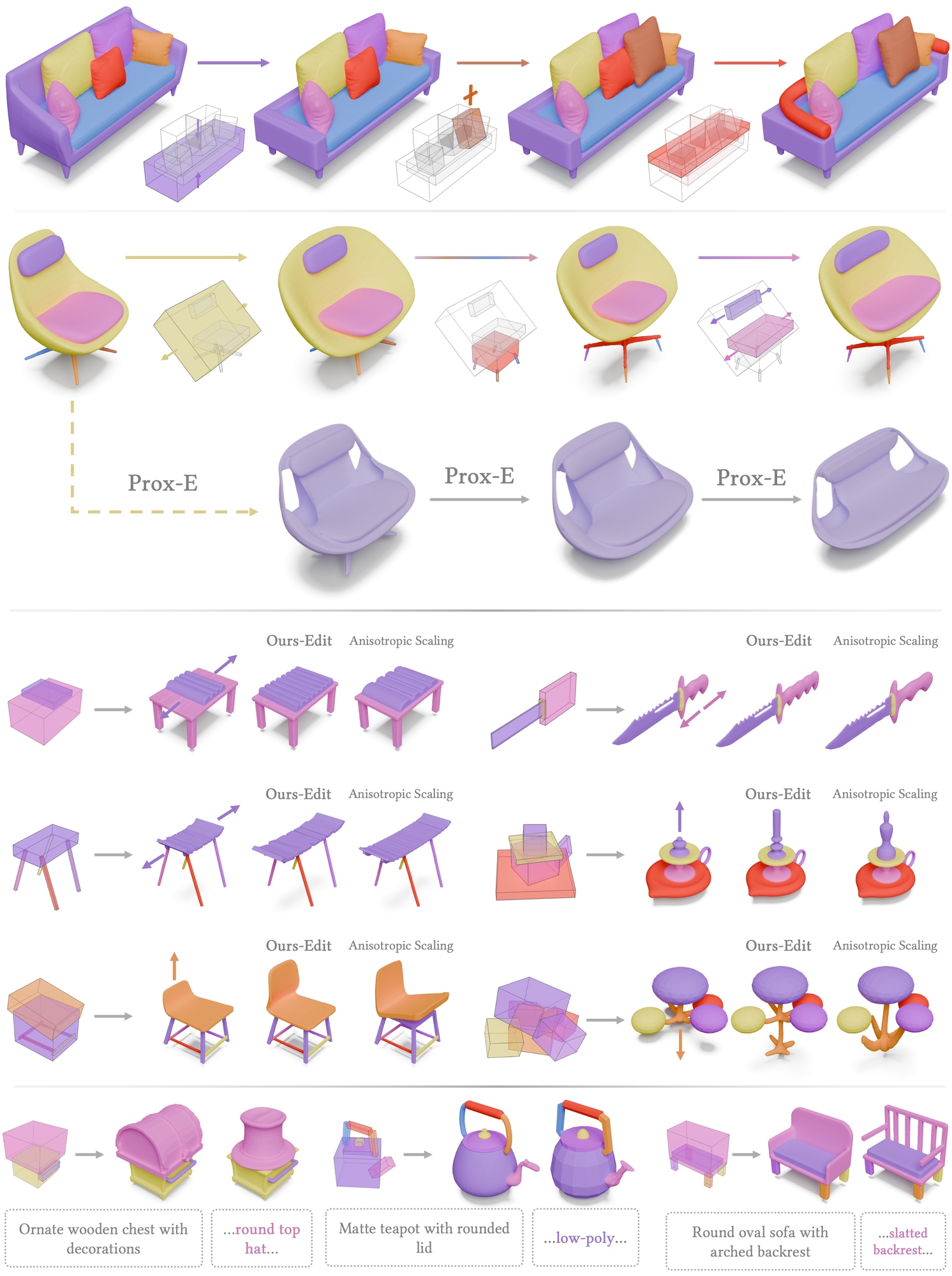}
    \vspace{0.5em}
    \caption{\textbf{Part-Level Editing.}
    We demonstrate our method's versatility through diverse editing operations.
    In the first two rows, we showcase edit sequences starting from a base shape. We iteratively add parts by creating new boxes or substitute them by rescaling without style preservation.
    In the third row, we compare with Prox-E~\cite{proxe_sella_2026} on the same edit chain, using the variant that obtains its primitives by inverting the input layout (Prox-E$_{\textsc{PrimInvert}}$). Since it cannot keep unedited parts fixed, it regenerates the whole object at each step: parts that were not edited change as well, and the shape drifts further from the original along the chain.
    In the following three rows, we show examples of identity-preserving edits, where we rescale parts while preserving their geometric identity, as described in~\cref{sec:method-editing}. We compare our method with Anisotropic Scaling, and provide a quantitative evaluation against Anisotropic Scaling and Part Substitution in~\cref{tab:editing_evaluation} of the supplementary.
    Finally, the last row showcases text-based part editing: modifying the text prompt while keeping the part layout fixed yields localized edits that preserve the shape's overall structure and style, unlike the global text-driven edits shown in the middle of the last row.
    \vspace{-0.5em}
    }
    {\label{fig:part-level-editing}}
\end{figure*}

\begin{table}[t]
    \vspace{-1em}
\centering
\caption{\textbf{Guided Synthesis on Objaverse}. %
        We compare our method with proxy-guided methods --- Fantasia3D~\cite{fantasia3d_chen_2023}, Coin3D~\cite{dong2024coin3d} + ControlNet~\cite{controlnet_zhang_2023}, Spice-E~\cite{spice_sella_2024} and SpaceControl~\cite{spacecontrol_fedele_2026} --- which, except for Spice-E$^\dagger$, support open-vocabulary text-to-3D guided synthesis similarly to ours, and with the image-conditioned part synthesis methods PartCrafter~\cite{PartCrafter_Lin_2025} and OmniPart~\cite{OmniPart_Yang_2025}.
        For image-conditioned part synthesis methods, which do not natively support 3D part layout control, we first render the 3D layout and generate a matching 2D image of the object by prompting Nano Banana Pro~\cite{nanobananapro_2025}, which is then provided as input.
        {\footnotesize
        \textcolor{gray!70!black}{$^\dagger$Spice-E evaluation was computed on a smaller subset containing only table, chair, and airplane categories, since it only supports these classes.}
        }}
\arrayrulecolor{black!25}%
\setlength{\arrayrulewidth}{0.3pt}%
\renewcommand{\arraystretch}{1.05}%
\maxresizebox{\textwidth}{%
\small
\setlength{\tabcolsep}{6pt}%
\begin{tabular}{@{}l l l cc ccc c@{}}
\arrayrulecolor{black}\toprule
& & &
\multicolumn{2}{c}{\textbf{FID} $\downarrow$} &
\multicolumn{3}{c}{\textbf{IoU} $\uparrow$} &
\multirow{2}{*}{\textbf{CLIP}$\uparrow$} \\
\arrayrulecolor{black!25}\cmidrule(lr){4-5} \cmidrule(lr){6-8}
\textbf{Method} & \textbf{Paradigm} & \textbf{Conditioning} &
            Object & Part &
            Object & Voxel & Part & \\
\arrayrulecolor{black}\midrule
\multicolumn{9}{@{}l@{}}{\textit{Proxy-Guided Synthesis}} \\
\arrayrulecolor{black!25}\midrule
Fantasia3D~\cite{fantasia3d_chen_2023}\,%
& SDS~\cite{dreamfusion_poole_2023} & Proxy Mesh & 77.64 & \iconog & 0.259 & 0.164 & \iconog & 27.22 \\
Coin3D~\cite{dong2024coin3d}\,%
& SDS~\cite{dreamfusion_poole_2023} & Proxy Mesh & 95.14 & \iconog & 0.168 & 0.095 & \iconog & 25.10 \\
Spice-E~\cite{spice_sella_2024}\,%
\textcolor{gray}{$^\dagger$}
& Neural Field Diff.~\cite{jun2023shape} & Proxy Mesh & \textcolor{gray}{90.30} & \textcolor{gray}{\iconog} & \textcolor{gray}{0.640} & \textcolor{gray}{0.237} & \textcolor{gray}{\iconog} & \textcolor{gray}{26.40} \\
\arrayrulecolor{black!45}\midrule
SpaceControl~\cite{spacecontrol_fedele_2026}\,%
& Structured Latents~\cite{trellis_xiang_2024} & Superquadrics & 42.96 & \iconog & 0.807 & \textbf{0.776} & \iconog & 26.73 \\

\arrayrulecolor{black}\midrule
\multicolumn{9}{@{}l@{}}{\textit{Image-Conditioned Part Synthesis}} \\
\arrayrulecolor{black!25}\midrule

PartCrafter~\cite{PartCrafter_Lin_2025}\,%
+ Nano Banana~\cite{nanobananapro_2025}\vphantom{\iconog}
& {VecSet~\cite{3DS2VS_zhang_2023}} & Image & 21.10 & 31.54 & 0.408 & 0.203 & 0.148 & 28.81 \\
OmniPart~\cite{OmniPart_Yang_2025}\,%
+ Nano Banana~\cite{nanobananapro_2025}\vphantom{\iconog}
& {Structured Latents~\cite{trellis_xiang_2024}} & Image & 12.73 & 16.38 & 0.513 & 0.269 & 0.262 & 29.64 \\

\arrayrulecolor{black!45}\midrule

\textbf{Ours}\vphantom{\iconog}
& VecSet~\cite{3DS2VS_zhang_2023} & Part Layout (\cref{fig:method_setting}) & \textbf{10.58} & \textbf{9.24} & \textbf{0.812} & 0.550 & \textbf{0.819} & \textbf{31.36} \\

\arrayrulecolor{black}\bottomrule
\end{tabular}%
        }
\arrayrulecolor{black}%
\label{tab:complete_guided_synthesis}
\vspace{-0.5em}
\vspace{1.2em}
\centering
        \vspace{1em}
        \caption{\textbf{GPT-Evaluation of Shape Generation and Editing}. %
                        We compare our method against each baseline by prompting GPT-4o
                        with the target layout, the text prompt, and both outputs, asking which one is better
                        along three axes, with ties allowed; each comparison is evaluated
                        under both A/B orderings to cancel position bias.
                        We report our win rate in \% (higher is better $\uparrow$),
                        with \textbf{Overall} averaging the three axes.
                        \textbf{Quality} is overall realism, \textbf{Layout Adherence} agreement with the
                        target layout, \textbf{Prompt Adherence} conformance to the text instruction, and
                        \textbf{Identity Preservation} fidelity to the style, details and untouched parts
                        of the initial shape, which GPT-4o is additionally shown for editing.
                        }
        \label{tab:gpt_eval}
        \setlength{\tabcolsep}{6pt}
        \renewcommand{\arraystretch}{1.2}
        \maxresizebox{0.90\textwidth}{%
        \small
        \begin{tabular}{lccc|c}
        \arrayrulecolor{black}\toprule
        \multicolumn{5}{@{}l@{}}{\textbf{Controllable Shape Generation}} \\
        \arrayrulecolor{black}\midrule
        Ours vs. & Quality $\uparrow$ & Layout Adherence $\uparrow$ & Prompt Adherence $\uparrow$ & Overall $\uparrow$ \\
        \arrayrulecolor{black!45}\midrule
        \multicolumn{5}{@{}l@{}}{\textit{Proxy-Guided Synthesis}} \\
        \arrayrulecolor{black!25}\midrule
        Fantasia3D~\cite{fantasia3d_chen_2023}   & \wc{90.0}  & \wb{86.7} & \wa{78.3}  & \wb{85.0} \\
        Coin3D~\cite{dong2024coin3d}       & \wc{100.0} & \wc{94.2} & \wc{91.7} & \wc{95.3} \\
        Spice-E~\cite{spice_sella_2024}      & \wc{93.3}  & \wa{68.3} & \wc{90.0}  & \wb{83.9} \\
        SpaceControl~\cite{spacecontrol_fedele_2026} & \wc{90.0}  & \wz{54.2} & \wb{86.7}  & \wa{76.9} \\
        \arrayrulecolor{black!45}\midrule
        \multicolumn{5}{@{}l@{}}{\textit{Image-Conditioned Part Synthesis}} \\
        \arrayrulecolor{black!25}\midrule
        PartCrafter~\cite{PartCrafter_Lin_2025}\,+ Nano Banana~\cite{nanobananapro_2025} & \wc{85.0} & \wb{75.0} & \wb{76.0} & \wb{78.7} \\
        OmniPart~\cite{OmniPart_Yang_2025}\,+ Nano Banana~\cite{nanobananapro_2025} & \wa{60.5} & \wz{54.0} & \wa{59.5} & \wz{58.0} \\
        \arrayrulecolor{black}\midrule
        \addlinespace[0.35em]
        \multicolumn{5}{@{}l@{}}{\textbf{Shape Editing}} \\
        \arrayrulecolor{black}\midrule
        Ours vs. & Quality $\uparrow$ & Layout Adherence $\uparrow$ & Identity Preservation $\uparrow$ & Overall $\uparrow$ \\
        \arrayrulecolor{black!25}\midrule
        Prox-E$_{\textsc{PrimInvert}}$~\cite{proxe_sella_2026}       & \wb{87.0} & \wa{65.2} & \wc{91.3} & \wb{81.2} \\
        Prox-E$_{\textsc{SuperDec}}$~\cite{proxe_sella_2026} & \wc{91.3} & \wa{67.4} & \wc{93.5} & \wb{84.1} \\
        \arrayrulecolor{black}\bottomrule
        \end{tabular}%
}

\end{table}

\section{Experiments}
\label{sec:experiments}
We evaluate our proposed method on 1,000 part layout and prompt pairs unseen during training, extracted from the Objaverse~\cite{objaverse_deitke_2023} dataset, following the preprocessing pipeline described in~\cref{sec:dataset_creation}.

\subsection{Compositional Synthesis and Editing}

\point{Synthesis} In~\cref{fig:prim-guided-synth}, we show diverse shape generations for given target layouts and text prompts, where we demonstrate strong alignment between our generated shapes and the provided layout constraints and prompts. As shown in the top rows of~\cref{fig:prim-guided-synth}, our method successfully generates varied objects, such that synthesized parts clearly correspond to the input bounding boxes and the overall structure matches the user's intent.\\[-0.75em]

\point{Editing} In~\cref{fig:part-level-editing}, we demonstrate our method's versatility through diverse editing operations. In the first two rows, we showcase edit sequences starting from a base shape: we iteratively add parts by creating boxes or substitute them by rescaling without style preservation. In the third row, we compare with Prox-E~\cite{proxe_sella_2026} on the same edit chain. Since it cannot keep unedited parts fixed, it regenerates the whole object at each step, so the parts are no longer separated and the shape drifts from the original. In rows four to six, we show examples of \textit{identity-preserving edits}, where we rescale parts while preserving their geometric identity, as described in~\cref{sec:method-editing}. We compare our method with anisotropic scaling, and provide a quantitative evaluation against anisotropic scaling and part substitution in~\cref{tab:editing_evaluation} of the supplementary. Finally, the last row showcases \textit{text-based part editing}: modifying the text prompt while keeping the part layout fixed yields localized edits that preserve the shape's overall structure and style, unlike the global text-driven edits shown in the middle of the last row.

\newpage \subsection{Comparisons}
\point{Metrics} We evaluate generation quality and controllability using several metrics. To measure visual quality, we compute the Fréchet Inception Distance (FID)~\cite{gans_heusel_2017, inception_szegedy_2015} on rendered normal maps at both the whole-object level (\textbf{Object-FID}) and for individual, normalized parts (\textbf{Part-FID}). Spatial alignment is quantified by Intersection-over-Union (IoU)~\cite{minobb_jylanki_2015}. \textbf{Object-IoU} measures the overlap between the generated object's bounding box and the union of all control boxes. We also report \textbf{Voxel-IoU} on a $64^3$ grid, which is useful for evaluating baselines that do not produce part segmentations, and \textbf{Part-IoU}, which averages the bounding-box overlap at the individual part level. Finally, to measure text-shape alignment, we compute the CLIP-Score~\cite{clip_radford_2021, clipscore_hessel_2021} between the input prompt and rendered views.\\[-1em]

\point{Baselines} To our knowledge, there is no prior work that accepts granular part-based oriented controls in combination with text prompts as we do. We therefore compare against two families of synthesis methods, reported as separate blocks in~\cref{tab:complete_guided_synthesis}, and against a dedicated editing baseline.

\textit{Proxy-guided synthesis} methods accept coarse, non-part-separated shape guidance: Spice-E~\cite{spice_sella_2024} (diffusion structural guidance), Fantasia3D~\cite{fantasia3d_chen_2023} (SDS initialization), and Coin3D~\cite{dong2024coin3d} combined with ControlNet~\cite{controlnet_zhang_2023} (proxy mesh conditioning). We additionally compare to SpaceControl~\cite{spacecontrol_fedele_2026}, which accepts superquadric conditioning by instantiating one superquadric per control box.

\textit{Image-conditioned part synthesis} methods decompose a single input image into separate 3D parts: PartCrafter~\cite{PartCrafter_Lin_2025} and OmniPart~\cite{OmniPart_Yang_2025}. Unlike our method, these approaches take no 3D layout control, so we must first express the target layout in the only modality they accept, an image synthesized with Nano Banana Pro~\cite{nanobananapro_2025}. This comparison quantifies the cost of routing 3D structural intent through a 2D image: the conditioning image is an imperfect and ambiguous proxy for the layout, and any structure it fails to convey --- in particular occluded parts and precise part extents --- cannot be recovered downstream.

\textit{Editing baselines.} For part-level editing, we compare against Prox-E~\cite{proxe_sella_2026}, which edits shapes through coarse primitive-based abstractions, in two variants that differ only in how its primitives are obtained: Prox-E$_{\textsc{PrimInvert}}$ isolates the editing mechanism under matched control, while Prox-E$_{\textsc{SuperDec}}$ reflects how the method is used in practice. We report both comparisons in the editing block of~\cref{tab:gpt_eval}.

We detail the evaluation protocol we follow for every baseline --- metric coverage, layout-to-image conditioning, part matching, and the two Prox-E variants --- in~\cref{sec:baseline_protocol} of the supplementary.\\[-0.5em]

As shown in~\cref{tab:complete_guided_synthesis}, our method achieves the best Object-FID (10.58), Part-FID (9.24), Object-IoU (0.812), Part-IoU (0.819) and CLIP-Score (31.36), losing only Voxel-IoU to SpaceControl (0.776), which densely fills its superquadric proxies rather than synthesizing meaningful part geometry. The image-conditioned part synthesis methods fail in the opposite way: strong image priors give them competitive visual quality (12.73 Object-FID for OmniPart), but they follow the requested structure poorly, reaching at most 0.513 Object-IoU and 0.262 Part-IoU against our 0.812 and 0.819. This gap is a direct consequence of routing 3D intent through a 2D image, and controlling parts natively in 3D, as we do, removes this bottleneck.\\[-0.5em]

\point{GPT Evaluation} To complement our quantitative metrics with a holistic perceptual assessment, we conduct an automated evaluation using GPT-4o~\cite{gpt4o_hurst_2024} as a judge. We report both evaluations in~\cref{tab:gpt_eval}, and describe the full protocol, prompts, and sample grids in~\cref{sec:gpt4_evaluation}.

For generation, our method is preferred over every baseline on every axis, with Overall win rates from 95.3\% against Coin3D down to 58.0\% against OmniPart, where Layout Adherence is the tightest axis (54.0\%). For editing, our edits win 81.2\% and 84.1\% overall against the two Prox-E variants, with the largest gap on Identity Preservation (91.3\% and 93.5\%); the advantage holds even under matched control, so it comes from our editing mechanism rather than from the control signal.\\[-0.5em]

\point{Inference Time} Our method offers competitive inference time compared to prior work, scaling gracefully with the number of parts $N_p$; we report a full comparison against baselines in~\cref{sec:inference_time} of the supplementary.

\section{Limitations}
Our method has several limitations that motivate future work. First, slight artifacts can appear in synthesized parts due to the out-of-distribution nature of part segments for the VAE, requiring a filtering stage; fine-tuning the VAE on part-decomposed data could help.
Second, the method is capped at 16 parts per shape, but can be trained on higher part counts by leveraging more memory-efficient attention schemes~\cite{linearattention_katharopoulos_2020, mamba_gu_2023, sanavideo_chen_2026}.
Finally, physical stability is not guaranteed: missing contacts or asymmetries between parts can compromise structural integrity, which could be addressed by fine-tuning against a stability reward from a physical simulator~\cite{GPLD3D_dong2024, brickgpt_pun2025}.

\section{Conclusion}

We introduced \paper, the first method to synthesize part-separated 3D shapes from coarse oriented part layouts, thereby enabling compositional editing of individual object parts.
Our method relies on a conditioning technique that ensures strong adherence to input part layouts, a diffusion transformer architecture that fuses part conditioning with global context, and a data processing pipeline that enables training on large 3D shape datasets without manual part annotations.
It supports a versatile set of editing capabilities, including substitution, addition, and style-preserving resizing of individual parts, and outperforms existing approaches as measured by objective metrics and LLM-based evaluations.
In the future, richer structured guidance, such as proxy meshes with animation rigs and physical affordances, could enable the synthesis of photorealistic, simulation-ready 3D assets. Beyond 3D, similar proxies inferred from videos could guide the generation of physically grounded video variations.

\bibliography{citations}
\bibliographystyle{iclr2027_conference}

\onecolumn
\clearpage
\setcounter{section}{0}
\appendix

\begin{center}
\Large{CompoSE: Supplementary Material}
\end{center}

\vspace{0.5em}

\renewcommand{\toclinesep}{0.55em plus 1fill}

\tocline{release_statement}{A}{Release Statement}{18}
\tocline{llm_disclosure}{B}{LLM Use Disclosure}{18}
\tocline{ui_interface}{C}{User Interface}{18}
\tocline{method-editing}{D}{Part-Aware Editing}{19}
\tocline{eval_details}{E}{Evaluation Details}{19}
\tocsubline{baseline_protocol}{E.1}{Baselines}{19}
\tocsubline{gpt4_evaluation}{E.2}{GPT-4o Evaluation}{20}
\tocline{additional_samples}{F}{Additional Samples}{22}
\tocsubline{additional_edits}{F.1}{Edit Chains and Identity-Preserving Edits}{22}
\tocsubline{additional_part_synthesis_results}{F.2}{Controllable Part Synthesis}{23}
\tocsubline{additional_control_comparisons}{F.3}{Comparison with Controllable Baselines}{28}
\tocsubline{additional_part_synthesis_comparisons}{F.4}{Comparison with Part Synthesis Baselines}{30}
\tocline{failure_cases}{G}{Failure Cases}{42}
\tocline{impl_details}{H}{Implementation Details}{43}
\tocsubline{architecture_details}{H.1}{Architecture Details}{43}
\tocsubline{cfgca}{H.2}{CFG Control Annealing}{44}
\tocsubline{layout_optimization}{H.3}{Layout Optimization and Artifact Filtering}{46}
\tocsubline{training}{H.4}{Training Details}{48}
\tocline{dataset}{I}{Dataset}{49}
\tocsubline{dataset_creation}{I.1}{Dataset Creation}{49}
\tocsubline{dataset_stats}{I.2}{Statistics}{51}
\tocline{additional_ablations}{J}{Additional Ablations}{52}
\tocsubline{cfg_scale_ablation}{J.1}{SpaceControl Classifier-Free Guidance Scale}{52}
\tocsubline{identity_preservation_ablation}{J.2}{Identity-Preservation under Part Rescaling}{52}
\tocsubline{model_ablation_study}{J.3}{Model Ablation Study}{53}
\tocsubline{inference_time}{J.4}{Inference Time}{54}

\begin{center}
    \vspace{-0.5em}
    \rule{0.6\textwidth}{0.4pt}
    \vspace{0.5em}
\end{center}
\clearpage

\FloatBarrier
\section{Release Statement}
\label{sec:release_statement}
Our code and dataset will be made publicly available upon publication.
\vspace{0.5em}

\FloatBarrier
\section{LLM Use Disclosure}
\label{sec:llm_disclosure}
We used large language models to aid and polish the writing of this paper, and a vision-language model (InternVL3-8B~\cite{chen2024internvl}) to generate the text captions of our synthetic training dataset, as described in~\cref{sec:dataset_creation}.
\vspace{0.5em}

\FloatBarrier
\section{User Interface}
\label{sec:ui_interface}

\begin{figure}[H]
    \vspace{-0.5em}
    \centering
    \includegraphics[width=0.75\textwidth]{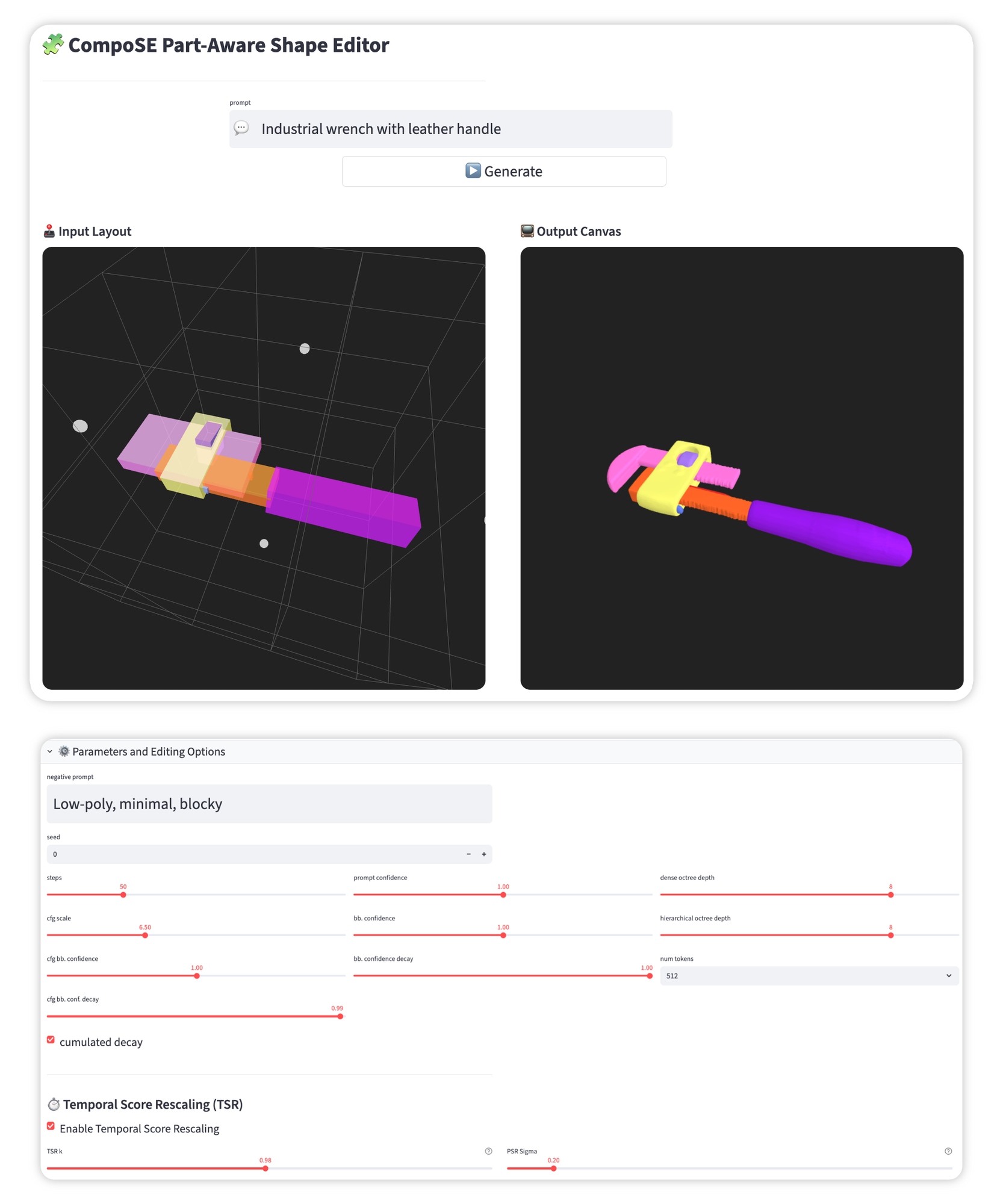}
    \caption{\textbf{CompoSE User Interface.}
    Our user interface allows users to input text prompts and define part layouts using bounding boxes.
    We show the layout editor and the output visualizer (top), and the parameter panel (bottom) for fine-tuning generation settings. 
    }%
    \label{fig:compose_ui}
\end{figure}

For the purposes of easily interacting with our method, we have developed a simple user interface (UI) that allows users to input text prompts and define part layouts by manipulating boxes, using the Streamlit framework\footnote{\url{https://streamlit.io/}}. The UI provides direct feedback on the generated 3D shapes, enabling users to visualize and refine their designs interactively.
We show the interface we have developed in \cref{fig:compose_ui}.\\

\FloatBarrier
\clearpage
\section{Part-Aware Editing}
\label{sec:method-editing}

When performing part-aware editing operations (see \cref{fig:part-level-editing}), we freeze part latents to control which parts are updated and which parts are preserved.
Specifically, when a user edits only a subset of parts, we freeze the latents of all other parts by restoring their latents from the \emph{reference shape} (i.e., the shape that has been synthesized by our model prior to any editing operations) at every diffusion time-step. This technique effectively locks the unedited parts in place, in the sense that they contribute to global consistency of the edited shapes, but do not themselves change their geometry or semantics.

In cases where the user wants to edit the geometry of a part while preserving its style and overall appearance, such as rescaling a part (\cref{fig:part-level-editing} rows 4--6), latent freezing is not sufficient.
To enable this type of appearance-preserving edit, we use a \emph{KV-reinjection} technique~\cite{loosecontrol_bhat2024,stableflow_avrahami2025,text2videozero_khachatryan2023}.
Specifically, during our model's forward pass for the reference shape, we cache the attention keys and values produced in the input attention layer and the text attention layer of every DiT block, and at every diffusion time-step.
Then, during editing, we recompute keys and values from the modified latents, but we blend them with the cached keys and values in a time-step-dependent manner. This weighted reinjection anchors the part's visual identity, ensuring that changes to the part's corresponding geometric primitive (e.g., rescaling) do not cause the appearance of the part to drift (\cref{fig:part-level-editing} rows 4--6).
As shown in \cref{fig:part-level-editing}, our method yields rescaled parts whose style, surface detail, and geometric character remain faithful to the reference shape.

Together, part-latent freezing and KV-reinjection provide a unified and flexible editing framework. Latent freezing controls \textit{where} the edit occurs, while KV-reinjection controls \textit{how much} of the original identity should be preserved. This combination enables expressive operations: adding parts, substituting them through text prompts, or performing identity-preserving geometric adjustments, all while maintaining global coherence and respecting the user-specified layout.

\newtcolorbox{promptboxnobreak}[1][]{%
  enhanced,
  unbreakable,
  colback=promptbgcolor,
  colframe=promptbordercolor,
  colbacktitle=prompttitlecolor,
  coltitle=white,
  fonttitle=\bfseries\sffamily,
  title=#1,
  boxrule=0.8pt,
  arc=3pt,
  left=8pt,
  right=8pt,
  top=10pt,
  bottom=10pt,
  attach boxed title to top left={yshift=-2mm, xshift=4mm},
  boxed title style={
    colframe=promptbordercolor,
    boxrule=0.6pt,
    arc=2pt,
  },
}

\FloatBarrier
\section{Evaluation Details}
\label{sec:eval_details}

\FloatBarrier
\subsection{Baselines}
\label{sec:baseline_protocol}

Here we detail the evaluation protocol we follow for each family of baselines.\\[-0.5em]

\point{Proxy-Guided Synthesis} Since Fantasia3D, Coin3D, and Spice-E do not synthesize decomposed parts, we only evaluate object-level metrics for these baselines; additionally, since Spice-E is fine-tuned only on table, chair, and airplane categories, we evaluate it on a filtered subset of these categories. Coin3D fails on roughly 10\% of test instances despite using the hyperparameters recommended by its authors; these failures are excluded from its reported metrics. For SpaceControl, we report a single guidance scale in~\cref{tab:complete_guided_synthesis} of the main paper, and provide a full sweep over this parameter, as well as a variant augmented with image conditioning via ControlNet~\cite{controlnet_zhang_2023}, as an ablation in~\cref{sec:cfg_scale_ablation}.\\[-0.5em]

\point{Image-Conditioned Part Synthesis} Since PartCrafter and OmniPart accept no 3D layout control, we must express the target layout in the only modality they accept: we render each test layout from a canonical viewpoint and prompt Nano Banana Pro~\cite{nanobananapro_2025} to synthesize an image of the object that follows both the rendered part arrangement and the text prompt, which we then provide as input. Because these methods output part-separated geometry, we evaluate them on part-level metrics as well. For fairness, we do not penalize them for producing a different part count than requested: we establish a one-to-one correspondence between generated and ground-truth parts via Hungarian matching, and average part-level metrics over the matched pairs only. We further score each generated asset under 25 canonical orientations and retain the best-scoring one, so that no method is penalized for a global rigid misalignment.\\[-0.5em]

\point{Editing Baselines} Both Prox-E~\cite{proxe_sella_2026} baselines run the method end to end, and differ only in how its primitives are obtained: Prox-E$_{\textsc{PrimInvert}}$ inverts the input primitive layout into the method's proxy representation, while Prox-E$_{\textsc{SuperDec}}$ follows the method's standard route and recovers primitives from the input shape through superquadric decomposition. In both cases the edit intent is expressed on those primitives and the edit itself is performed entirely by Prox-E.

\FloatBarrier
\subsection{GPT-4o Evaluation}
\label{sec:gpt4_evaluation}

Our use of a multimodal model as a judge is motivated by prior work demonstrating that such models are scalable, human-aligned evaluators for 3D~\cite{wu2024gpteval3d,maiti2025gen3deval} and vision-language generation~\cite{zhang2023gpt4v}. Rather than ranking all methods jointly, we run pairwise comparisons: for each test instance and each baseline, we compose a grid image showing the target layout alongside our output and the baseline's, and ask GPT-4o~\cite{gpt4o_hurst_2024} which of the two is better along three axes, allowing ties. For generation, the axes are \textbf{Quality} (overall realism), \textbf{Layout Adherence} (agreement with the target layout), and \textbf{Prompt Adherence} (conformance to the text instruction); for editing, the grid additionally shows the initial shape, and Prompt Adherence is replaced by \textbf{Identity Preservation} (fidelity to the style, details, and untouched parts of the initial shape). Every comparison is evaluated under both A/B orderings so that any positional bias cancels out, and we report our win rate in \%, counting each tie as half a win (i.e., $(\text{wins} + 0.5 \times \text{ties}) / \text{comparisons}$), with \textbf{Overall} averaging the three axes. For Layout Adherence, we explicitly instruct GPT-4o to treat the input boxes as loose bounding volumes rather than target shapes. Without this instruction, we found that the judge tends to reward outputs that simply fill each box with a block-like shape, which trivially maximizes spatial overlap while producing implausible geometry; the instruction aligns the criterion with the intended use of layouts as coarse guidance, consistent with our IoU-based metrics which measure bounding-box overlap rather than volumetric filling. The results are given in~\cref{tab:gpt_eval} of the main paper.

\begin{figure*}[h]
    \vspace{-0.5em}
    \centering
    \includegraphics[width=0.61\textwidth]{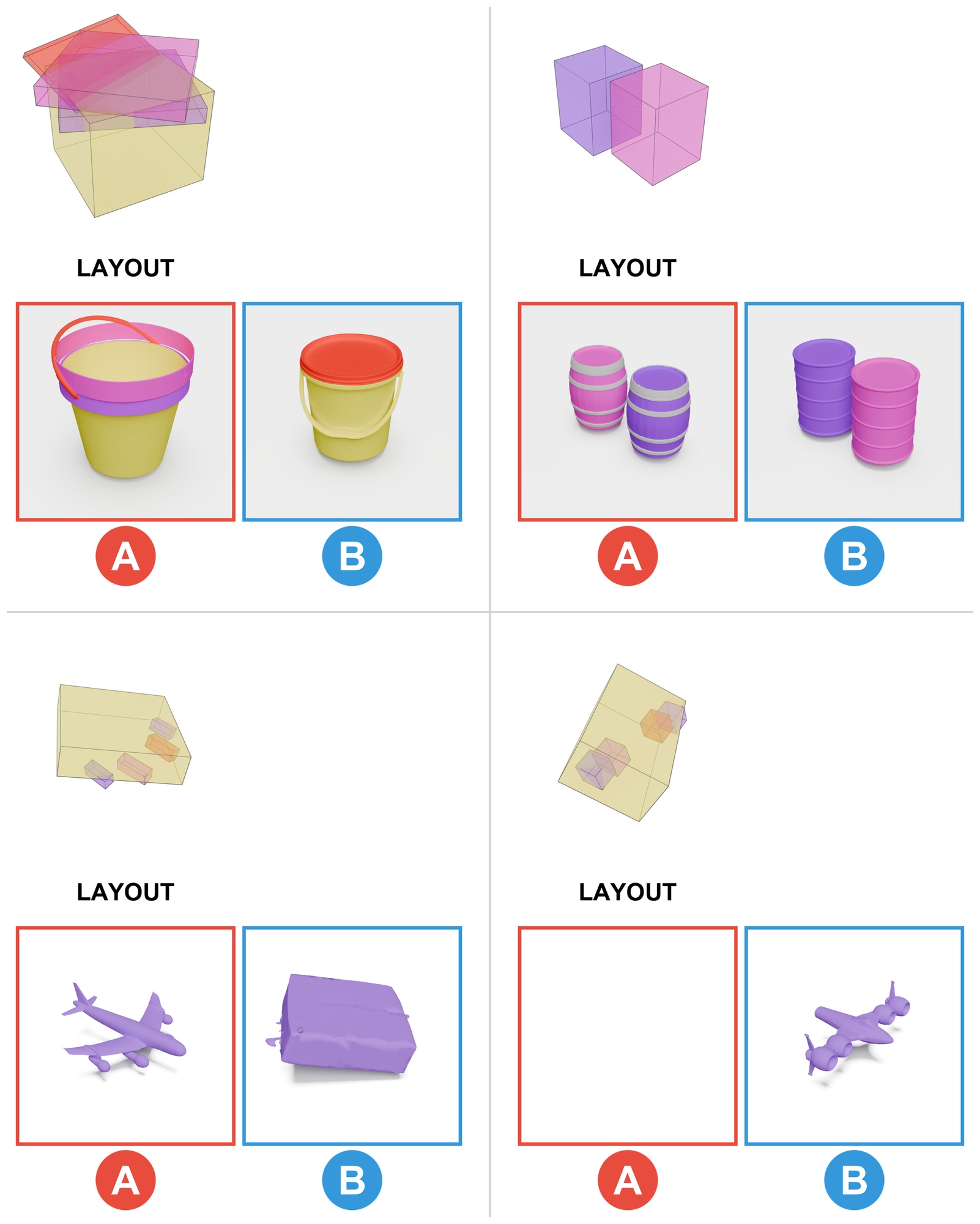}
    \caption{\textbf{Generated Grids for GPT-4o Pairwise Comparison.}
    We show samples of the image grids shown to GPT-4o for evaluating controllable generation, which contain the target layout (top), and the shapes generated by our method and one baseline (bottom).
    Each sample is evaluated under both A/B orderings of the two methods, so that any bias towards a spatial location cancels out.
    When a method fails (bottom-right sample, tile A), we replace the corresponding image with a white tile.
    }%
    \label{fig:fig_gpt_grids}
\end{figure*}

\point{Controllable Generation Evaluation}
We provide the complete prompt used for our GPT-4o controllable generation evaluation in \cref{fig:gpt4_eval_prompt}, alongside samples of the image grids shown to GPT-4o for evaluation in \cref{fig:fig_gpt_grids}.

\vfill
\vfill
\newpage

\begin{figure*}[h]
  \vspace{0.5em}
  \centering
  \begin{minipage}{\textwidth}
  \begin{promptboxnobreak}[GPT-4o Controllable Generation Evaluation Prompt]
  \scriptsize\ttfamily
  This image shows a 3D shape generation task.\\[2pt]
  Top-left (LAYOUT): the target bounding-box layout of the shape's parts, given as generation guidance.\\[2pt]
  Bottom row: two 3D shapes generated by two different methods, labeled A and B.\\[2pt]
  The intended text description of the shape is: \texttt{"\{text description\}"}\\[2pt]
  Compare A and B. For each question, answer \texttt{"A"}, \texttt{"B"}, or \texttt{"tie"} if neither is clearly better.\\[2pt]
  Ignore differences in color and rendering style.\\[6pt]
  \textbf{Q1 (Overall Quality):} Which generated shape has better overall quality and realism?\\[2pt]
  \textbf{Q2 (Layout Adherence):} Which generated shape better matches the target layout? Each part (identified by its color-coded bounding box) should be in the position, scale, and orientation shown in the layout. The boxes are only loose bounding volumes for each part, not target shapes to fill --- a part that is stretched or blocked out into a rectangular shape merely to fill its box completely should NOT be rewarded for this; correct part geometry within a roughly matching box is what matters. A shape that is completely distorted (broken, melted, missing parts, or no longer recognizable as the object) must always be judged worse, even if its rough proportions match the layout.\\[2pt]
  \textbf{Q3 (Prompt Adherence):} Which generated shape better conforms to the text description above?\\[6pt]
  Reply ONLY with a JSON object of the form\\
  \texttt{\{"1": "A", "2": "tie", "3": "B"\}},\\
  with no extra text, no markdown, no code fences.
  \end{promptboxnobreak}
  \end{minipage}
  \caption{Prompt provided to GPT-4o for comparing two generated 3D shapes against a target layout and text description. GPT-4o is asked to choose the better shape (or a tie) along three axes: overall quality (Q1), layout adherence (Q2), and prompt adherence (Q3). The model is constrained to respond with a strict JSON object containing one verdict per question.}
  \label{fig:gpt4_eval_prompt}
\end{figure*}

\FloatBarrier
\point{Shape Editing Evaluation}
For editing, we manually collect editing examples that each involve one to two parts, resulting in 200 evaluations in total. We provide the complete prompt used for our GPT-4o shape editing evaluation in \cref{fig:gpt4_edit_eval_prompt}.

\begin{figure*}[h]
  \vspace{0.5em}
  \centering
  \begin{minipage}{\textwidth}
  \begin{promptboxnobreak}[GPT-4o Shape Editing Evaluation Prompt]
  \scriptsize\ttfamily
  This image shows a 3D shape editing task.\\[2pt]
  Top-left (INITIAL): the initial 3D shape before editing.\\[2pt]
  Top-right (LAYOUT): the edit instruction, shown as the target bounding-box layout of the shape's parts; arrows indicate how parts should be moved or resized.\\[2pt]
  Bottom row: two edited 3D shapes produced by two different methods, labeled A and B.\\[2pt]
  Compare A and B. For each question, answer \texttt{"A"}, \texttt{"B"}, or \texttt{"tie"} if neither is clearly better.\\[2pt]
  Ignore differences in color and rendering style.\\[6pt]
  \textbf{Q1 (Overall Quality):} Which edited shape has better overall quality and realism?\\[2pt]
  \textbf{Q2 (Layout Adherence):} Which edited shape better matches the edit layout? A shape that is completely distorted (broken, melted, missing parts, or no longer recognizable as the object) must always be judged worse, even if its rough proportions match the layout.\\[2pt]
  \textbf{Q3 (Identity Preservation):} Which edited shape better preserves the identity of the initial shape (its style, details, and the parts not targeted by the edit)?\\[6pt]
  Reply ONLY with a JSON object of the form\\
  \texttt{\{"1": "A", "2": "tie", "3": "B"\}},\\
  with no extra text, no markdown, no code fences.
  \end{promptboxnobreak}
  \end{minipage}
  \caption{Prompt provided to GPT-4o for comparing two 3D shape edits against the initial shape and the target layout. GPT-4o is asked to choose the better edit (or a tie) along three axes: overall quality (Q1), layout adherence (Q2), and identity preservation (Q3). The model is constrained to respond with a strict JSON object containing one verdict per question.}
  \label{fig:gpt4_edit_eval_prompt}
\end{figure*}

\FloatBarrier
\clearpage
\section{Additional Samples}
\label{sec:additional_samples}

\FloatBarrier
\subsection{Edit Chains and Identity-Preserving Edits}
\label{sec:additional_edits}

We show additional results for our method's editing capabilities. The top two rows illustrate chained edits, where a sequence of localized part-level operations (e.g., substitution and resizing) is applied successively to the same base shape, with each step's target part highlighted alongside the resulting edit; the shape's identity and previously edited parts are preserved throughout the chain. The bottom three rows compare our identity-preserving part rescaling against Anisotropic Scaling and Part Substitution, with a zoomed-in inset highlighting how well each method preserves the original part's fine surface detail after resizing.

\vfill
\begin{figure*}[h]
    \vspace{0.5em}
    \centering
    \includegraphics[width=0.95\textwidth]{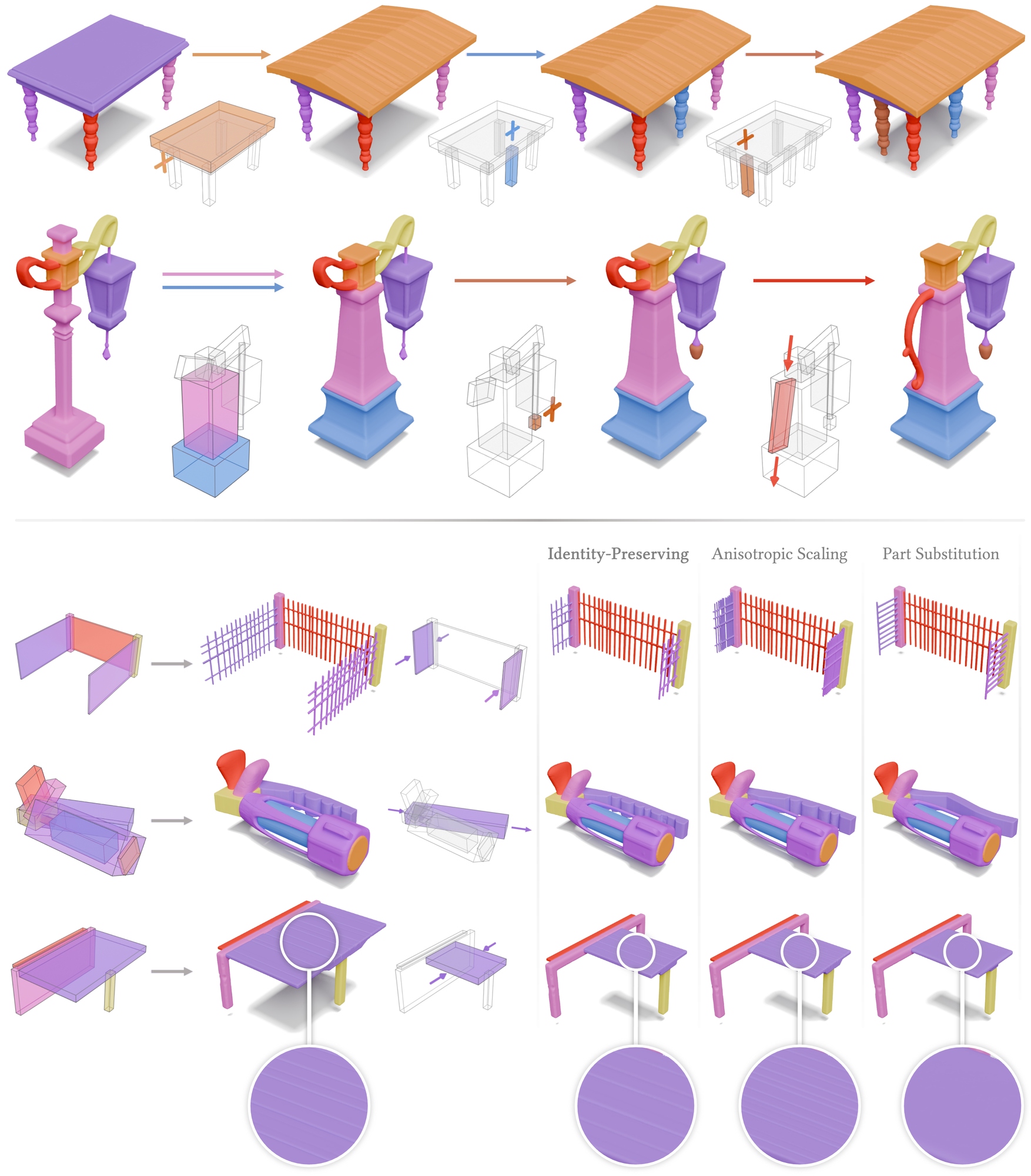}
    \label{fig:extra_edits}
\end{figure*}
\vfill
 
\FloatBarrier
\clearpage
\subsection{Controllable Part Synthesis}
\label{sec:additional_part_synthesis_results}

We provide additional qualitative results of our method's controllable part synthesis capabilities, sampling shapes with varying numbers of parts and levels of geometric detail from the input layout.

\FloatBarrier
\subsubsection{16-Part Generation at High Resolution (2048 Tokens)}
\label{sec:additional_part_synthesis_16p}

We show additional results generated with $D=2048$ tokens per part, using input layouts composed of up to 16 parts. This higher-token-resolution variant is able to capture finer geometric detail across a larger number of parts per shape.

\vspace{1em}

{\centering{\includegraphics[width=0.98\textwidth]{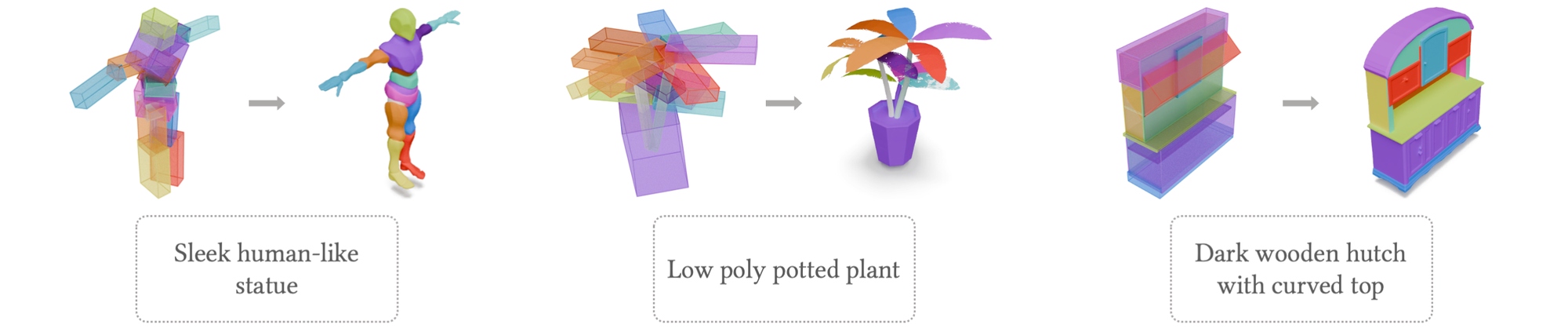}\par\vspace{0.3em}}}
{\centering{\includegraphics[width=0.98\textwidth]{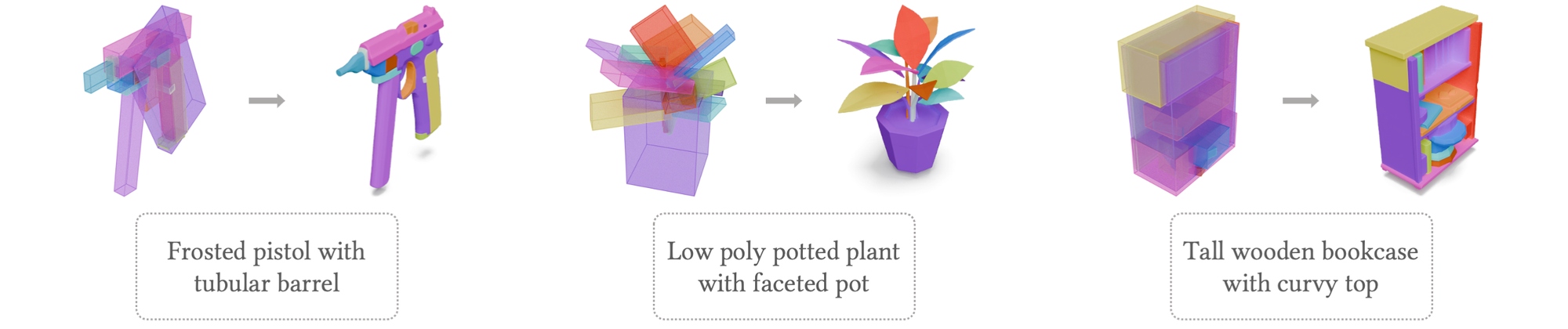}\par\vspace{0.3em}}}
{\centering{\includegraphics[width=0.98\textwidth]{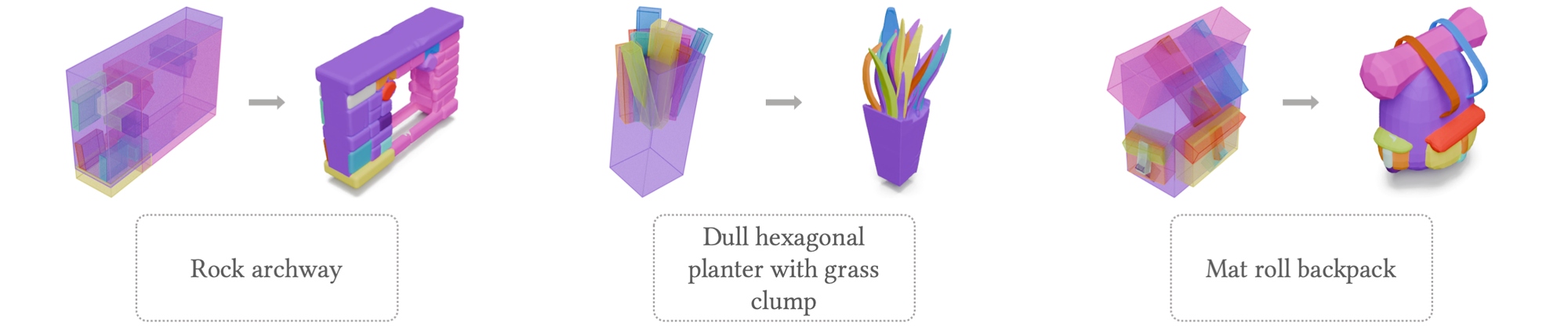}\par\vspace{0.3em}}}
{\centering{\includegraphics[width=0.98\textwidth]{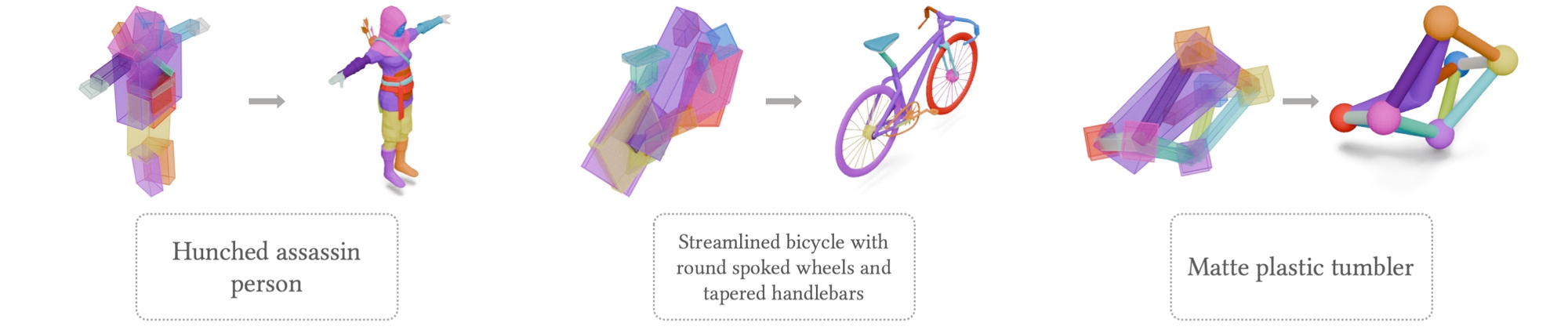}\par\vspace{0.3em}}}
{\centering{\includegraphics[width=0.98\textwidth]{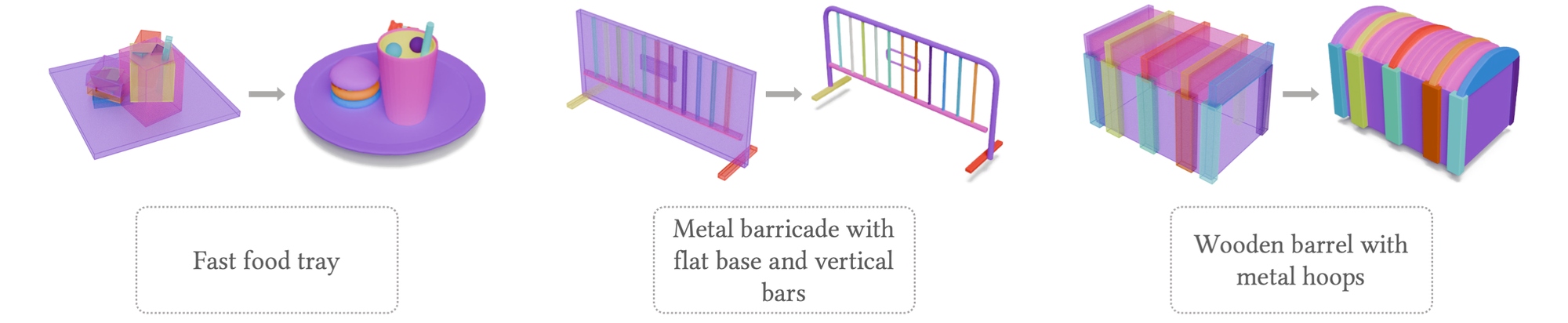}\par\vspace{0.3em}}}
{\centering{\includegraphics[width=0.98\textwidth]{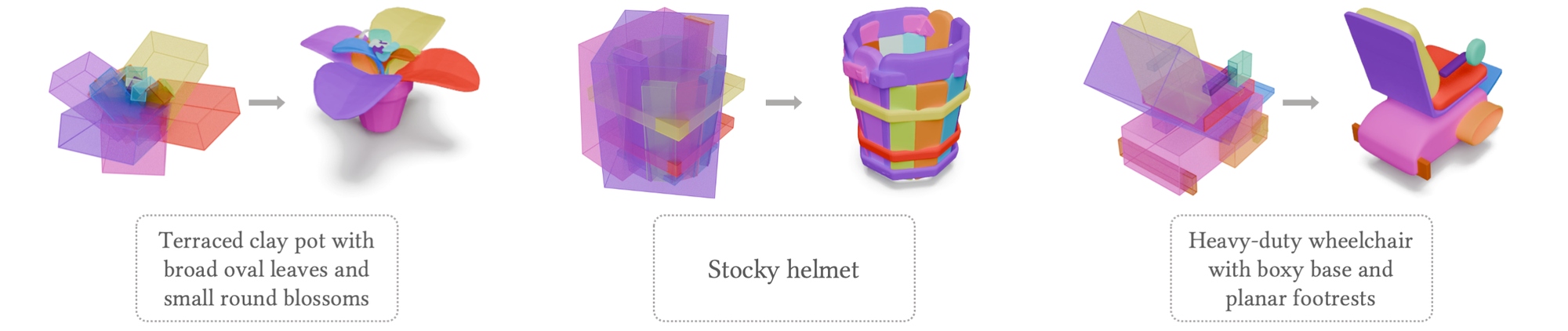}\par\vspace{0.3em}}}
{\centering{\includegraphics[width=0.98\textwidth]{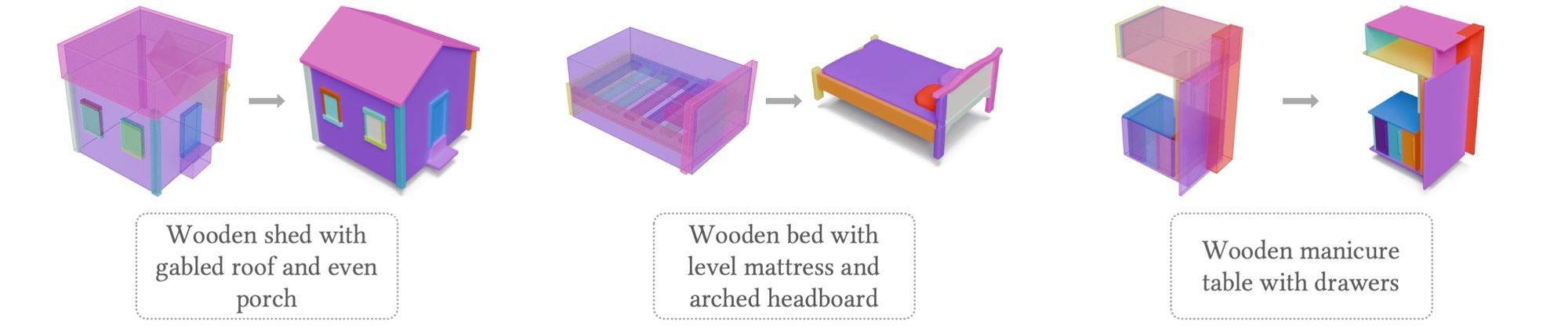}\par\vspace{0.3em}}}
{\centering{\includegraphics[width=0.98\textwidth]{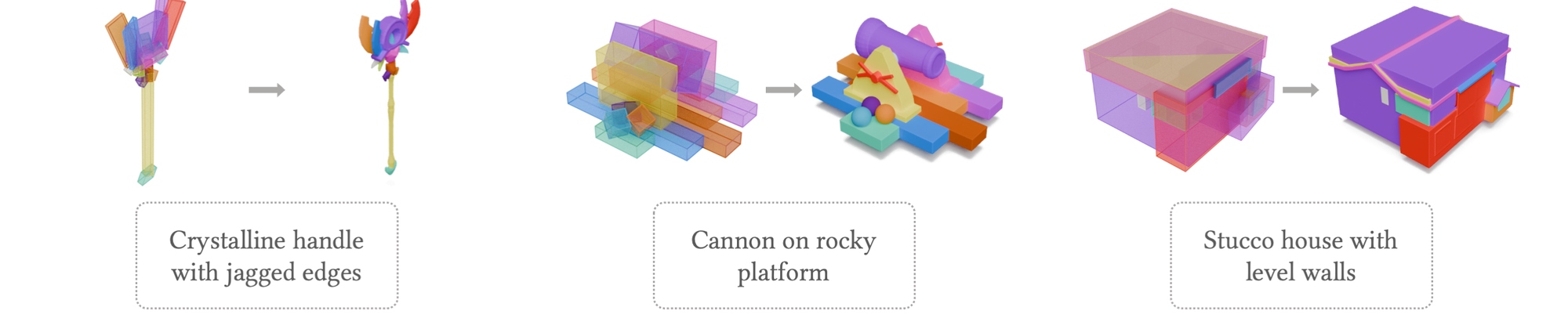}\par\vspace{0.3em}}}
{\centering{\includegraphics[width=0.98\textwidth]{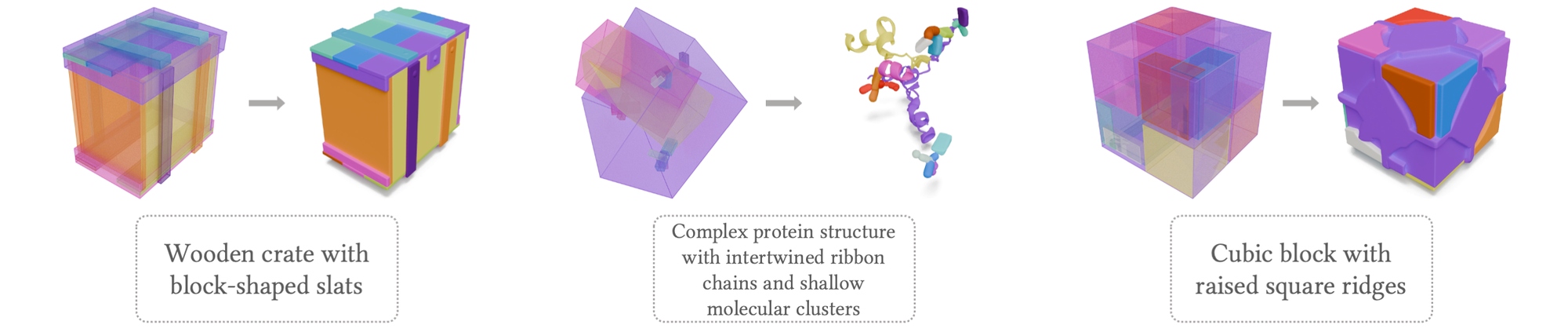}\par\vspace{0.3em}}}
{\centering{\includegraphics[width=0.98\textwidth]{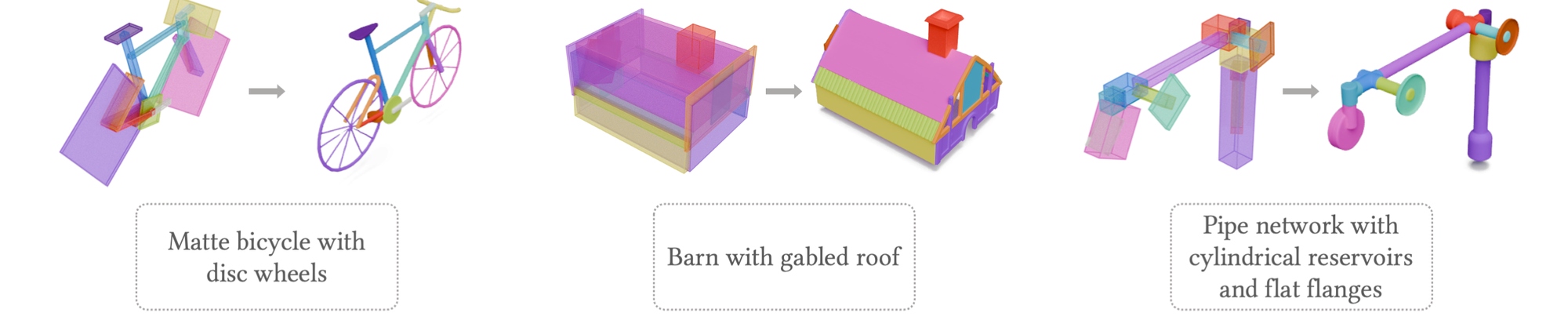}\par\vspace{0.3em}}}
{\centering{\includegraphics[width=0.98\textwidth]{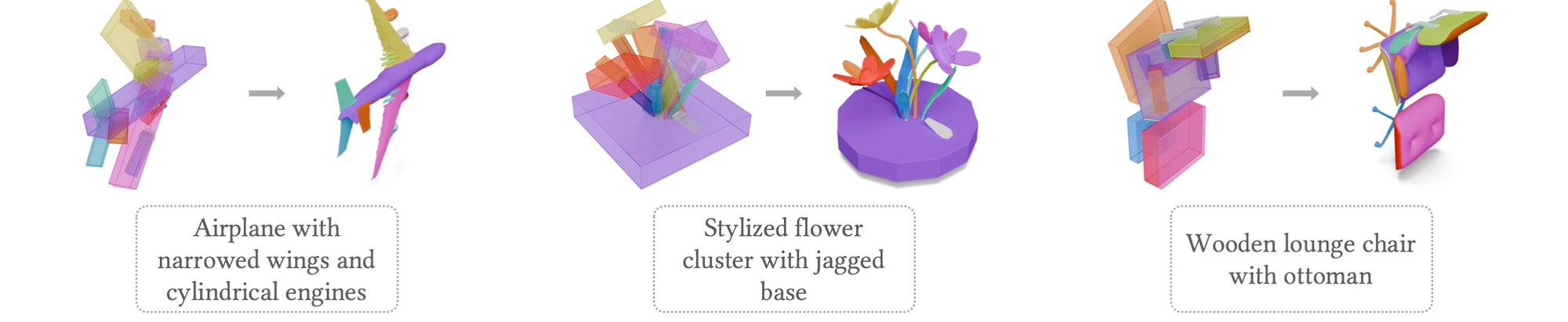}\par\vspace{0.3em}}}
{\centering{\includegraphics[width=0.98\textwidth]{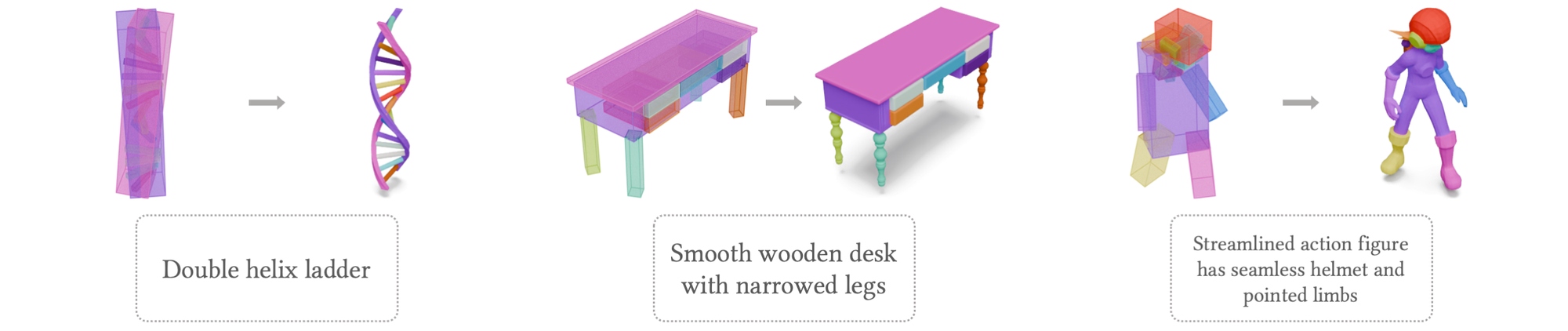}\par\vspace{0.3em}}}

\FloatBarrier
\subsubsection{8-Part Generation at Base Resolution (512 Tokens)}
\label{sec:additional_part_synthesis_8p}

We show additional results generated with our base model, using $D=512$ tokens per part and input layouts composed of up to 8 parts.

\vspace{1em}

{\centering{\includegraphics[width=0.98\textwidth]{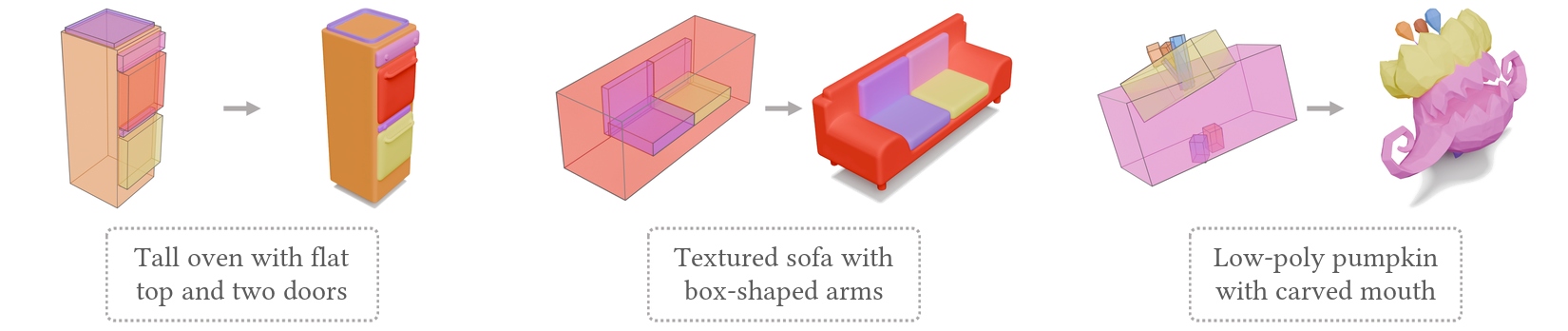}\par\vspace{0.3em}}}
{\centering{\includegraphics[width=0.98\textwidth]{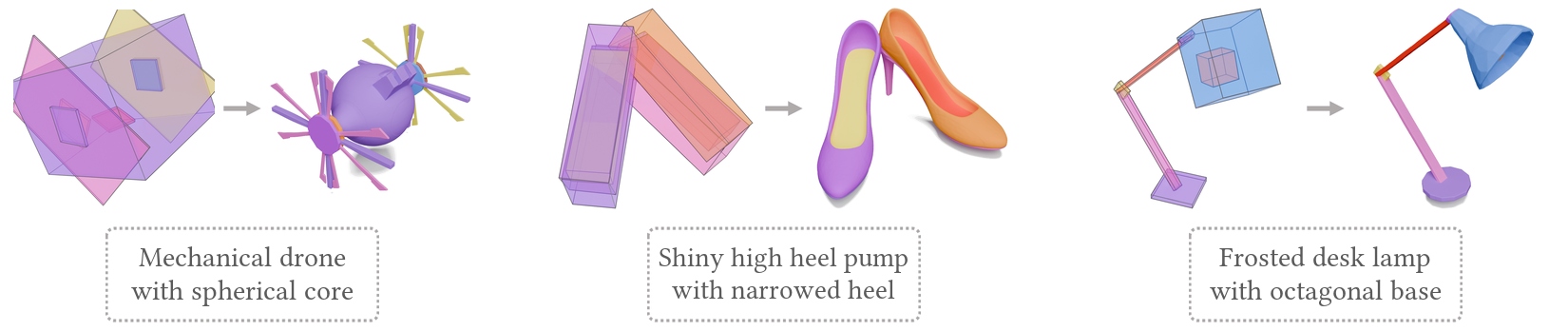}\par\vspace{0.3em}}}
{\centering{\includegraphics[width=0.98\textwidth]{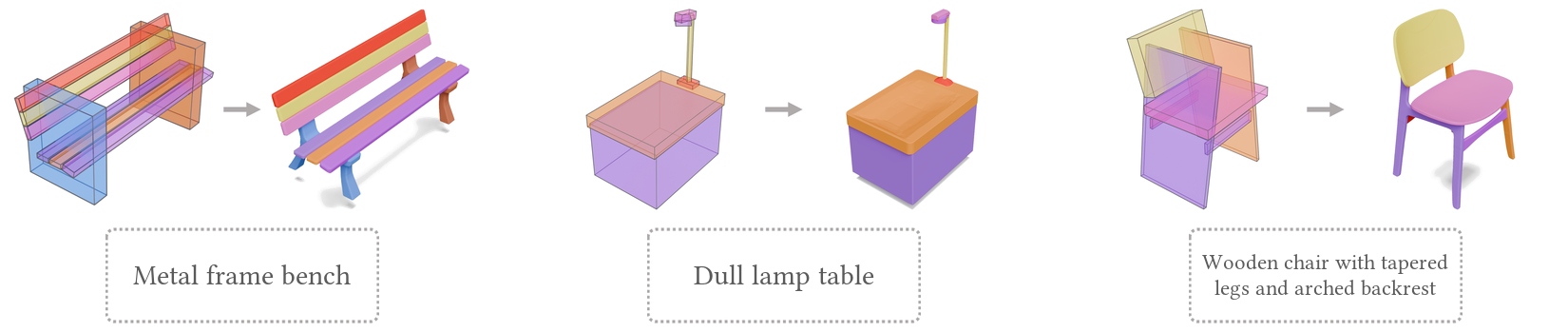}\par\vspace{0.3em}}}
{\centering{\includegraphics[width=0.98\textwidth]{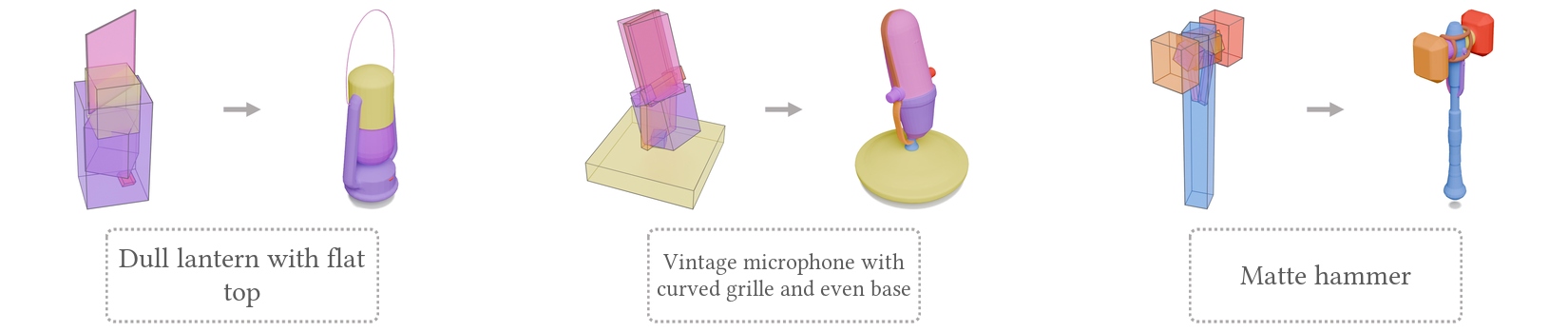}\par\vspace{0.3em}}}
{\centering{\includegraphics[width=0.98\textwidth]{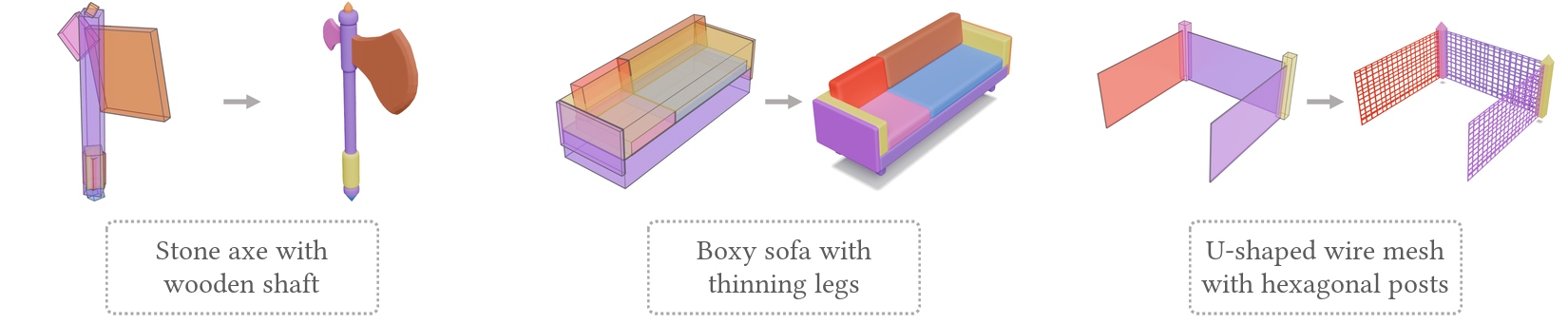}\par\vspace{0.3em}}}
{\centering{\includegraphics[width=0.98\textwidth]{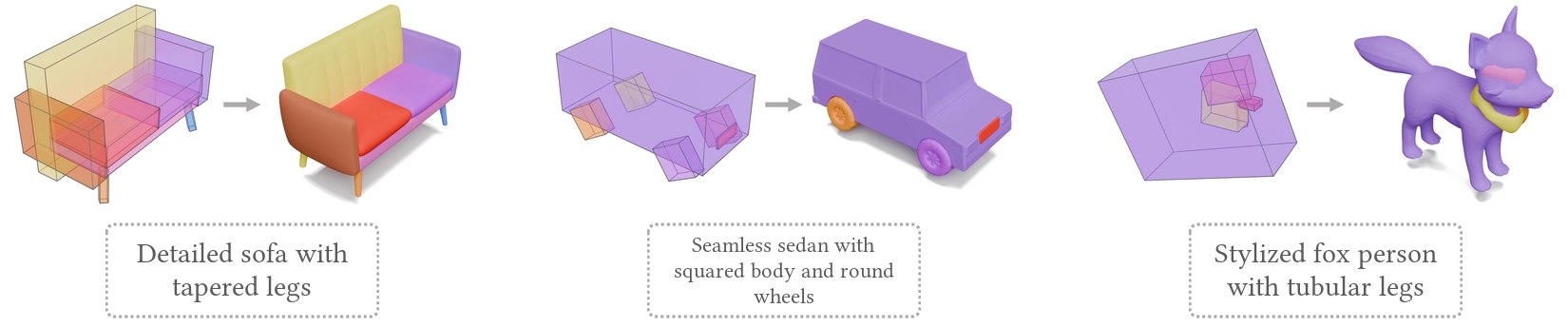}\par\vspace{0.3em}}}
{\centering{\includegraphics[width=0.98\textwidth]{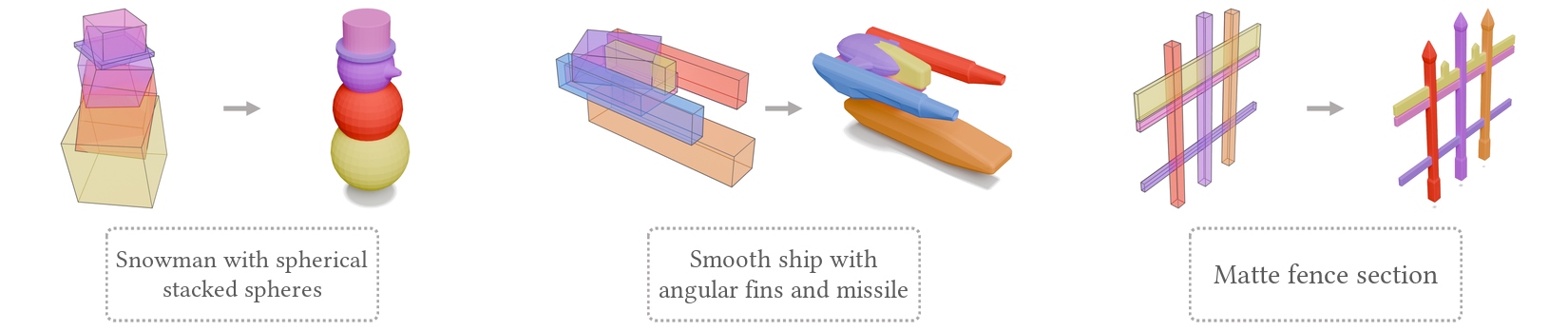}\par\vspace{0.3em}}}
{\centering{\includegraphics[width=0.98\textwidth]{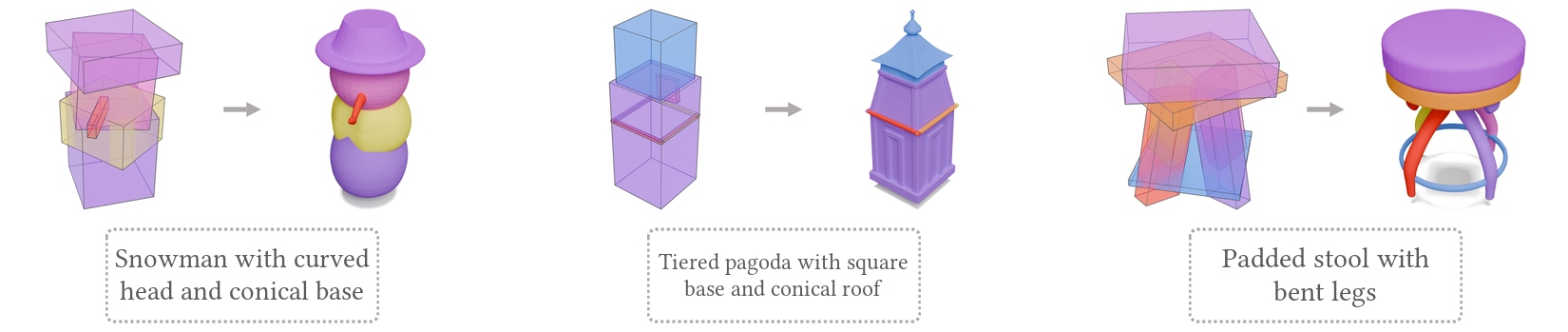}\par\vspace{0.3em}}}
{\centering{\includegraphics[width=0.98\textwidth]{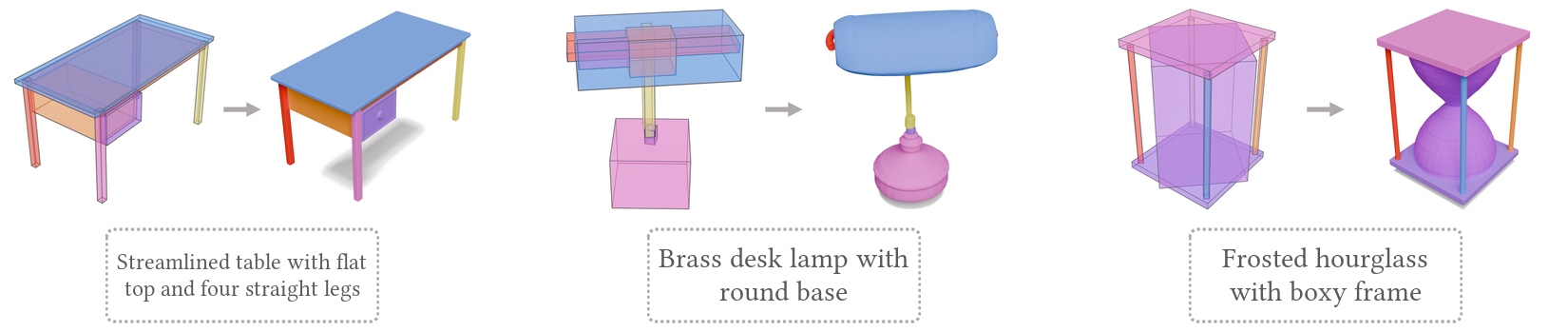}\par\vspace{0.3em}}}
{\centering{\includegraphics[width=0.98\textwidth]{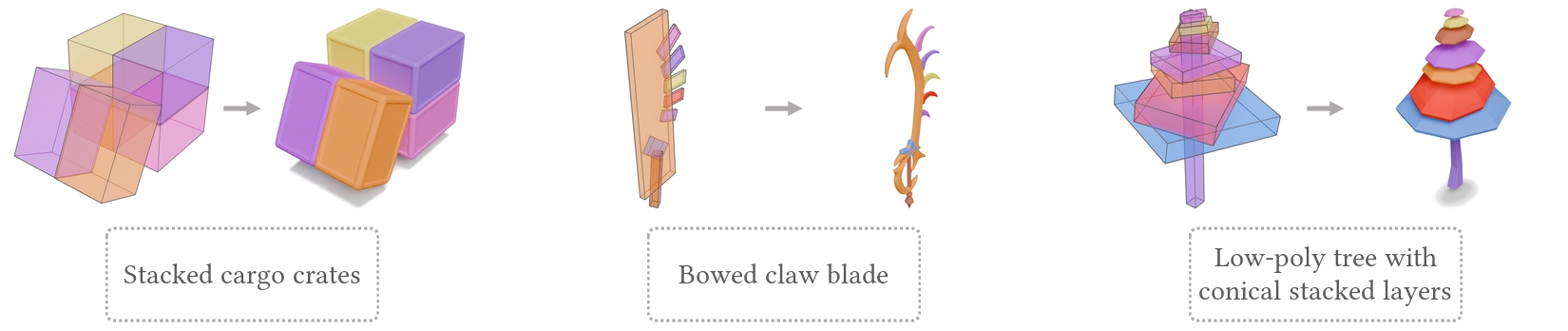}\par\vspace{0.3em}}}
{\centering{\includegraphics[width=0.98\textwidth]{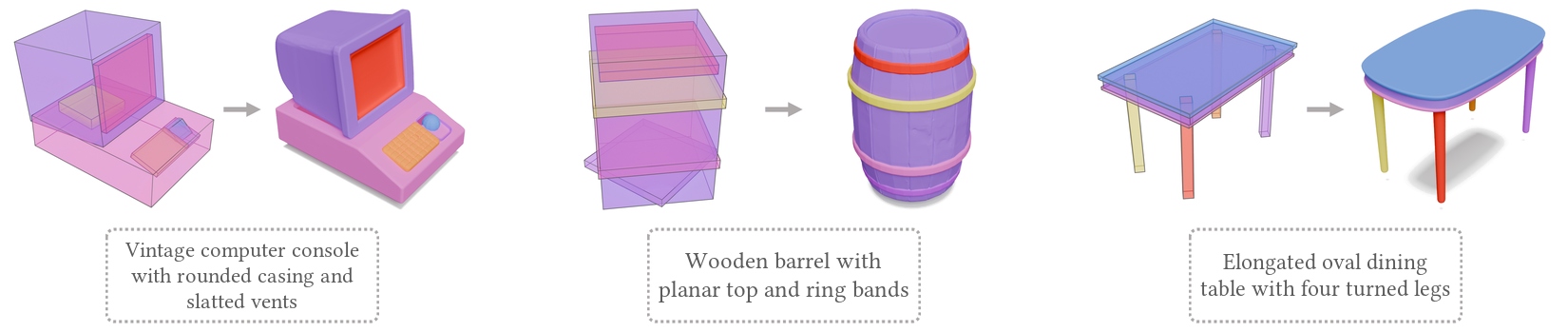}\par\vspace{0.3em}}}
{\centering{\includegraphics[width=0.98\textwidth]{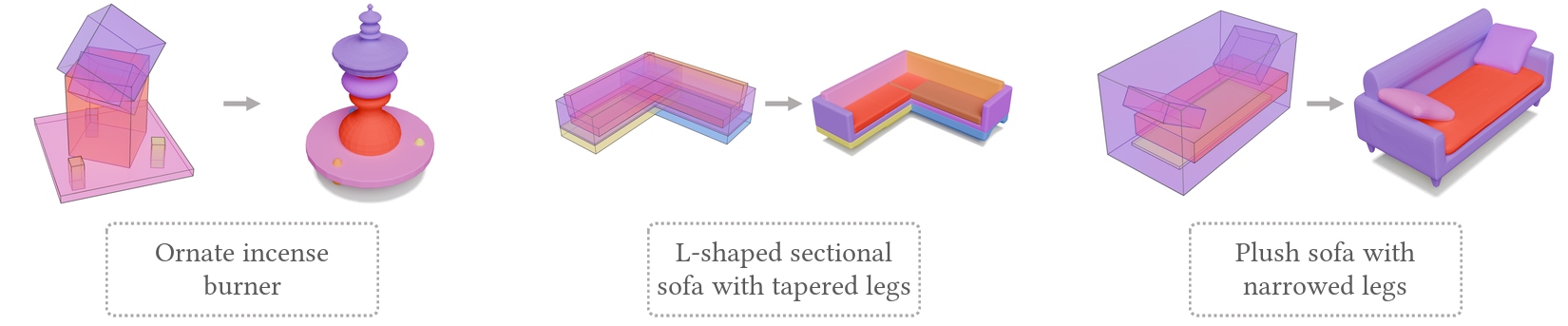}\par\vspace{0.3em}}}
{\centering{\includegraphics[width=0.98\textwidth]{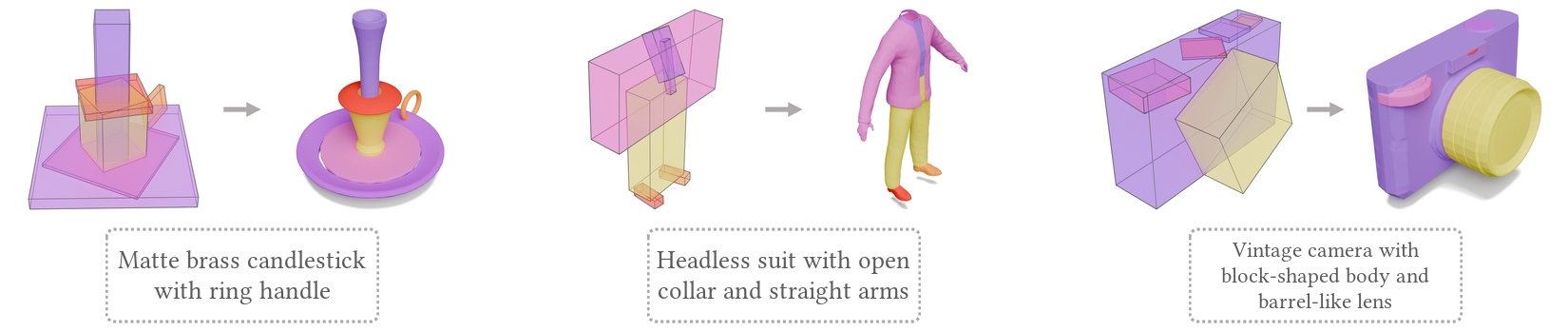}\par\vspace{0.3em}}}
{\centering{\includegraphics[width=0.98\textwidth]{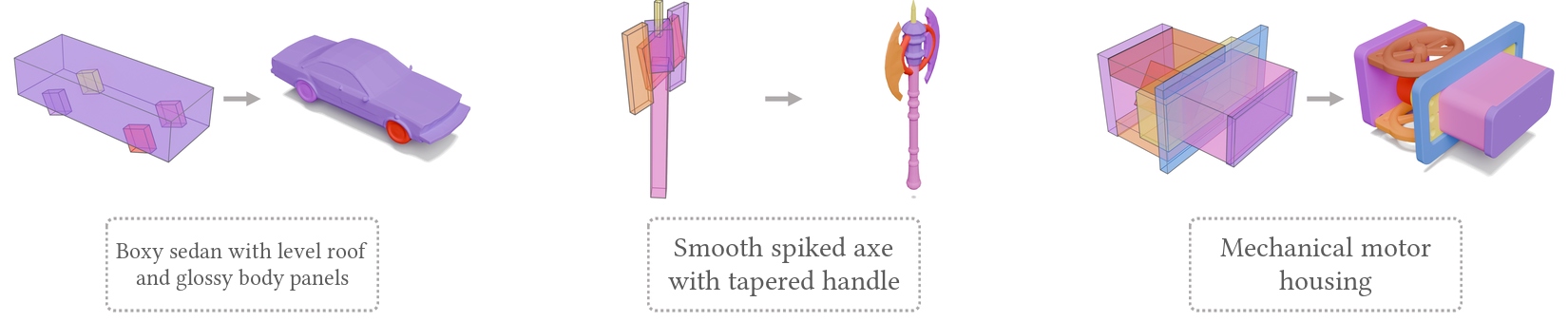}\par\vspace{0.3em}}}
{\centering{\includegraphics[width=0.98\textwidth]{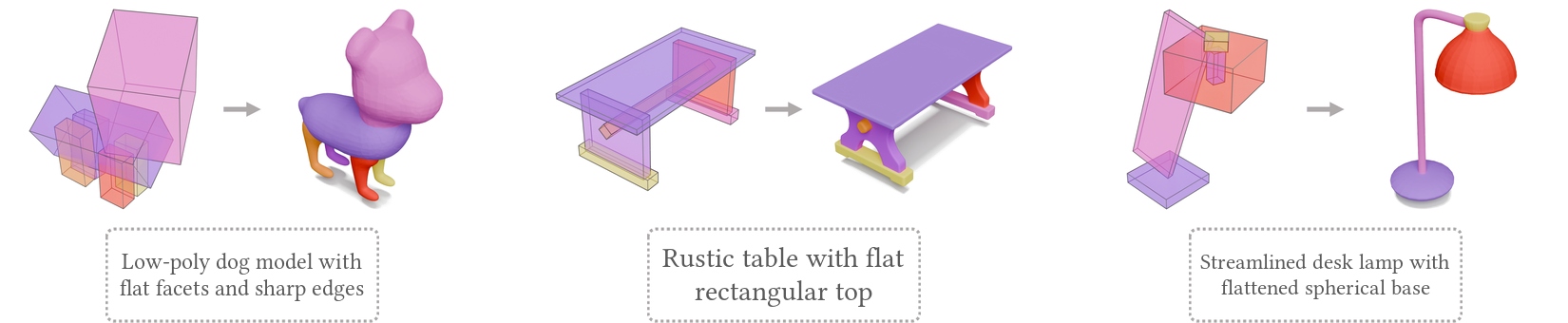}\par\vspace{0.3em}}}

\FloatBarrier
\clearpage
\subsection{Comparison with Controllable Baselines}
\label{sec:additional_control_comparisons}

{\centering{\includegraphics[width=0.96\textwidth]{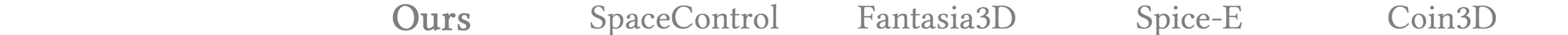}\par\vspace{0pt}}}
{\centering{\includegraphics[width=0.96\textwidth]{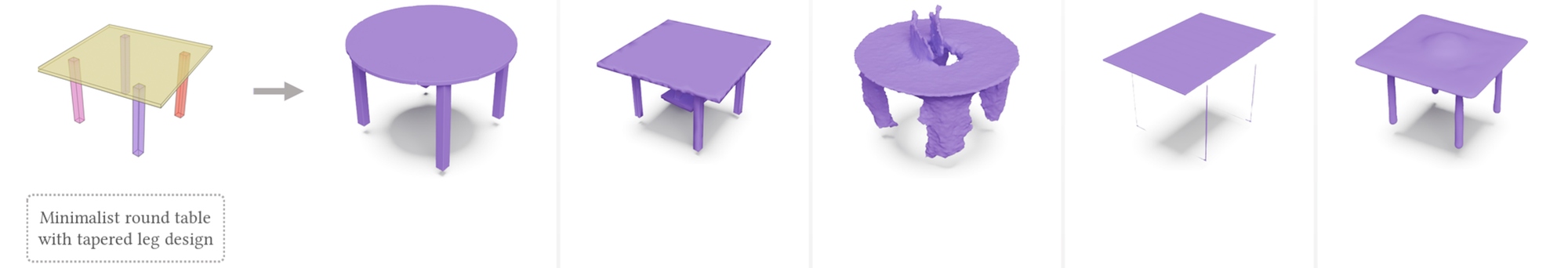}\par\vspace{0pt}}}
{\centering{\includegraphics[width=0.96\textwidth]{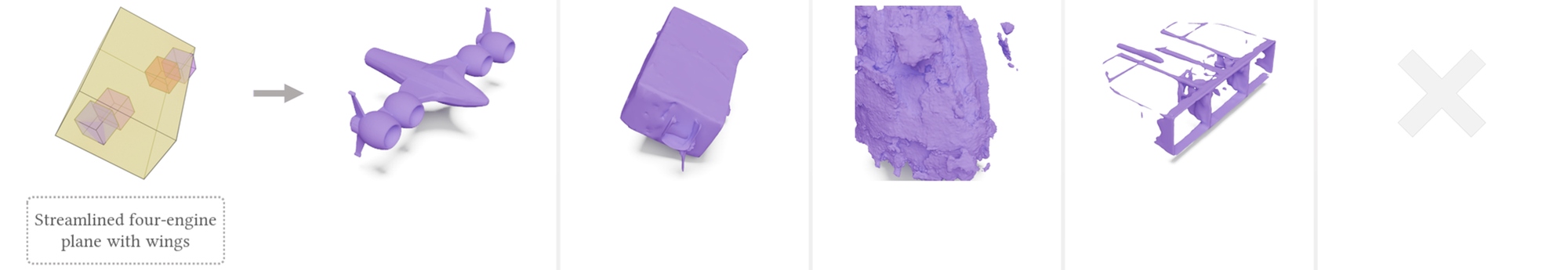}\par\vspace{0pt}}}
{\centering{\includegraphics[width=0.96\textwidth]{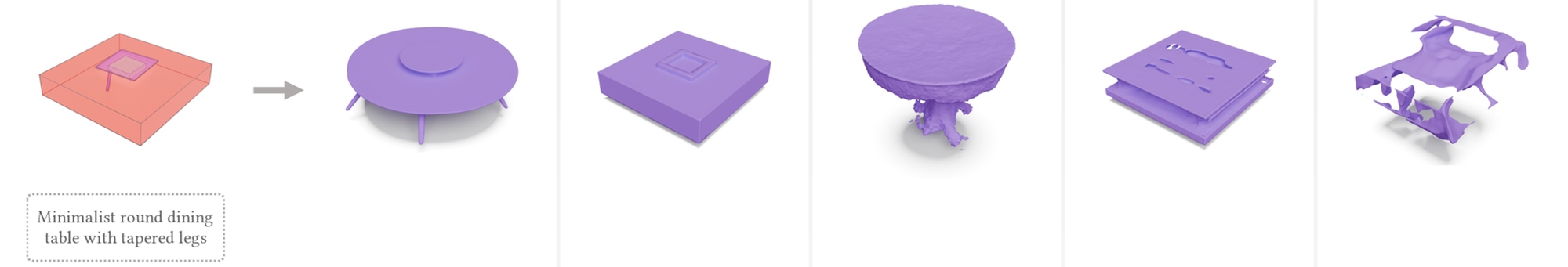}\par\vspace{0pt}}}
{\centering{\includegraphics[width=0.96\textwidth]{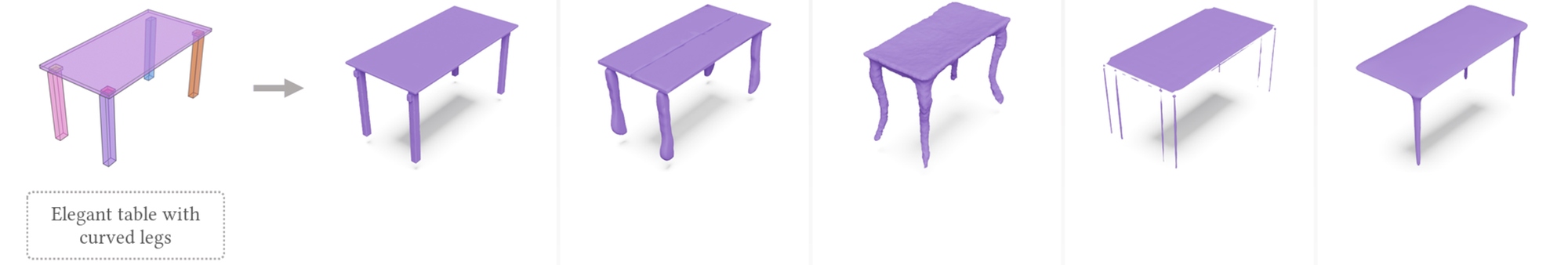}\par\vspace{0pt}}}
{\centering{\includegraphics[width=0.96\textwidth]{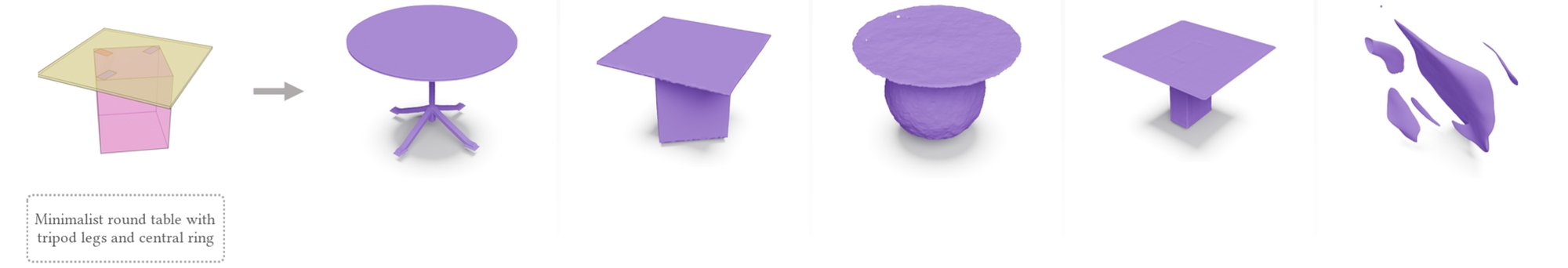}\par\vspace{0pt}}}
{\centering{\includegraphics[width=0.96\textwidth]{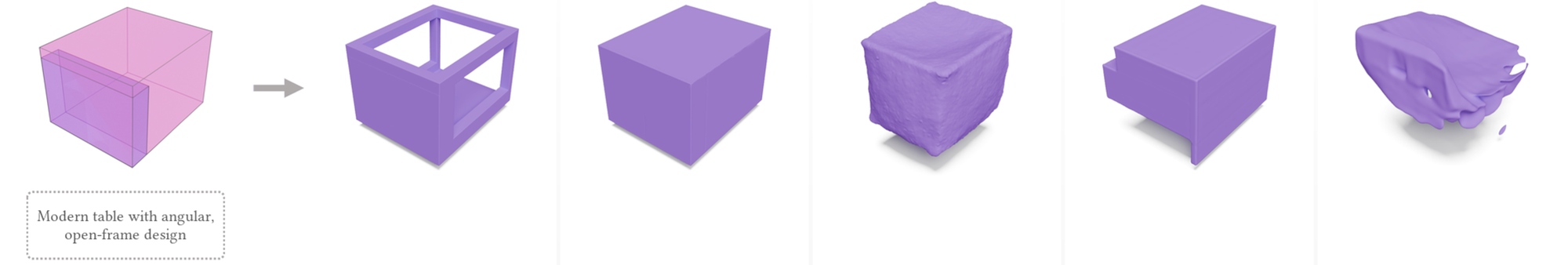}\par\vspace{0pt}}}
{\centering{\includegraphics[width=0.96\textwidth]{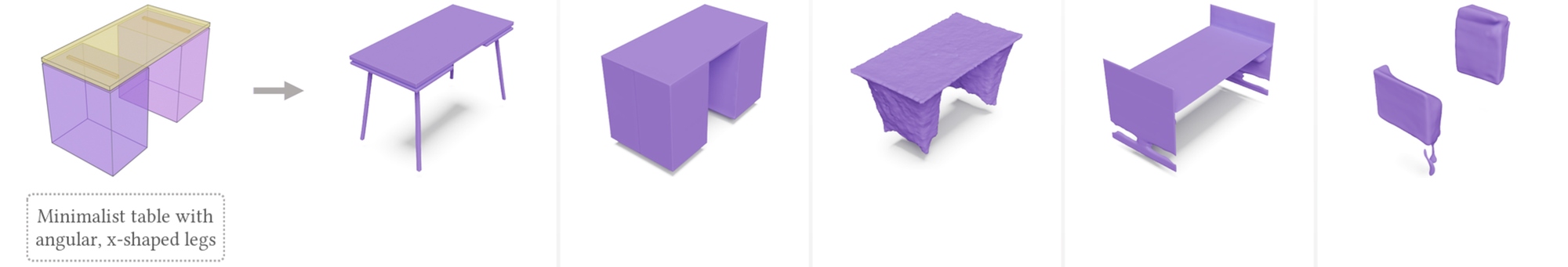}\par\vspace{0pt}}}
{\centering{\includegraphics[width=0.96\textwidth]{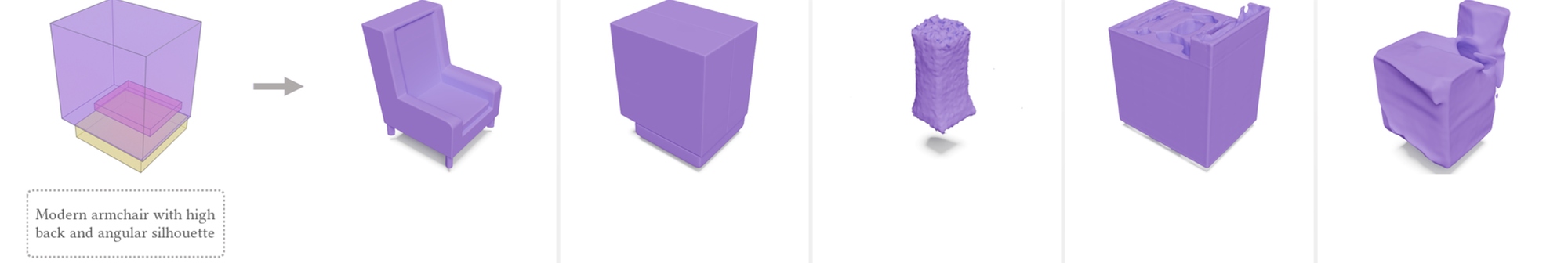}\par\vspace{0pt}}}
\clearpage
{\centering{\includegraphics[width=0.96\textwidth]{fig/rows_d3/header}\par\vspace{0pt}}}
{\centering{\includegraphics[width=0.96\textwidth]{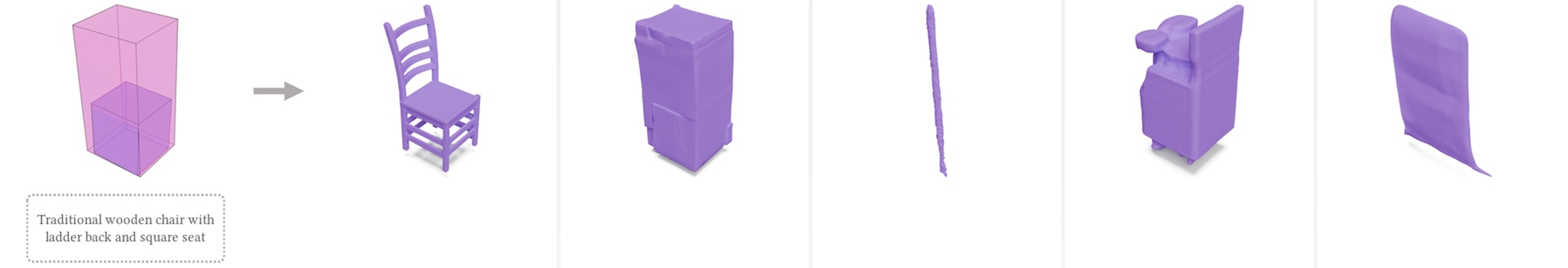}\par\vspace{0pt}}}
{\centering{\includegraphics[width=0.96\textwidth]{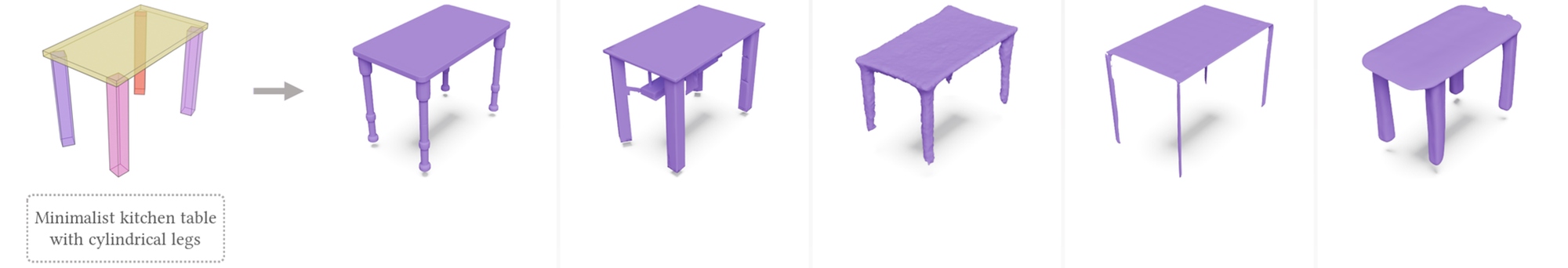}\par\vspace{0pt}}}
{\centering{\includegraphics[width=0.96\textwidth]{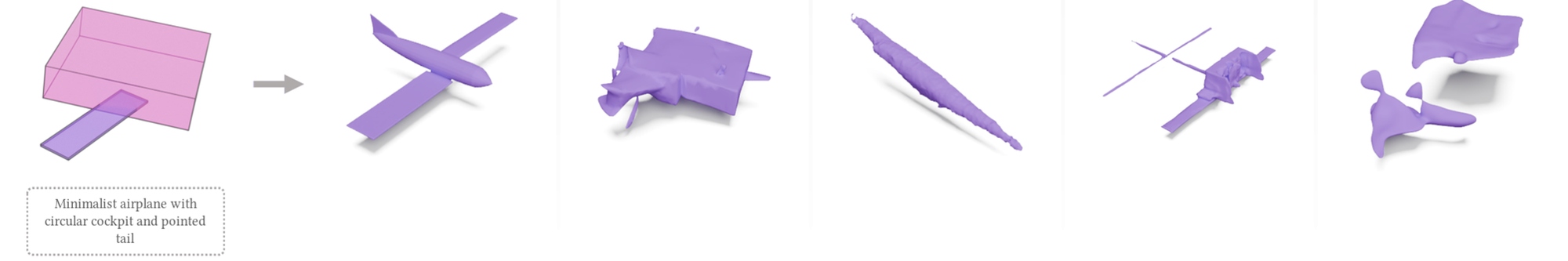}\par\vspace{0pt}}}
{\centering{\includegraphics[width=0.96\textwidth]{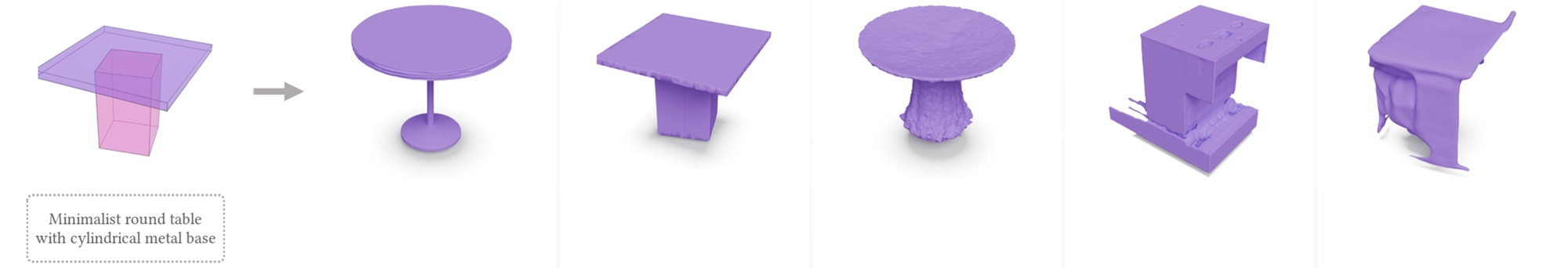}\par\vspace{0pt}}}
{\centering{\includegraphics[width=0.96\textwidth]{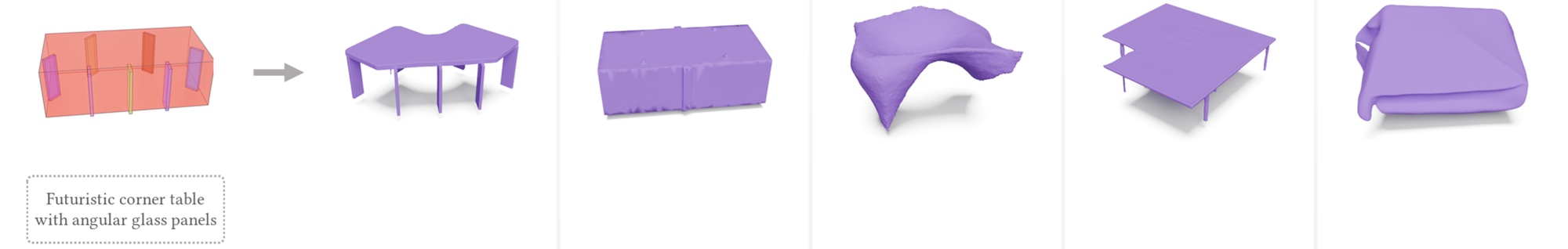}\par\vspace{0pt}}}
{\centering{\includegraphics[width=0.96\textwidth]{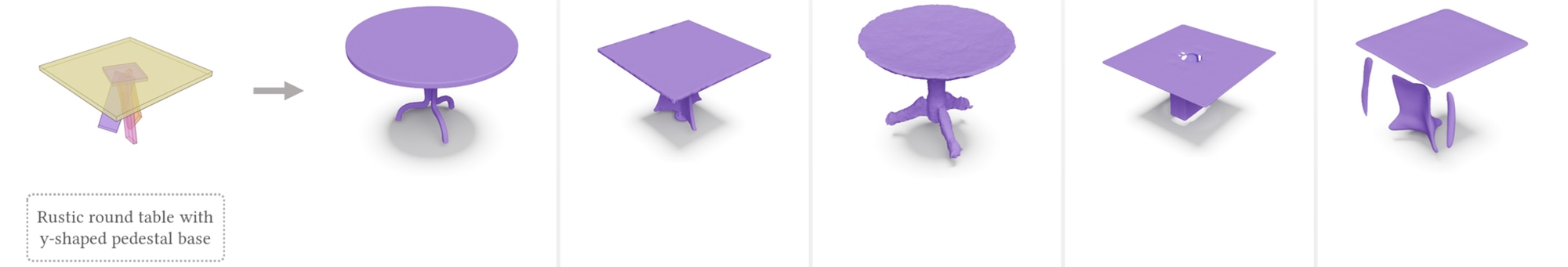}\par\vspace{0pt}}}
{\centering{\includegraphics[width=0.96\textwidth]{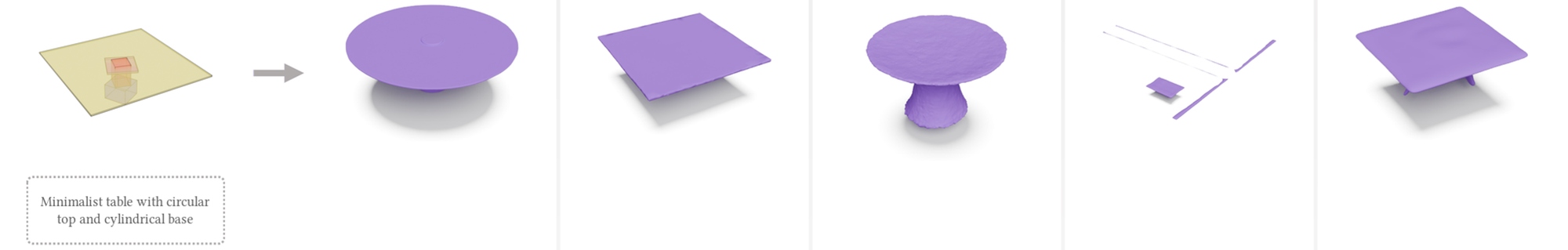}\par\vspace{0pt}}}
{\centering{\includegraphics[width=0.96\textwidth]{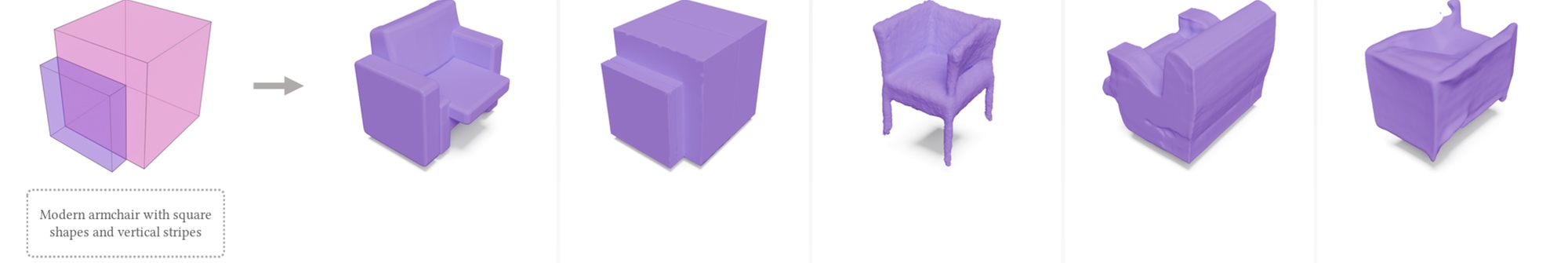}\par\vspace{0pt}}}
{\centering{\includegraphics[width=0.96\textwidth]{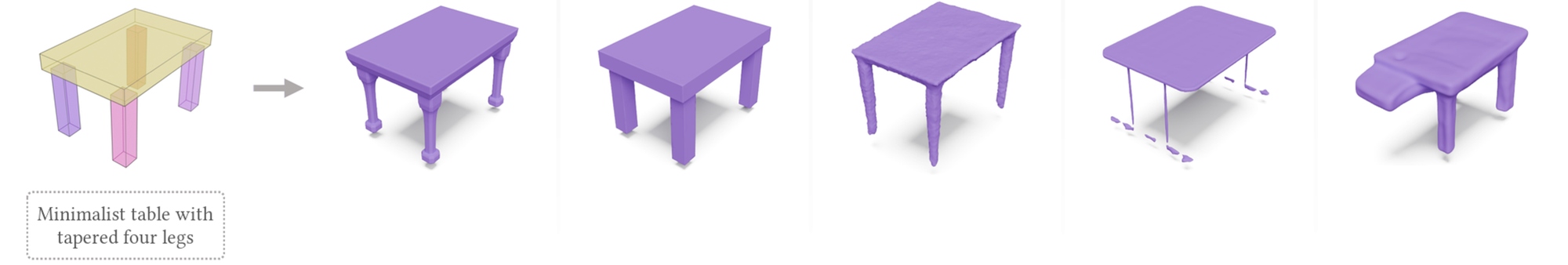}\par\vspace{0pt}}}

\FloatBarrier
\clearpage
\subsection{Comparison with Part Synthesis Baselines}
\label{sec:additional_part_synthesis_comparisons}

{\centering{\includegraphics[width=0.98\textwidth]{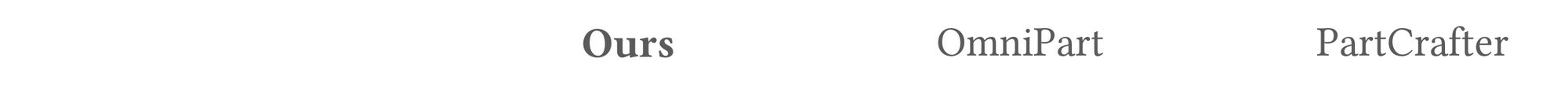}\par\vspace{0pt}}}
{\centering{\includegraphics[width=0.98\textwidth]{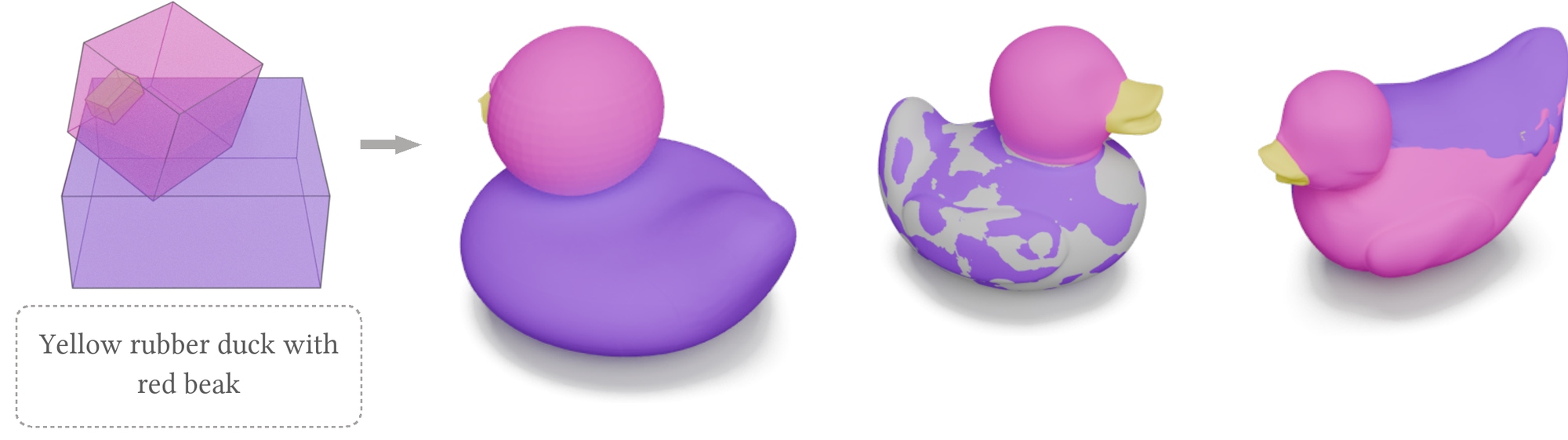}\par\vspace{0pt}}}
\vfill
{\centering{\includegraphics[width=0.98\textwidth]{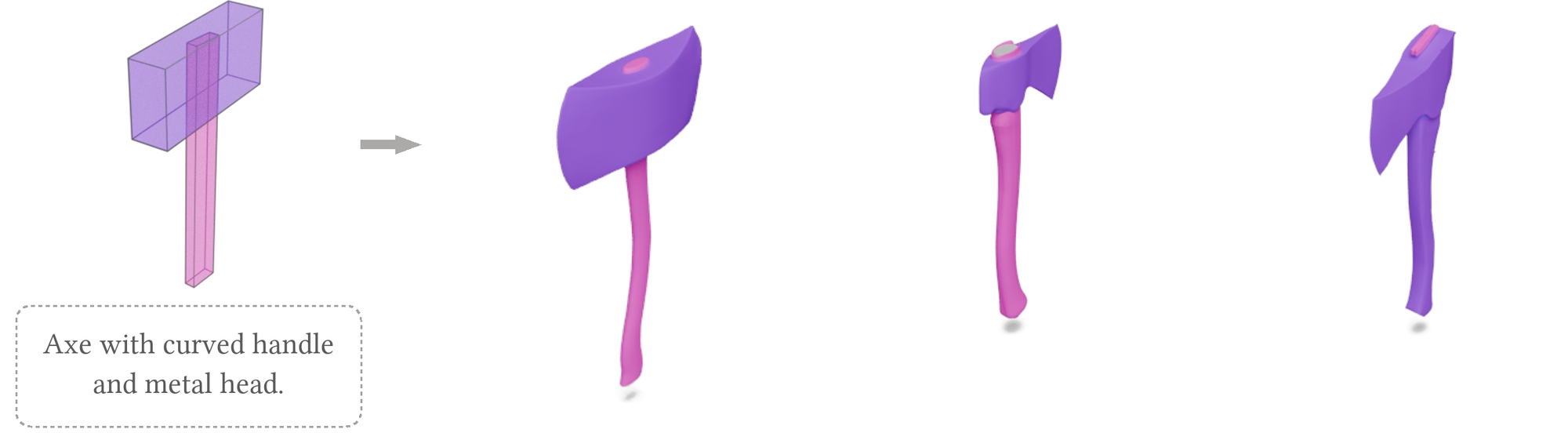}\par\vspace{0pt}}}
\vfill
{\centering{\includegraphics[width=0.98\textwidth]{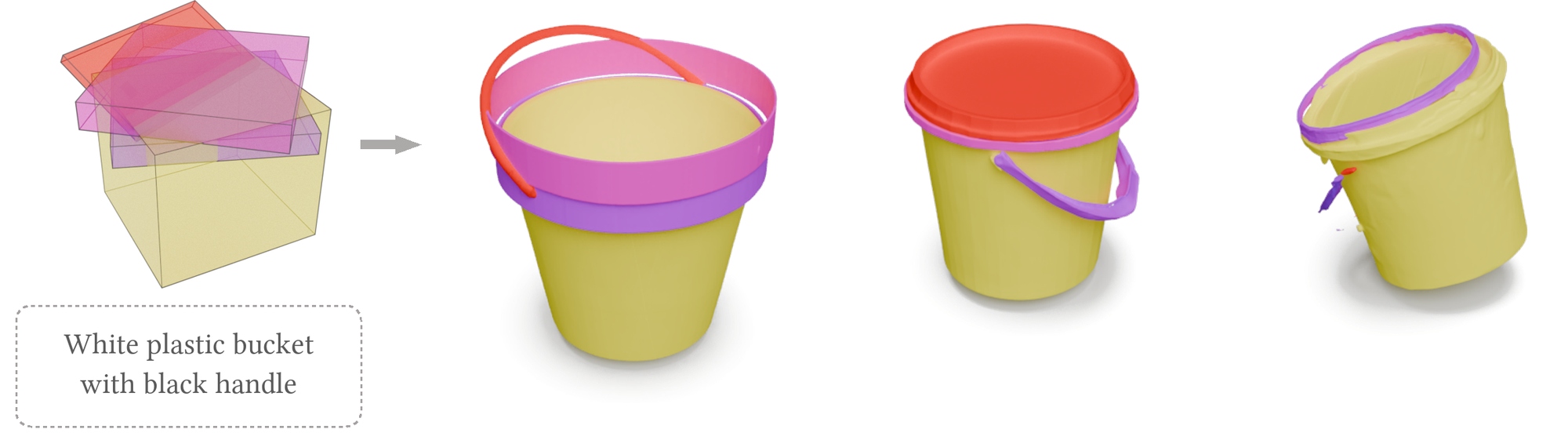}\par\vspace{0pt}}}
\vfill
{\centering{\includegraphics[width=0.98\textwidth]{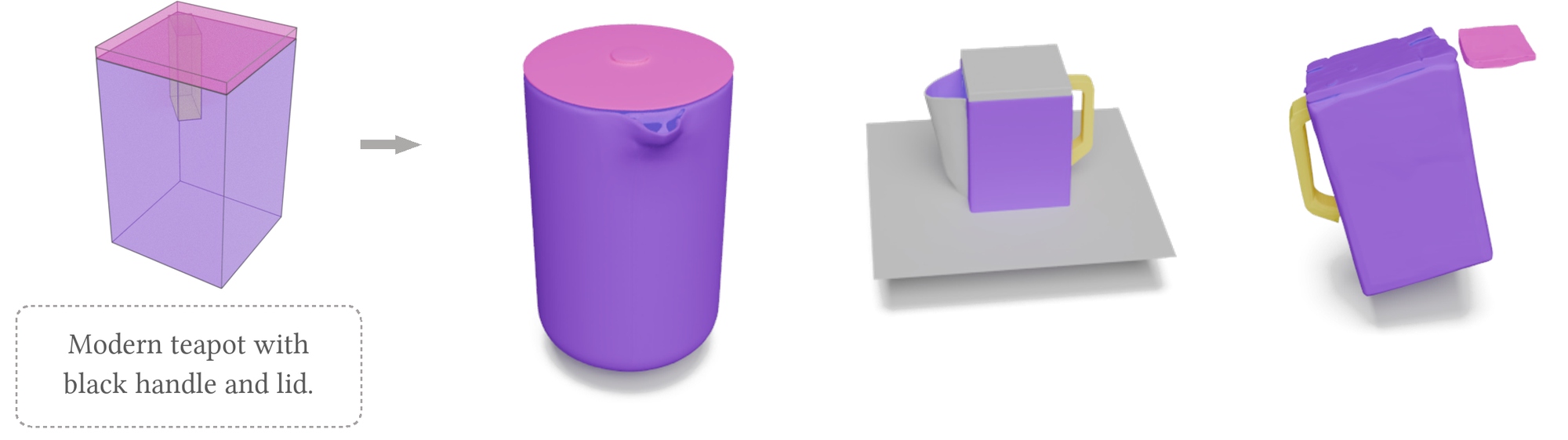}\par\vspace{0pt}}}
\vfill
{\centering{\includegraphics[width=0.98\textwidth]{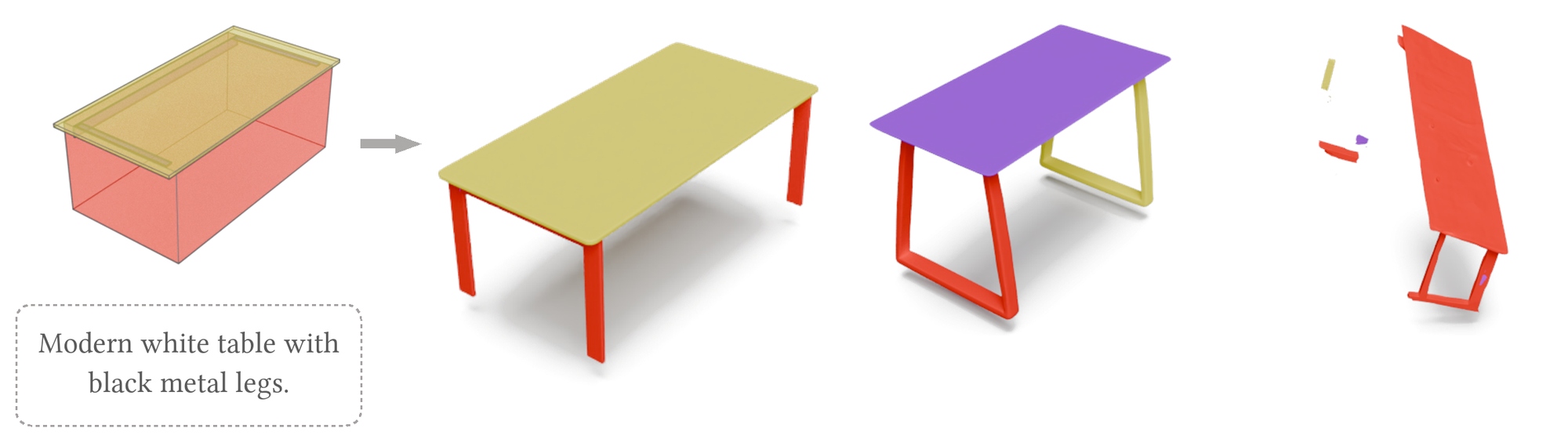}\par\vspace{0pt}}}
\clearpage
{\centering{\includegraphics[width=0.98\textwidth]{fig/rows_d4/header}\par\vspace{0pt}}}
{\centering{\includegraphics[width=0.98\textwidth]{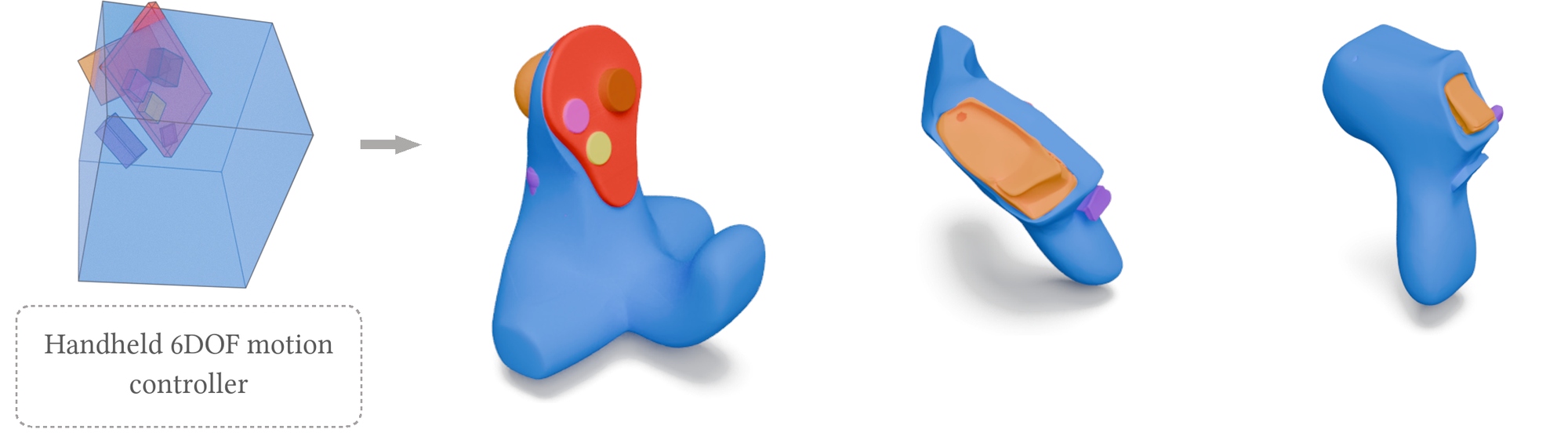}\par\vspace{0pt}}}
\vfill
{\centering{\includegraphics[width=0.98\textwidth]{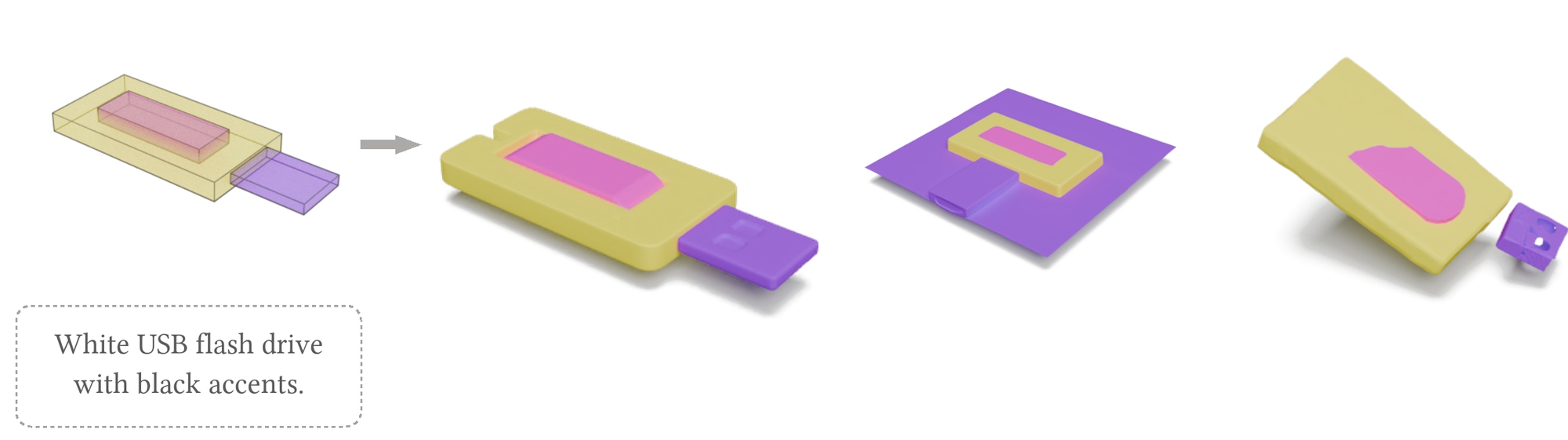}\par\vspace{0pt}}}
\vfill
{\centering{\includegraphics[width=0.98\textwidth]{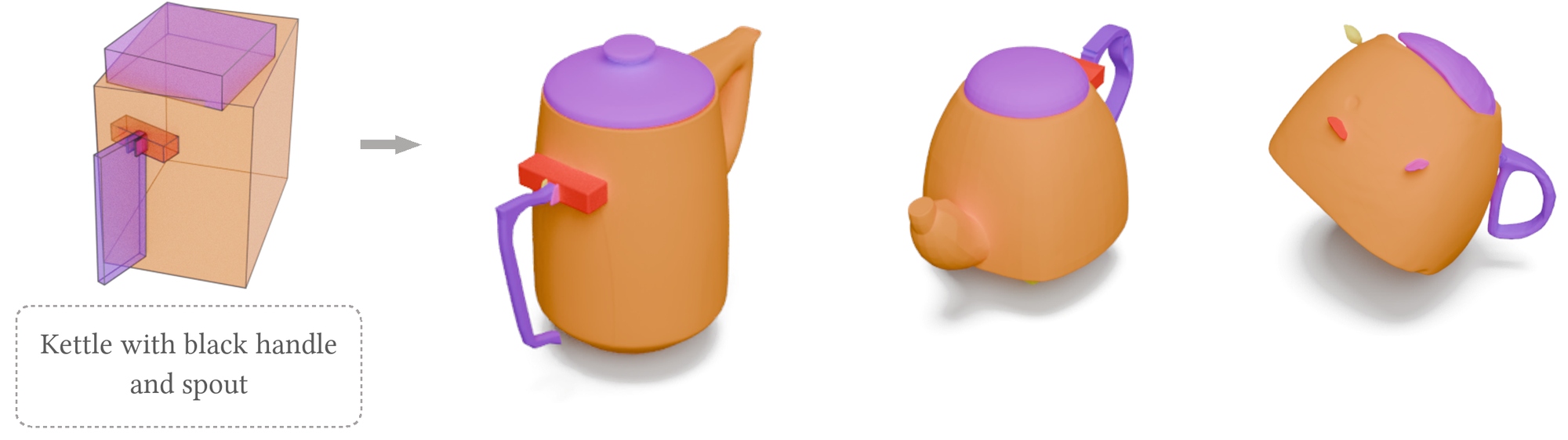}\par\vspace{0pt}}}
\vfill
{\centering{\includegraphics[width=0.98\textwidth]{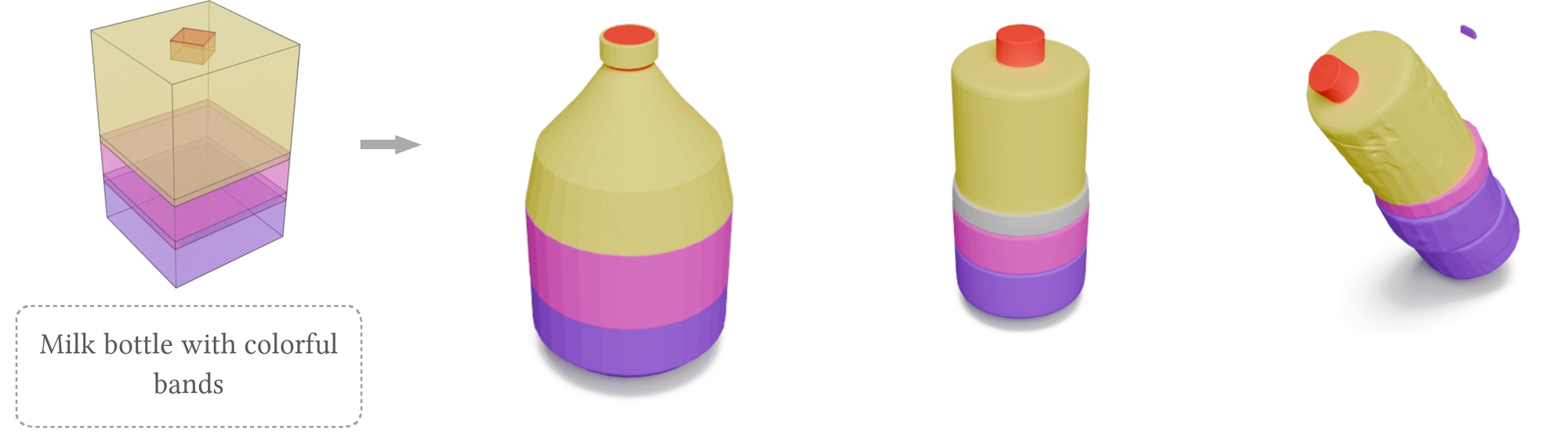}\par\vspace{0pt}}}
\vfill
{\centering{\includegraphics[width=0.98\textwidth]{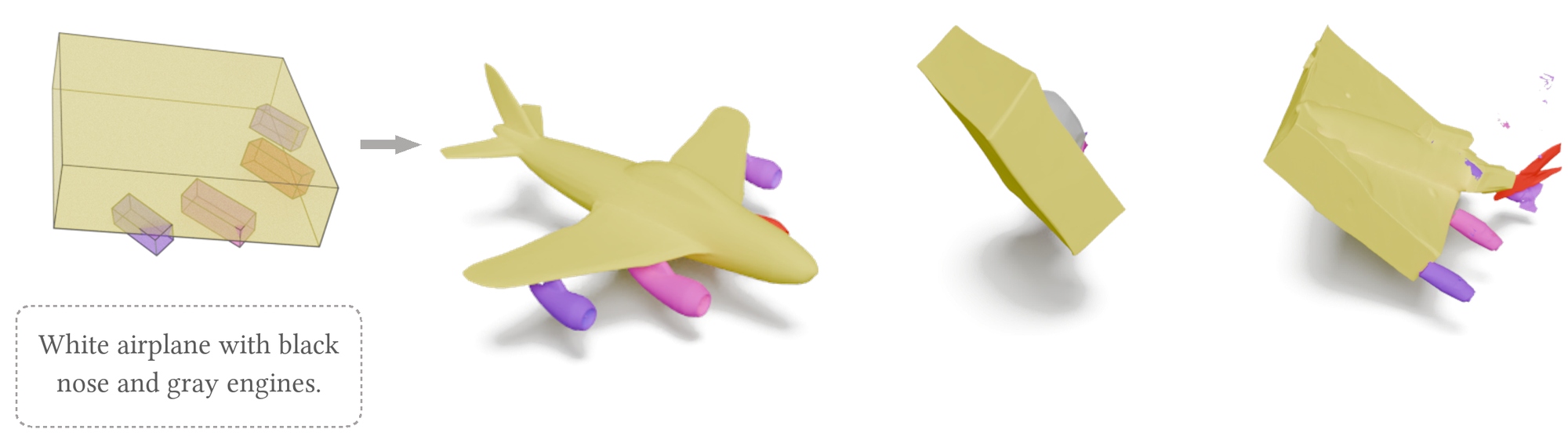}\par\vspace{0pt}}}
\clearpage
{\centering{\includegraphics[width=0.98\textwidth]{fig/rows_d4/header}\par\vspace{0pt}}}
{\centering{\includegraphics[width=0.98\textwidth]{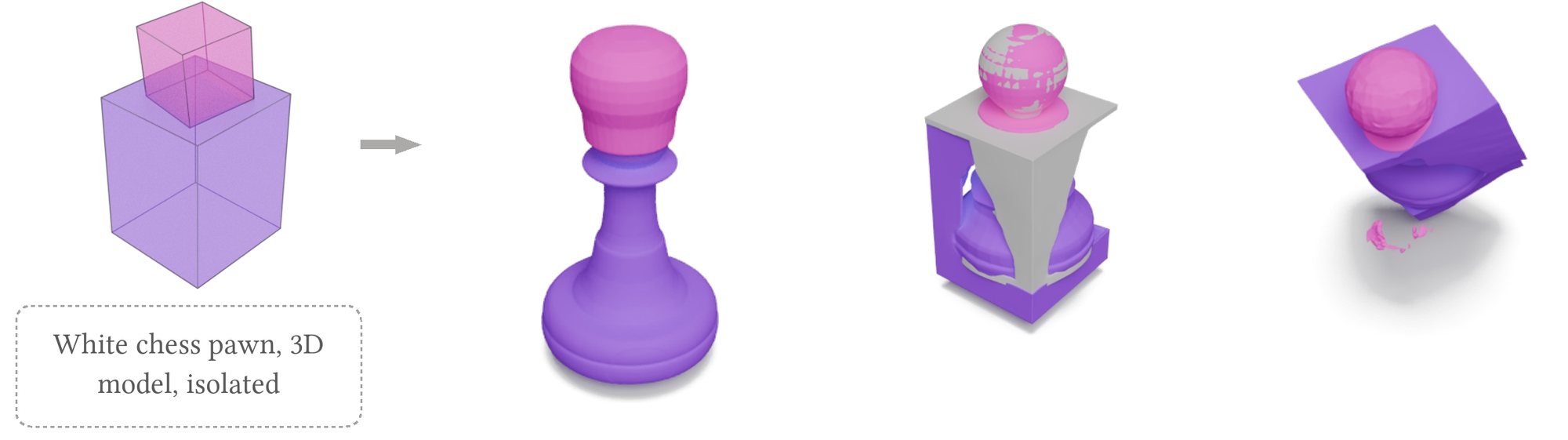}\par\vspace{0pt}}}
\vfill
{\centering{\includegraphics[width=0.98\textwidth]{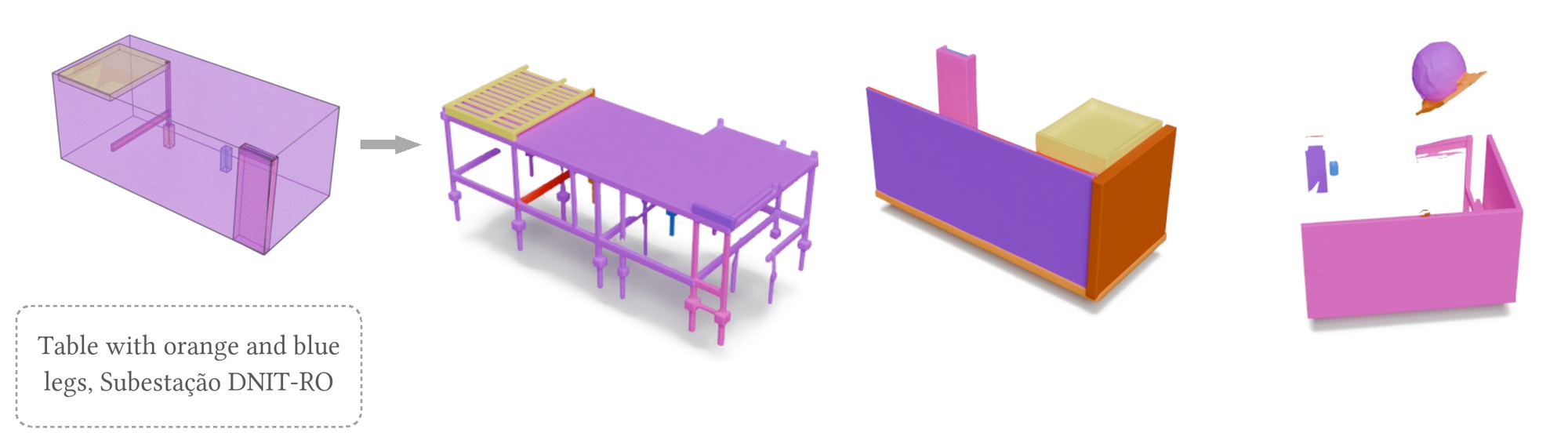}\par\vspace{0pt}}}
\vfill
{\centering{\includegraphics[width=0.98\textwidth]{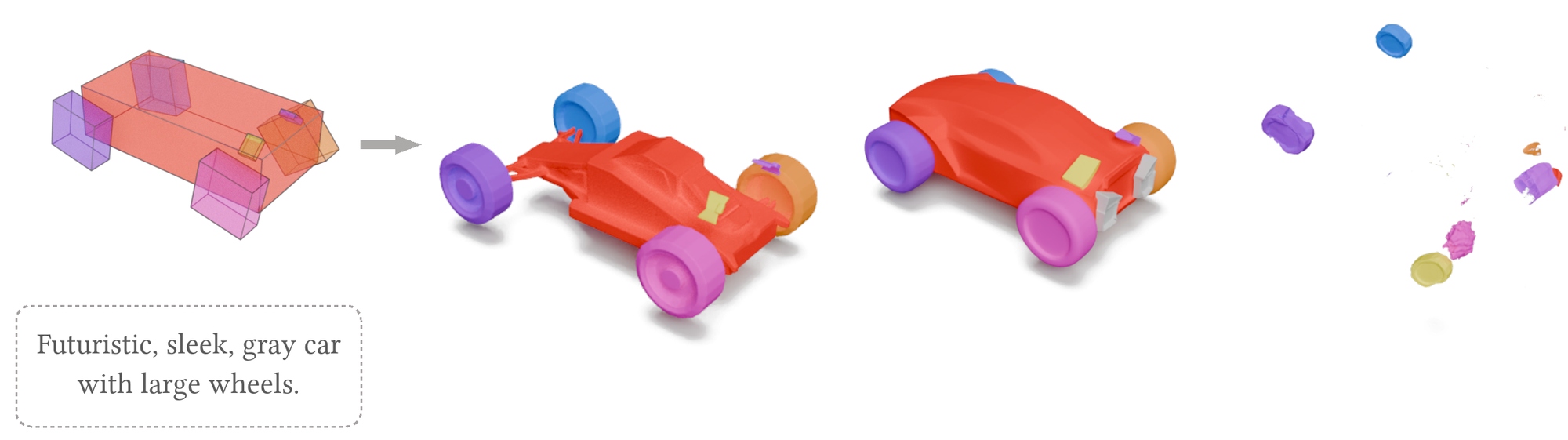}\par\vspace{0pt}}}
\vfill
{\centering{\includegraphics[width=0.98\textwidth]{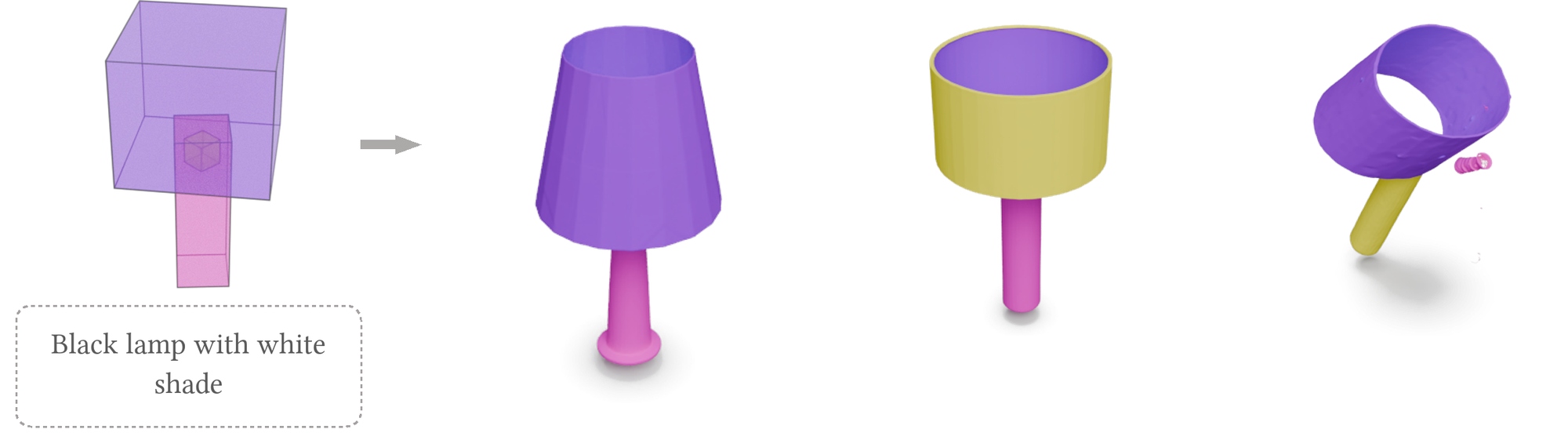}\par\vspace{0pt}}}
\vfill
{\centering{\includegraphics[width=0.98\textwidth]{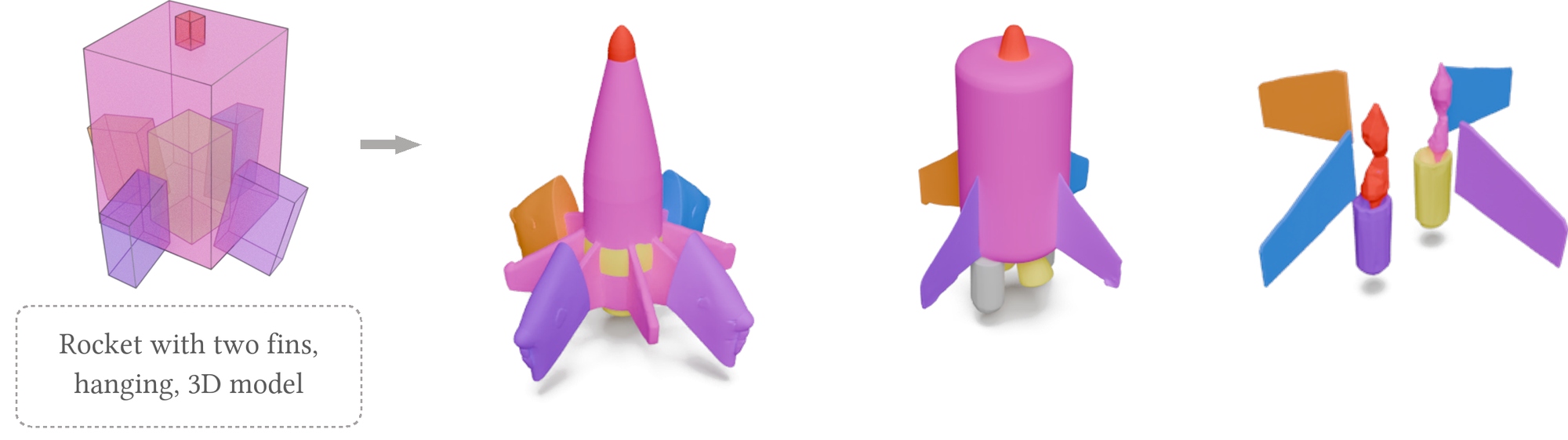}\par\vspace{0pt}}}
\clearpage
{\centering{\includegraphics[width=0.98\textwidth]{fig/rows_d4/header}\par\vspace{0pt}}}
{\centering{\includegraphics[width=0.98\textwidth]{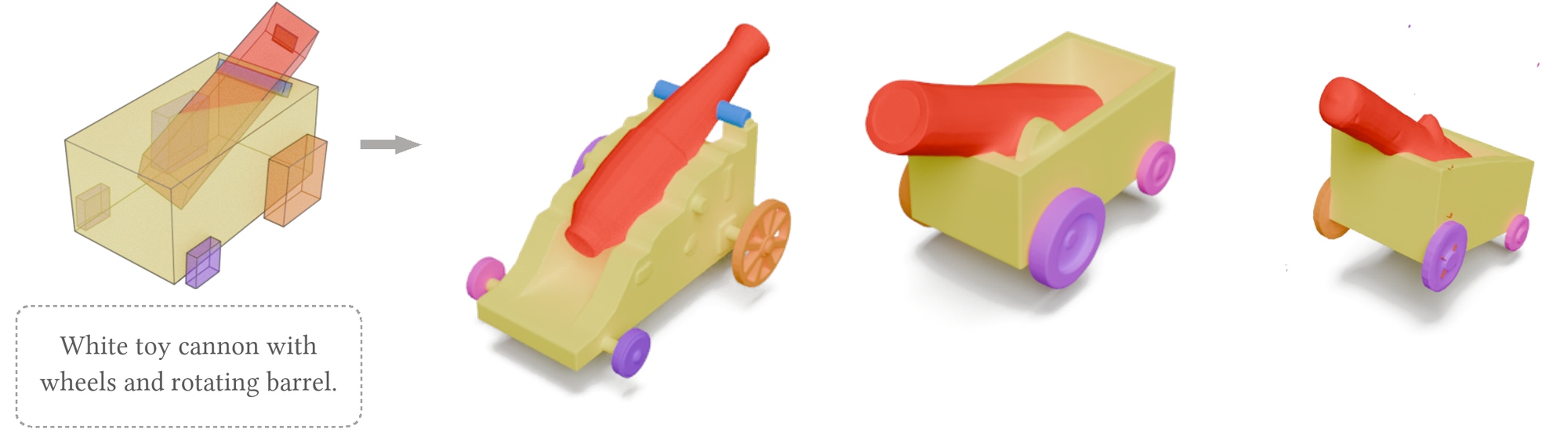}\par\vspace{0pt}}}
\vfill
{\centering{\includegraphics[width=0.98\textwidth]{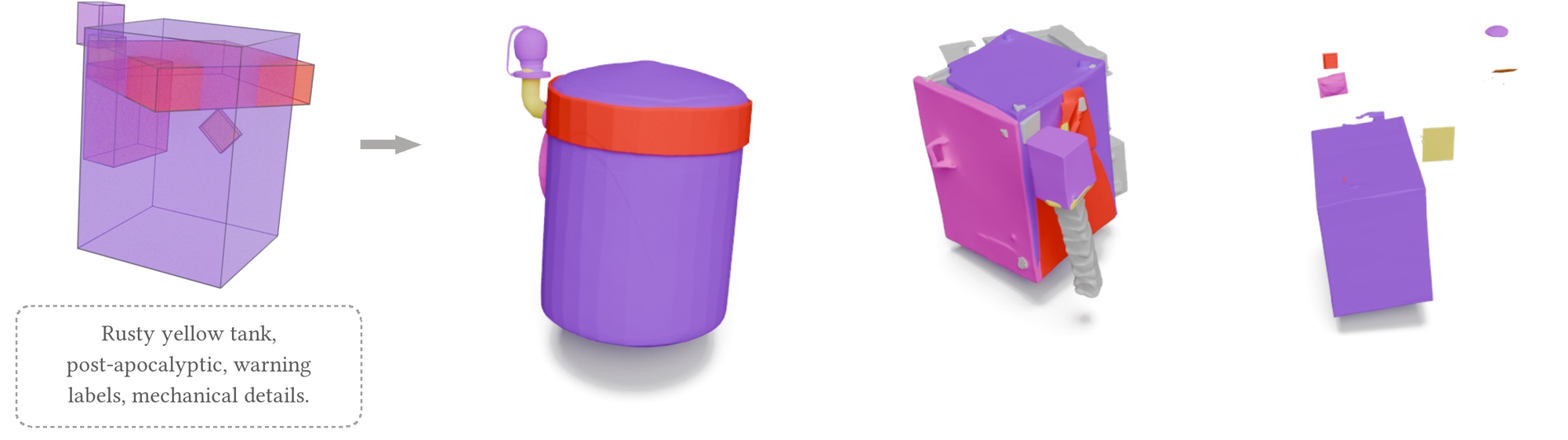}\par\vspace{0pt}}}
\vfill
{\centering{\includegraphics[width=0.98\textwidth]{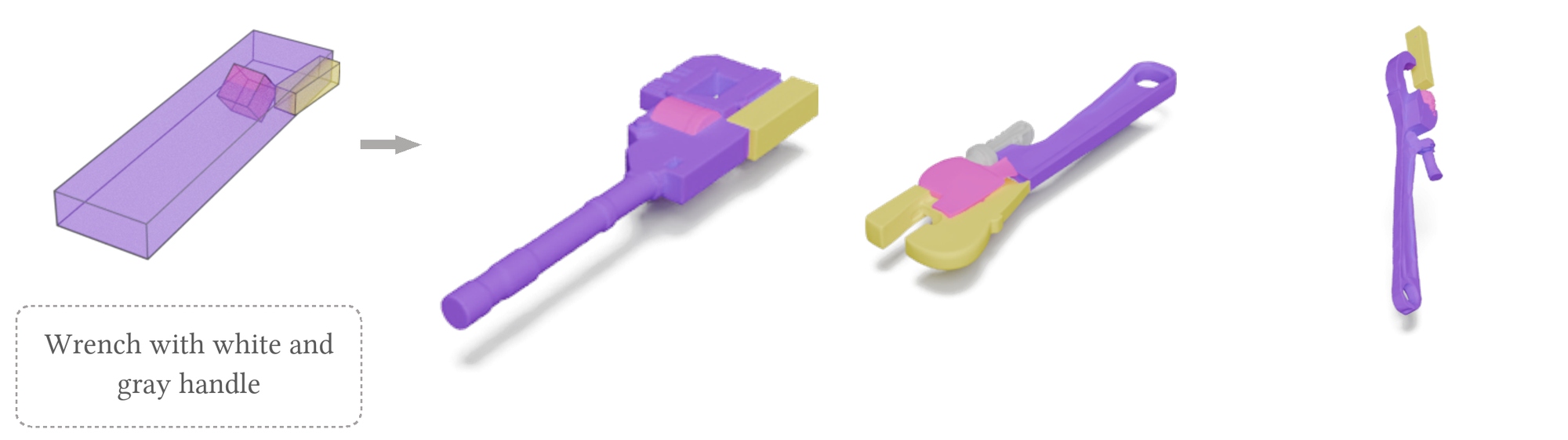}\par\vspace{0pt}}}
\vfill
{\centering{\includegraphics[width=0.98\textwidth]{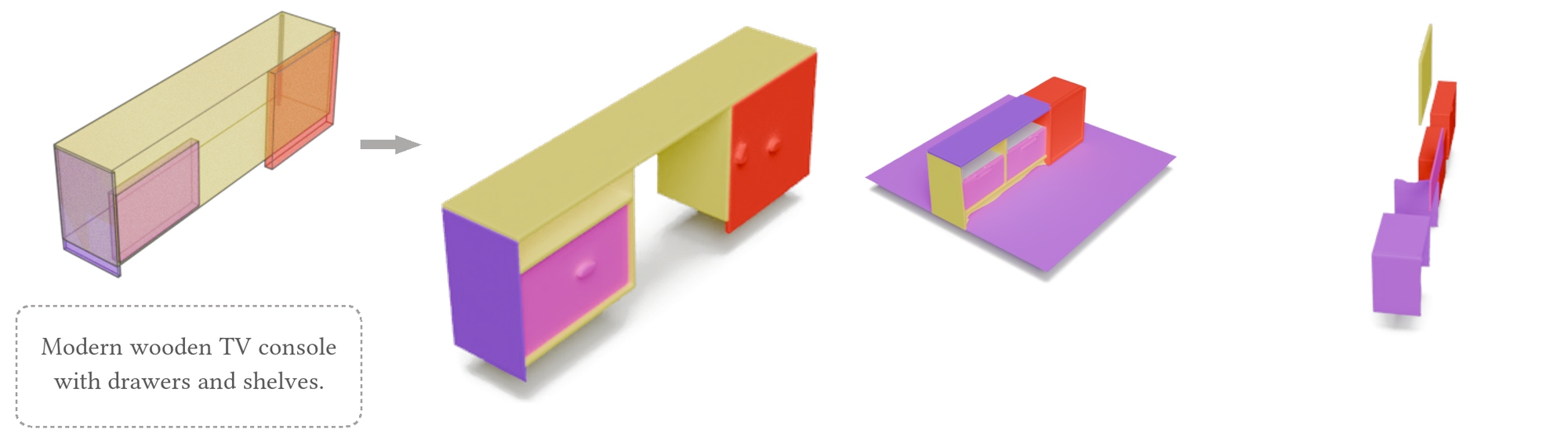}\par\vspace{0pt}}}
\vfill
{\centering{\includegraphics[width=0.98\textwidth]{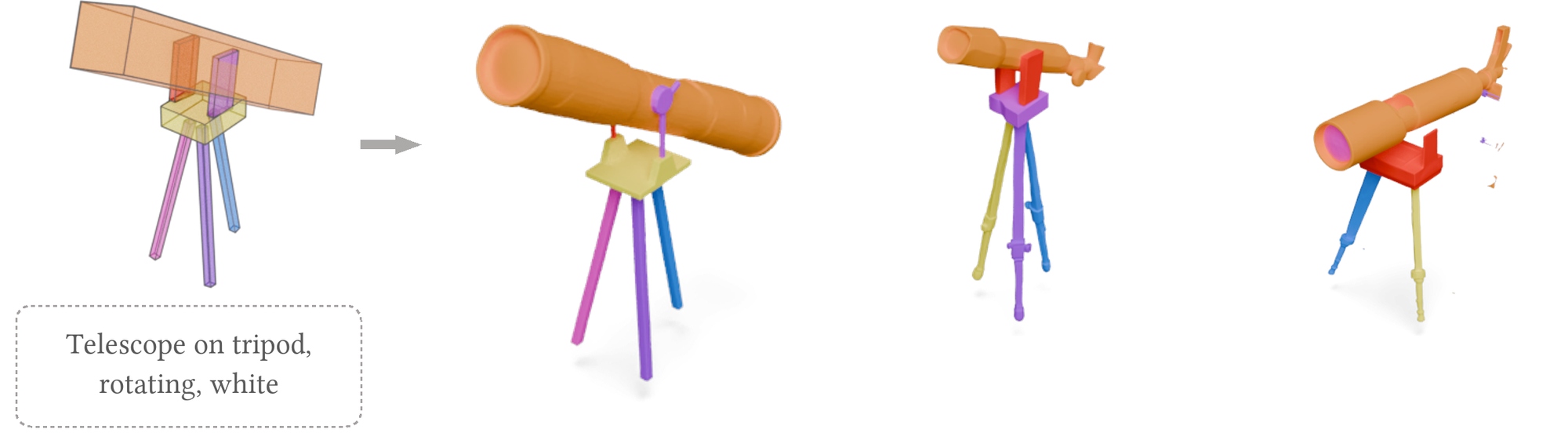}\par\vspace{0pt}}}
\clearpage
{\centering{\includegraphics[width=0.98\textwidth]{fig/rows_d4/header}\par\vspace{0pt}}}
{\centering{\includegraphics[width=0.98\textwidth]{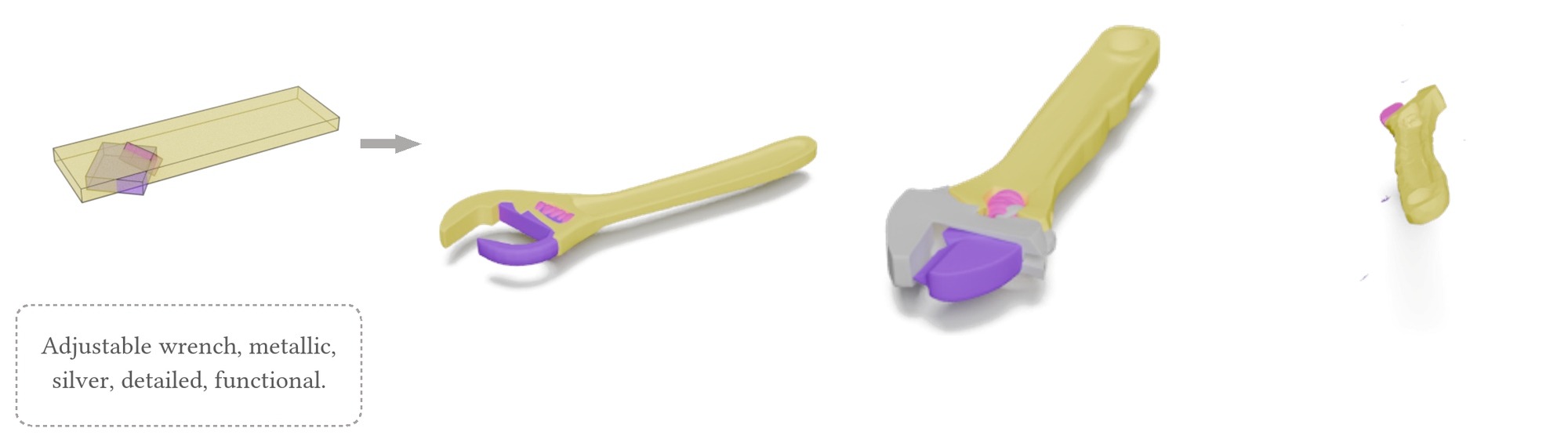}\par\vspace{0pt}}}
\vfill
{\centering{\includegraphics[width=0.98\textwidth]{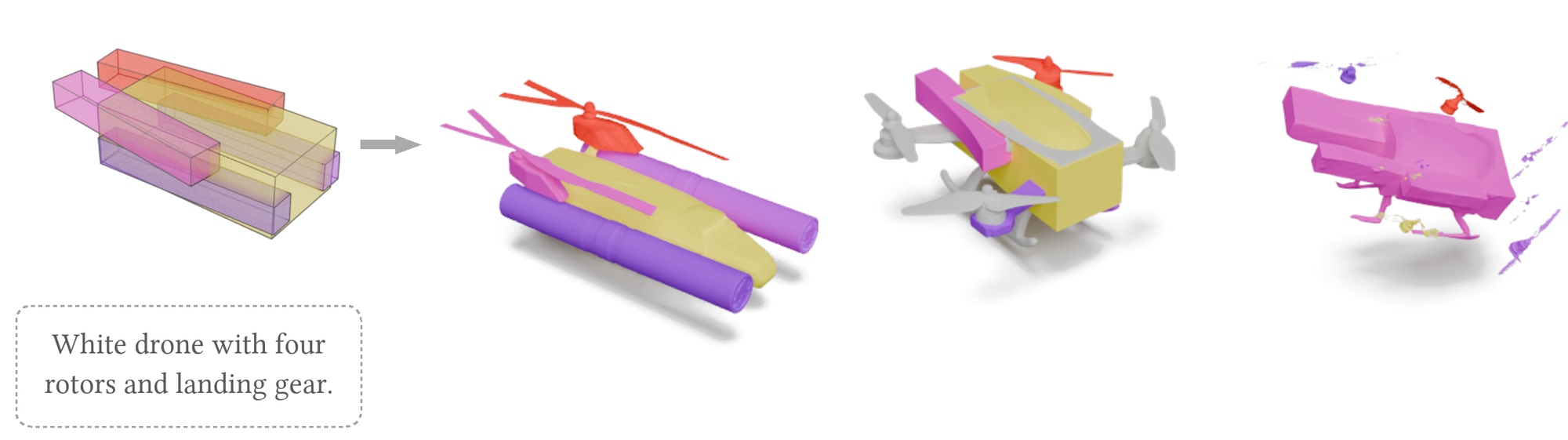}\par\vspace{0pt}}}
\vfill
{\centering{\includegraphics[width=0.98\textwidth]{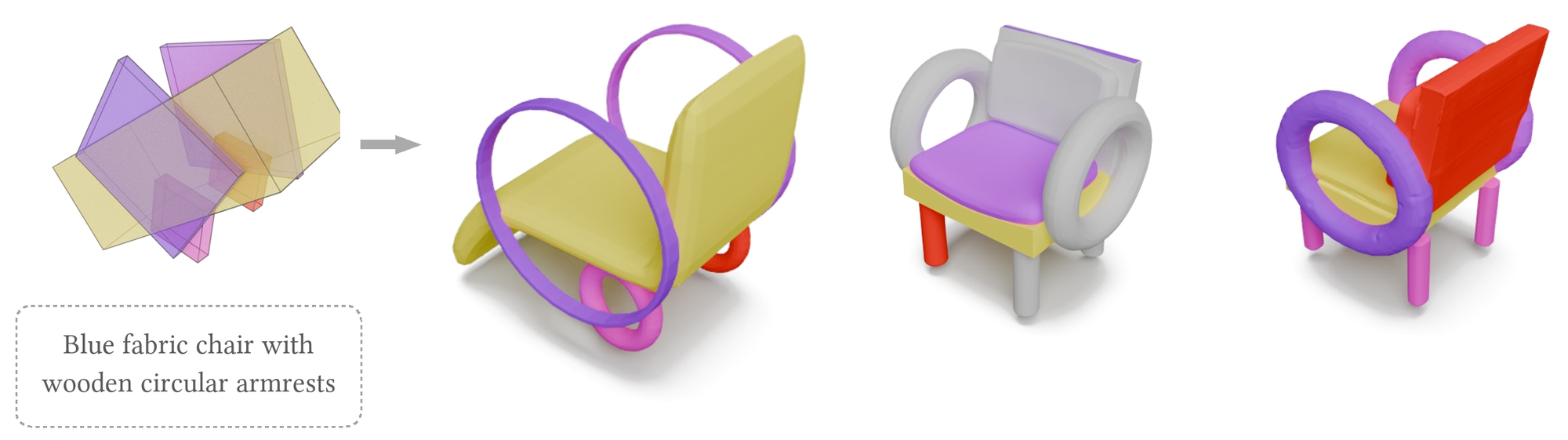}\par\vspace{0pt}}}
\vfill
{\centering{\includegraphics[width=0.98\textwidth]{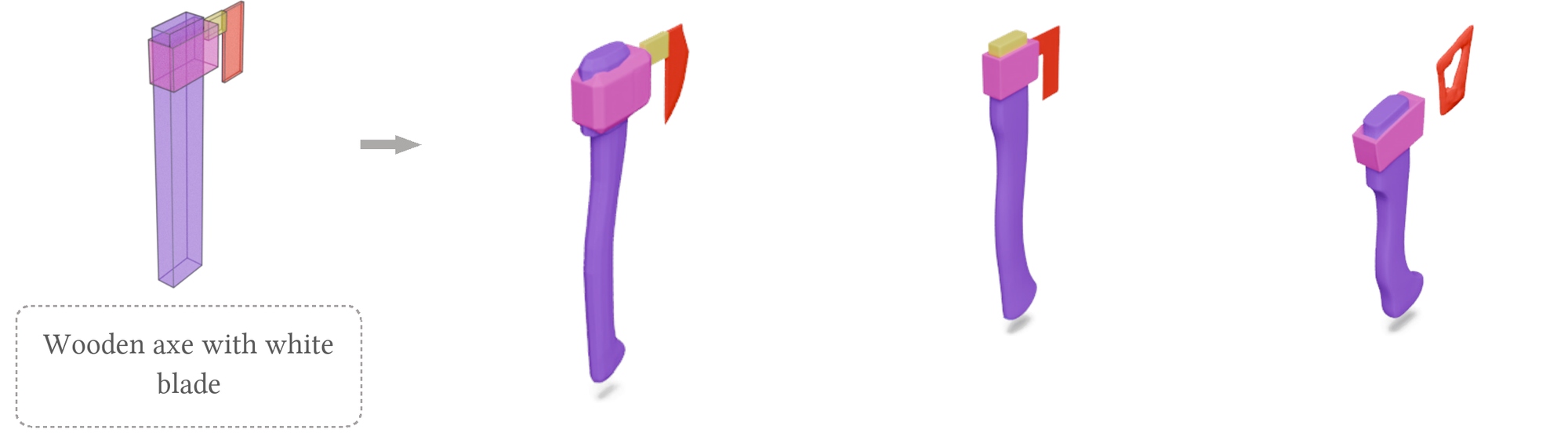}\par\vspace{0pt}}}
\vfill
{\centering{\includegraphics[width=0.98\textwidth]{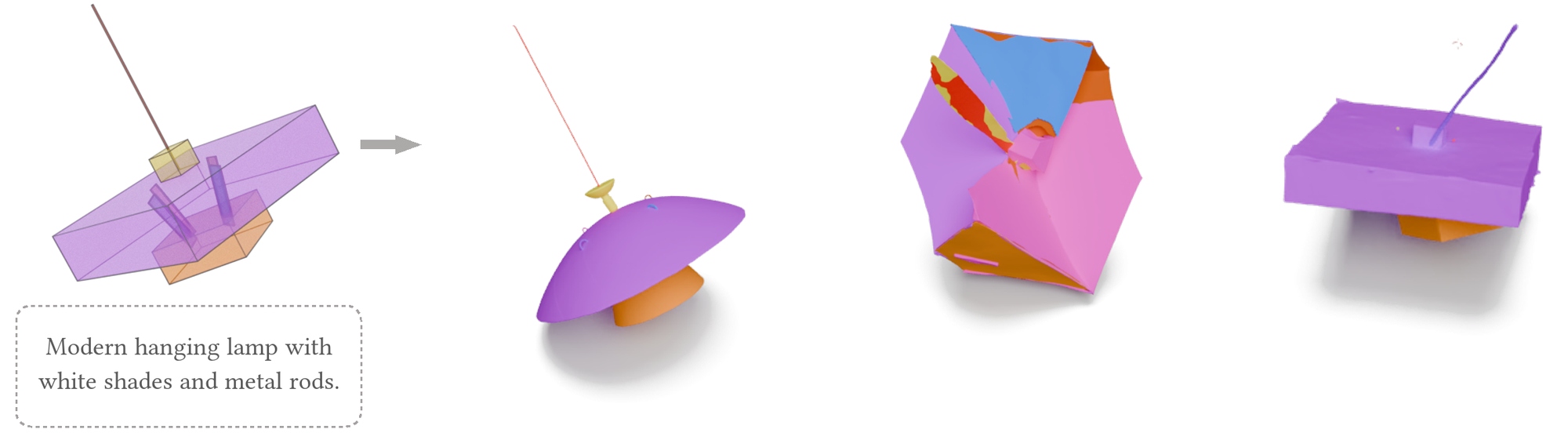}\par\vspace{0pt}}}
\clearpage
{\centering{\includegraphics[width=0.98\textwidth]{fig/rows_d4/header}\par\vspace{0pt}}}
{\centering{\includegraphics[width=0.98\textwidth]{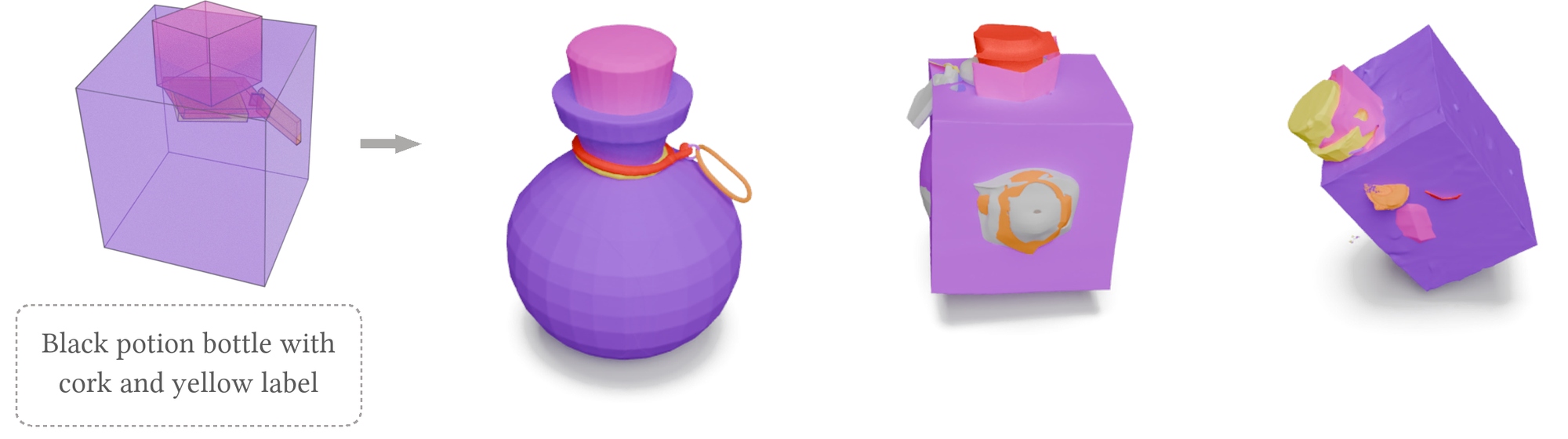}\par\vspace{0pt}}}
\vfill
{\centering{\includegraphics[width=0.98\textwidth]{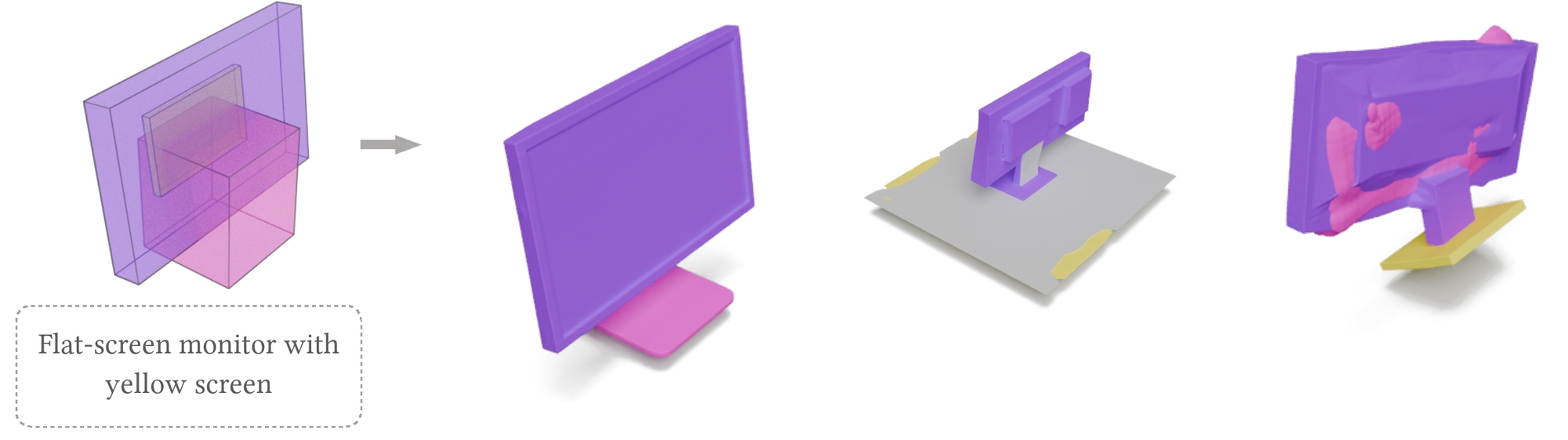}\par\vspace{0pt}}}
\vfill
{\centering{\includegraphics[width=0.98\textwidth]{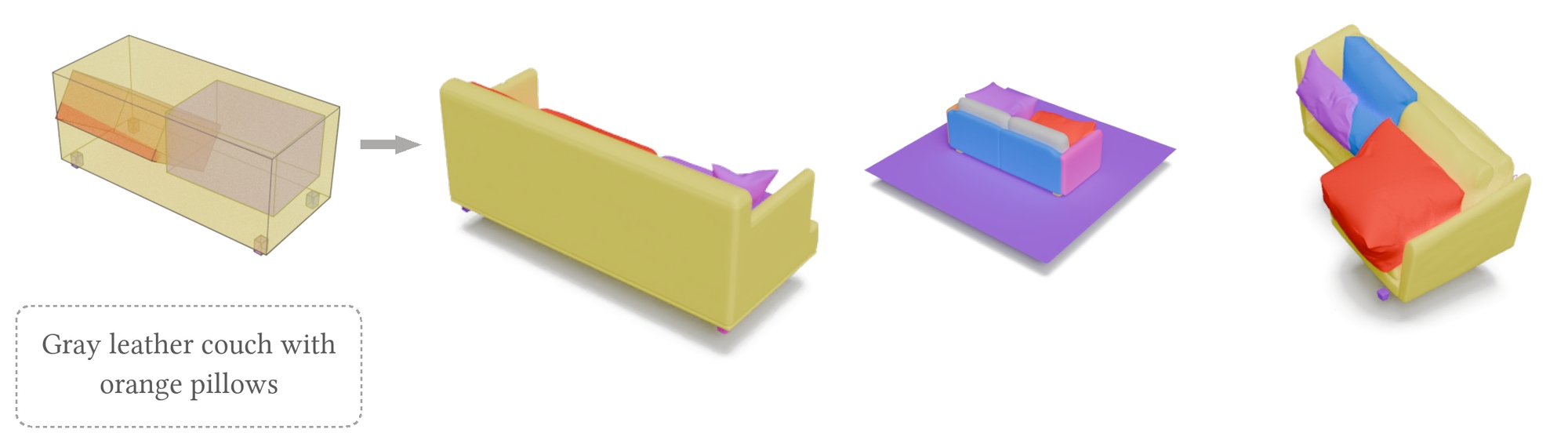}\par\vspace{0pt}}}
\vfill
{\centering{\includegraphics[width=0.98\textwidth]{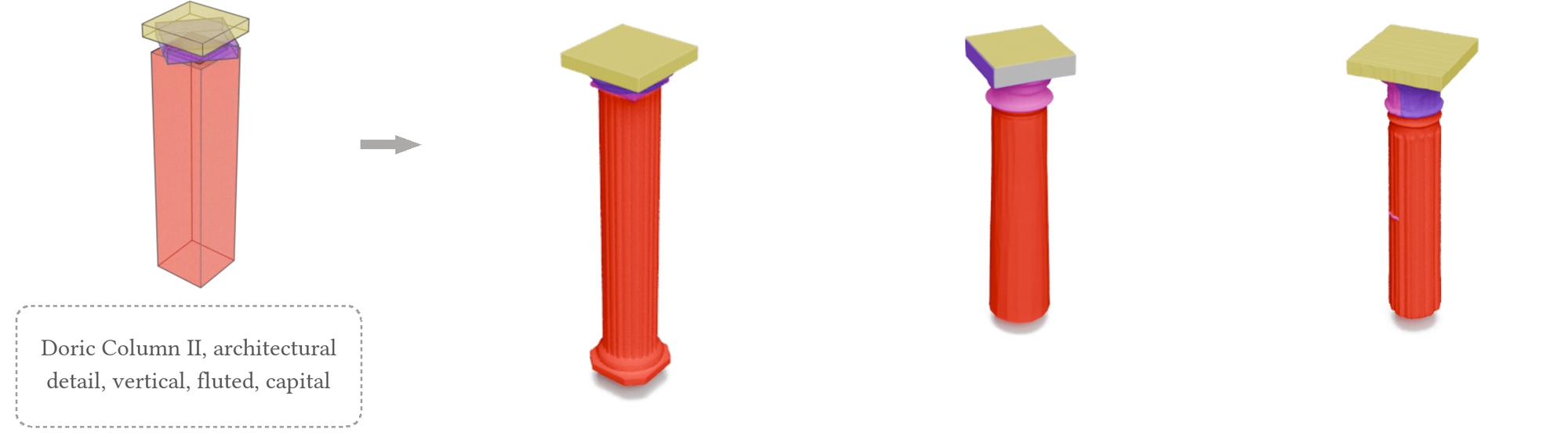}\par\vspace{0pt}}}
\vfill
{\centering{\includegraphics[width=0.98\textwidth]{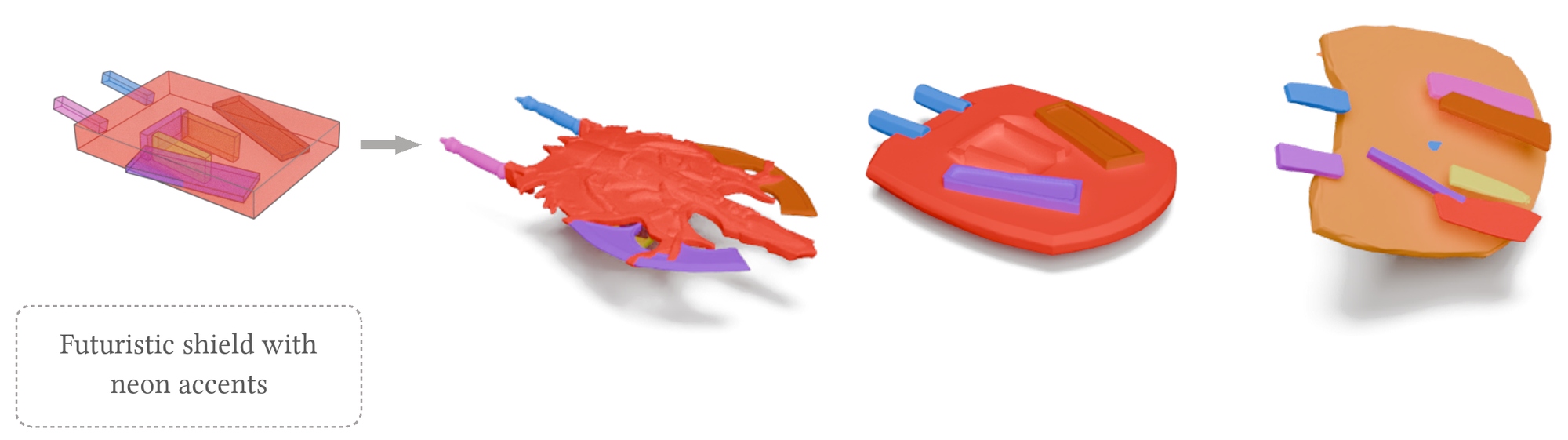}\par\vspace{0pt}}}
\clearpage
{\centering{\includegraphics[width=0.98\textwidth]{fig/rows_d4/header}\par\vspace{0pt}}}
{\centering{\includegraphics[width=0.98\textwidth]{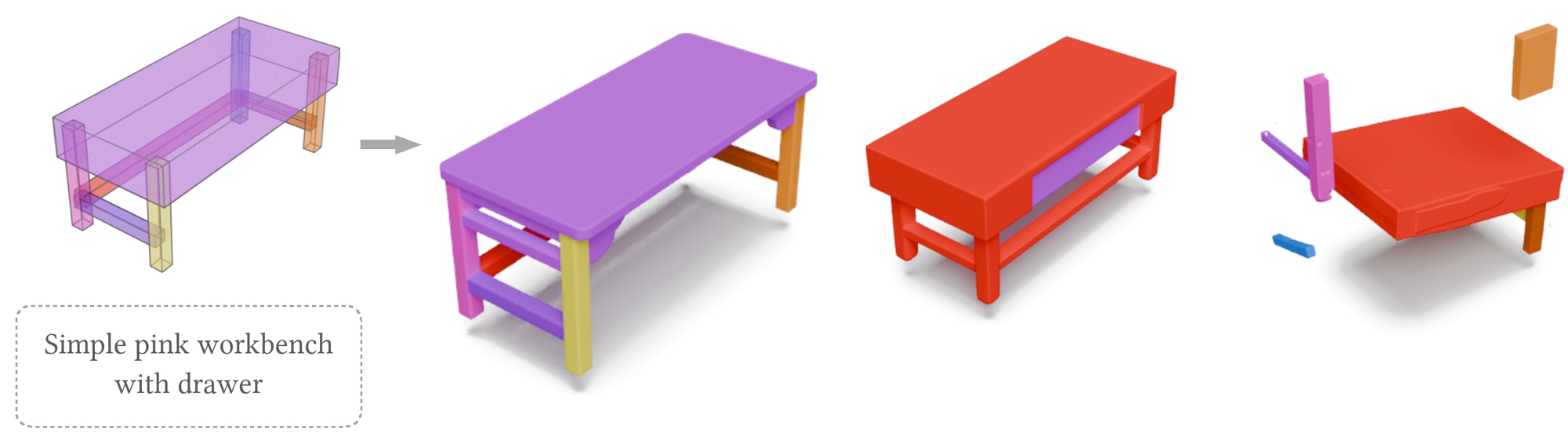}\par\vspace{0pt}}}
\vfill
{\centering{\includegraphics[width=0.98\textwidth]{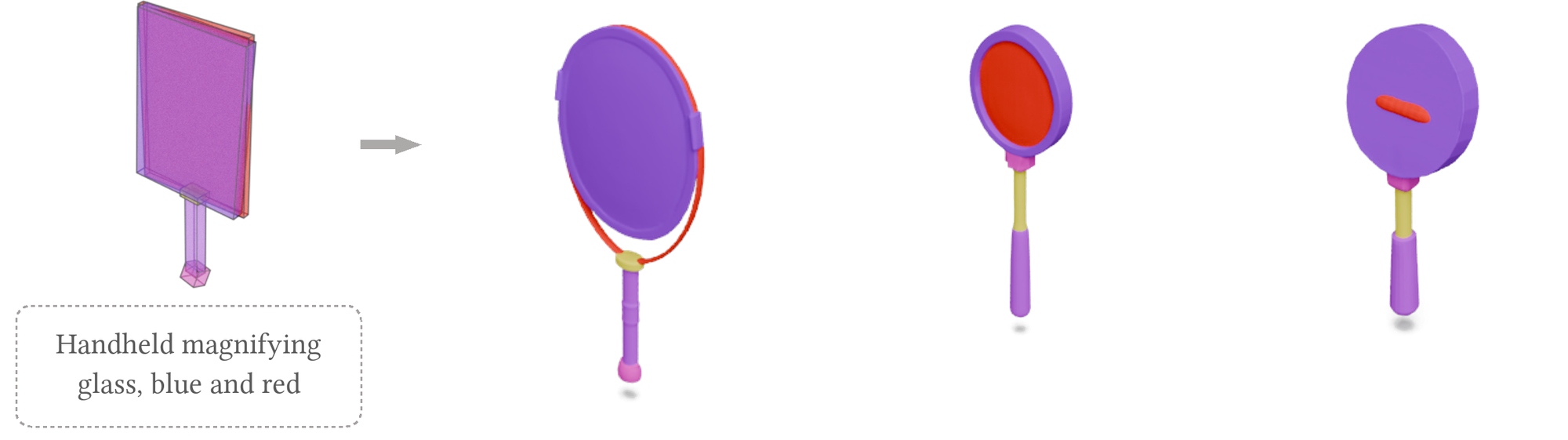}\par\vspace{0pt}}}
\vfill
{\centering{\includegraphics[width=0.98\textwidth]{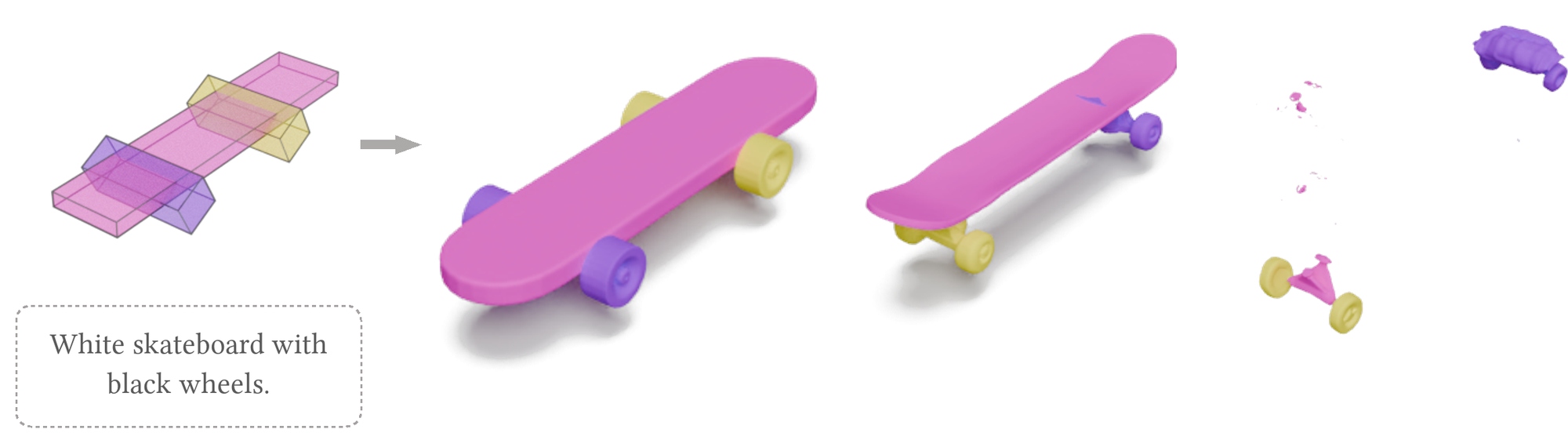}\par\vspace{0pt}}}
\vfill
{\centering{\includegraphics[width=0.98\textwidth]{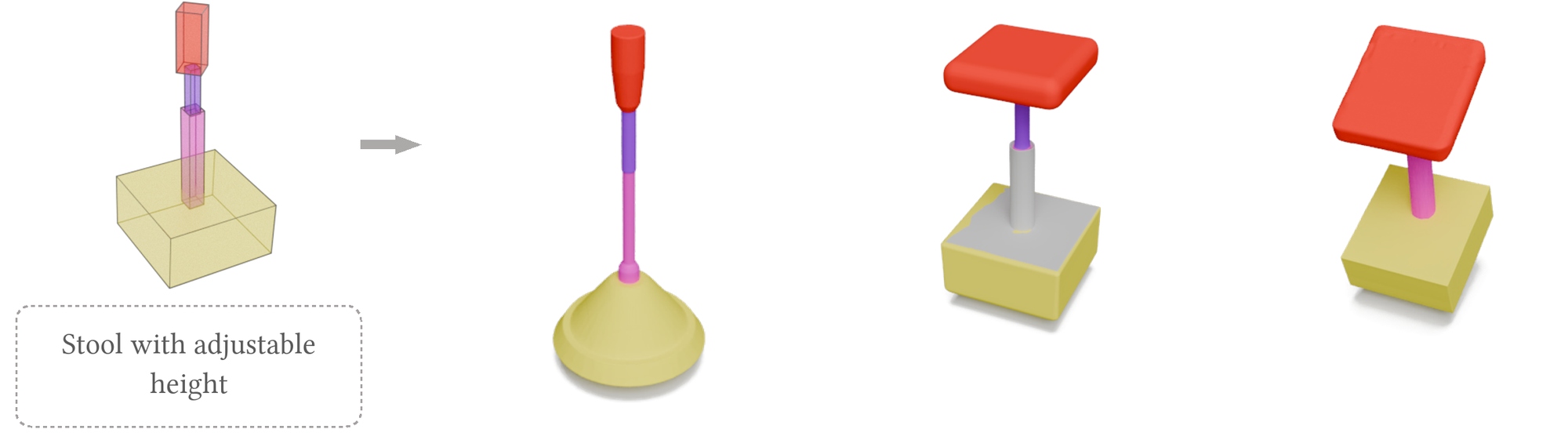}\par\vspace{0pt}}}
\vfill
{\centering{\includegraphics[width=0.98\textwidth]{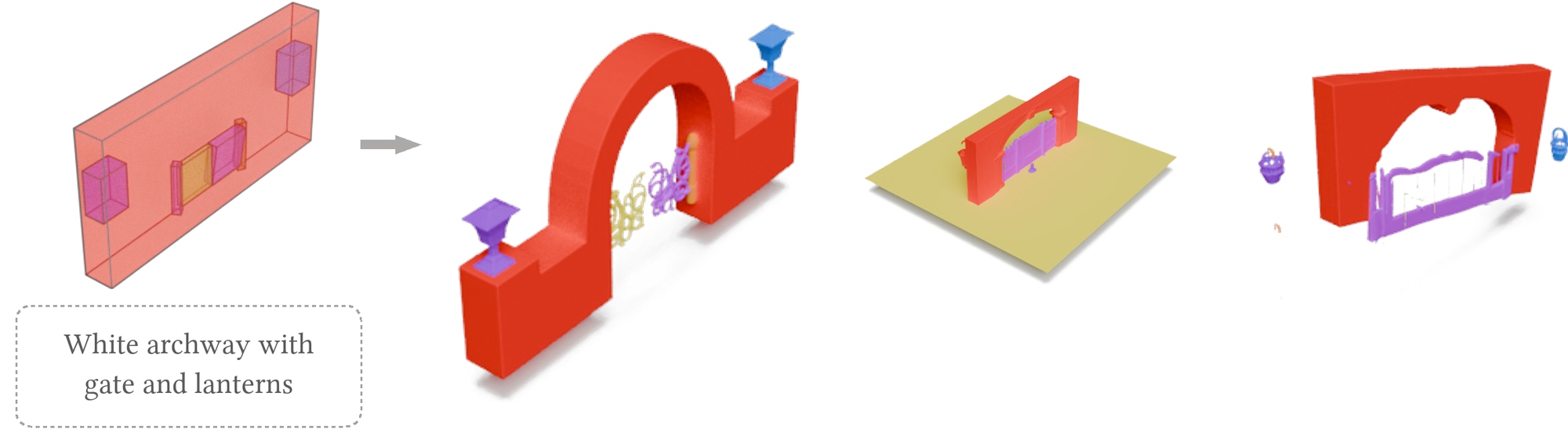}\par\vspace{0pt}}}
\clearpage
{\centering{\includegraphics[width=0.98\textwidth]{fig/rows_d4/header}\par\vspace{0pt}}}
{\centering{\includegraphics[width=0.98\textwidth]{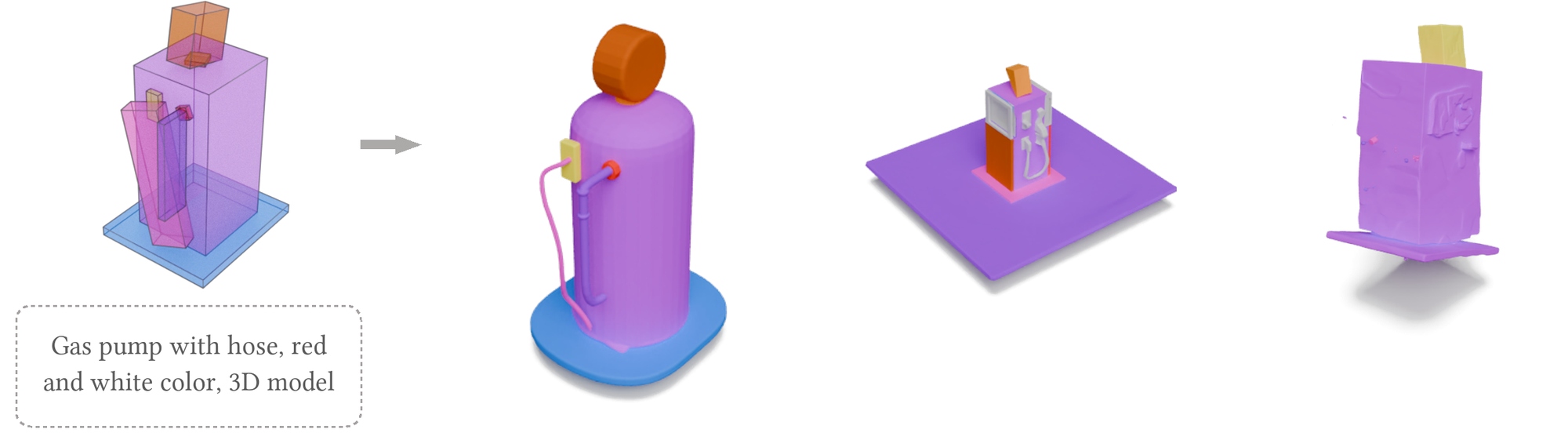}\par\vspace{0pt}}}
\vfill
{\centering{\includegraphics[width=0.98\textwidth]{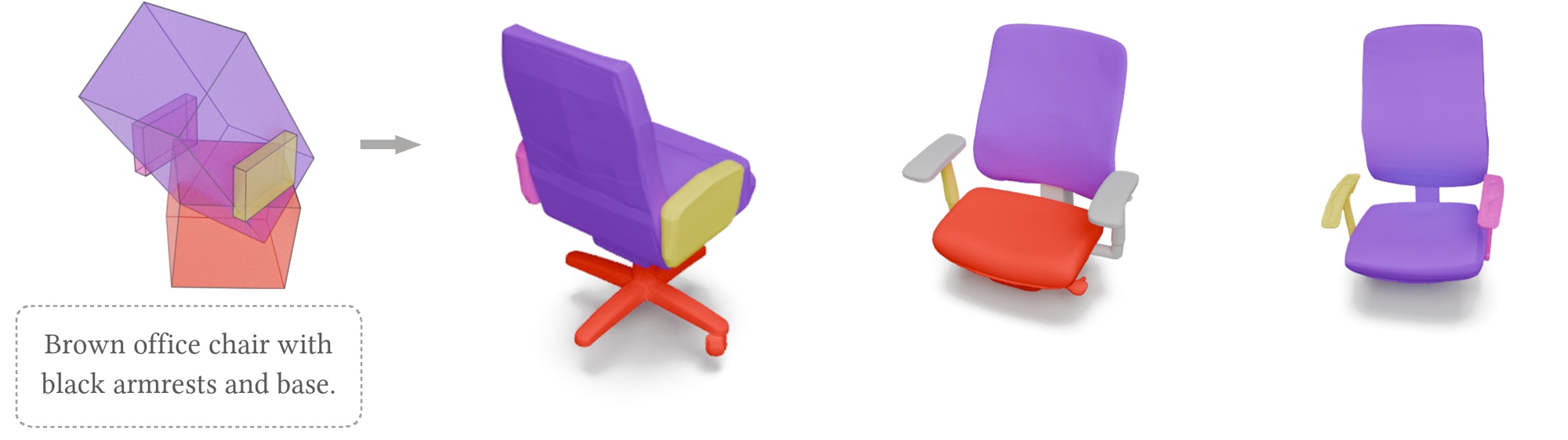}\par\vspace{0pt}}}
\vfill
{\centering{\includegraphics[width=0.98\textwidth]{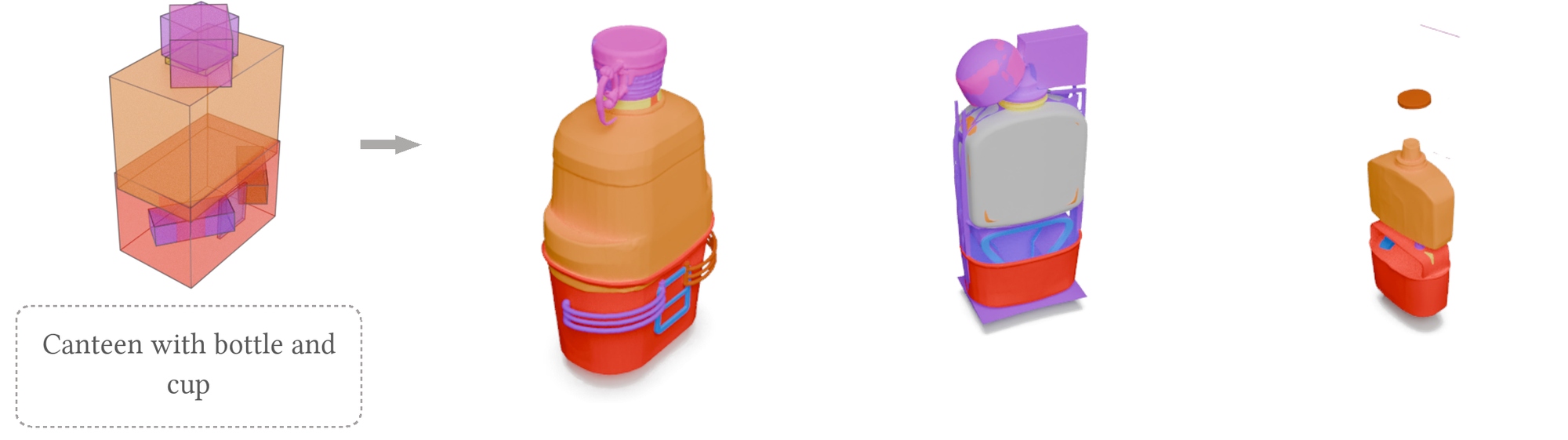}\par\vspace{0pt}}}
\vfill
{\centering{\includegraphics[width=0.98\textwidth]{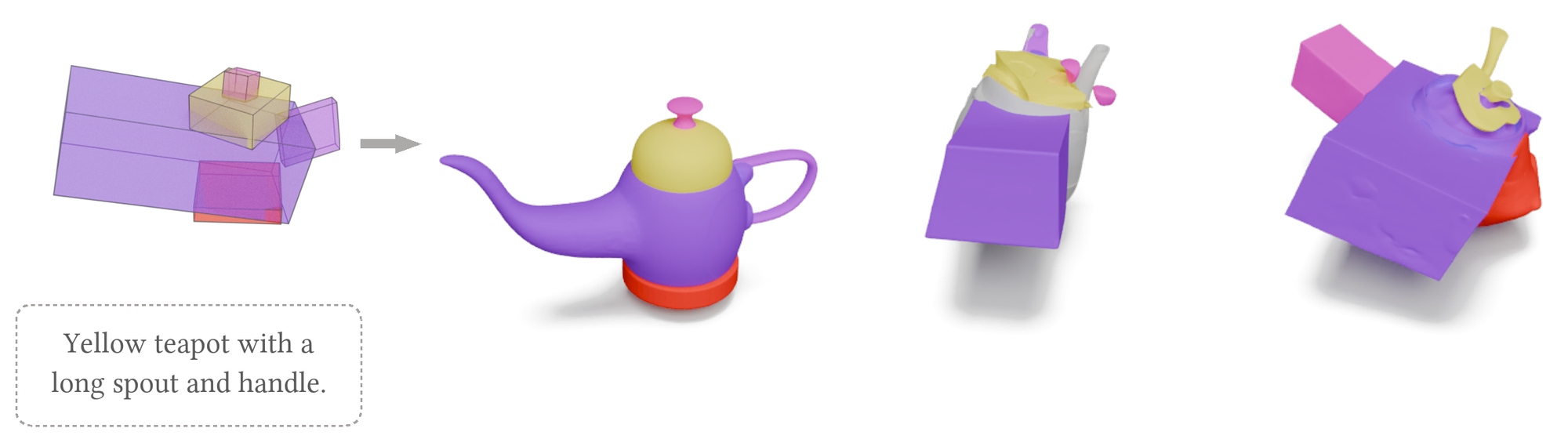}\par\vspace{0pt}}}
\vfill
{\centering{\includegraphics[width=0.98\textwidth]{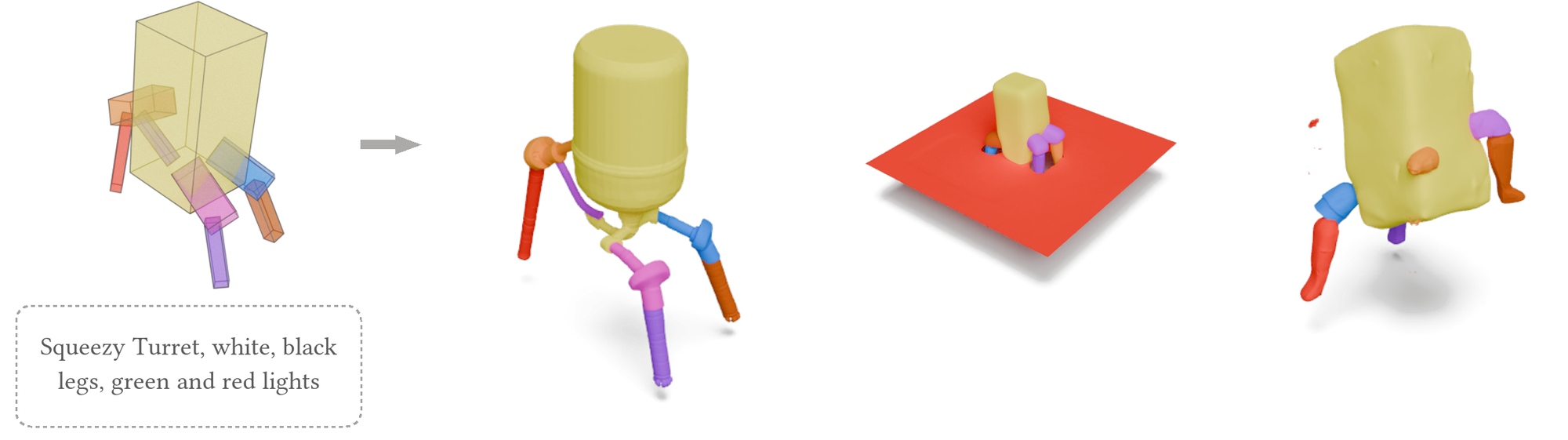}\par\vspace{0pt}}}
\clearpage
{\centering{\includegraphics[width=0.98\textwidth]{fig/rows_d4/header}\par\vspace{0pt}}}
{\centering{\includegraphics[width=0.98\textwidth]{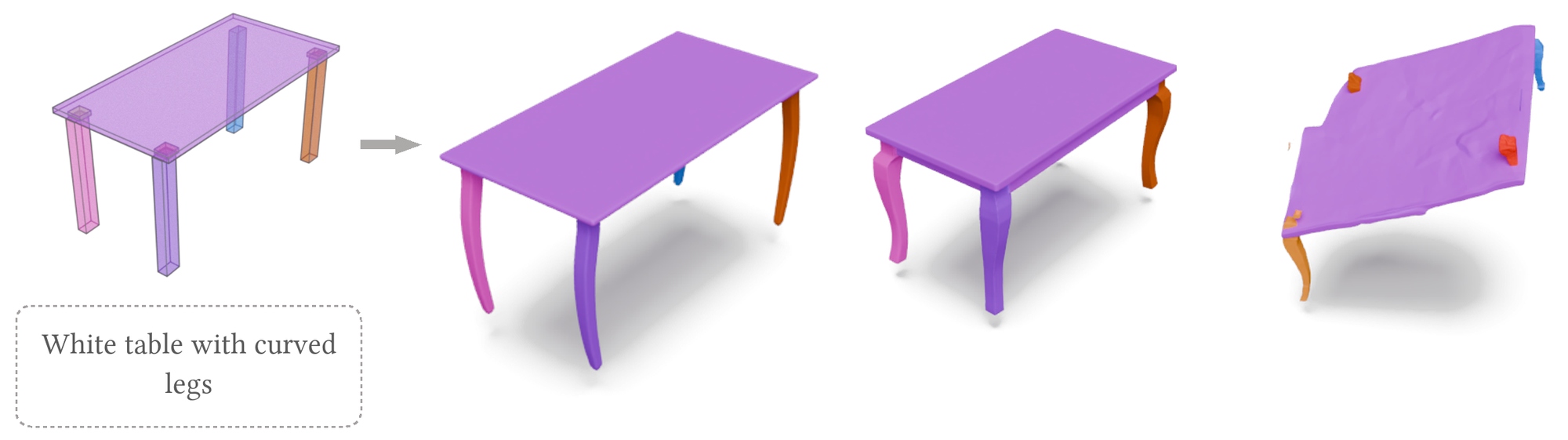}\par\vspace{0pt}}}
\vfill
{\centering{\includegraphics[width=0.98\textwidth]{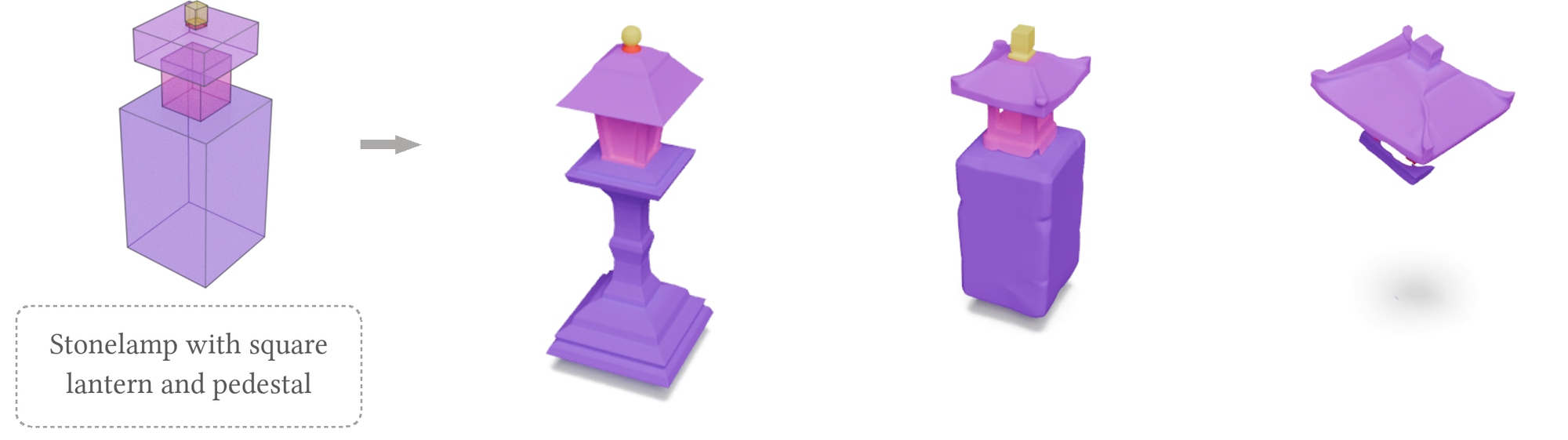}\par\vspace{0pt}}}
\vfill
{\centering{\includegraphics[width=0.98\textwidth]{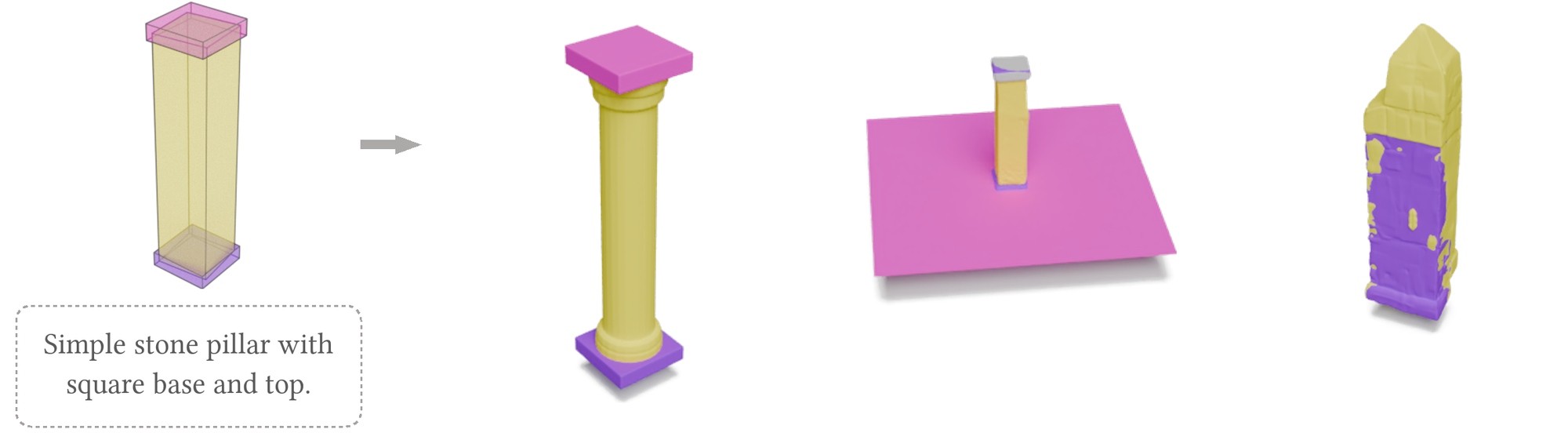}\par\vspace{0pt}}}
\vfill
{\centering{\includegraphics[width=0.98\textwidth]{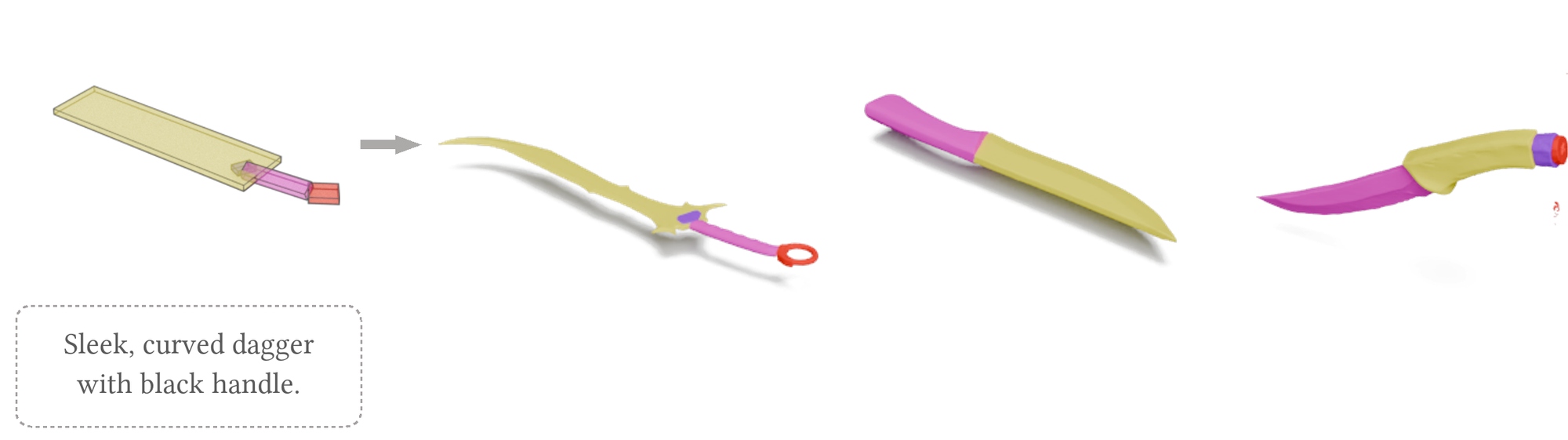}\par\vspace{0pt}}}
\vfill
{\centering{\includegraphics[width=0.98\textwidth]{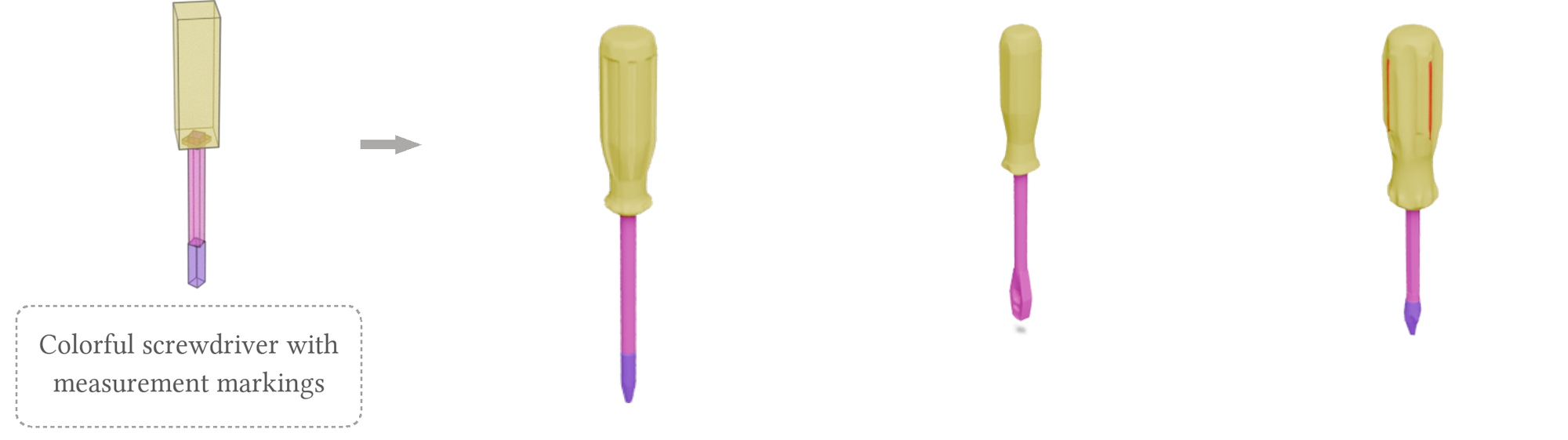}\par\vspace{0pt}}}
\clearpage
{\centering{\includegraphics[width=0.98\textwidth]{fig/rows_d4/header}\par\vspace{0pt}}}
{\centering{\includegraphics[width=0.98\textwidth]{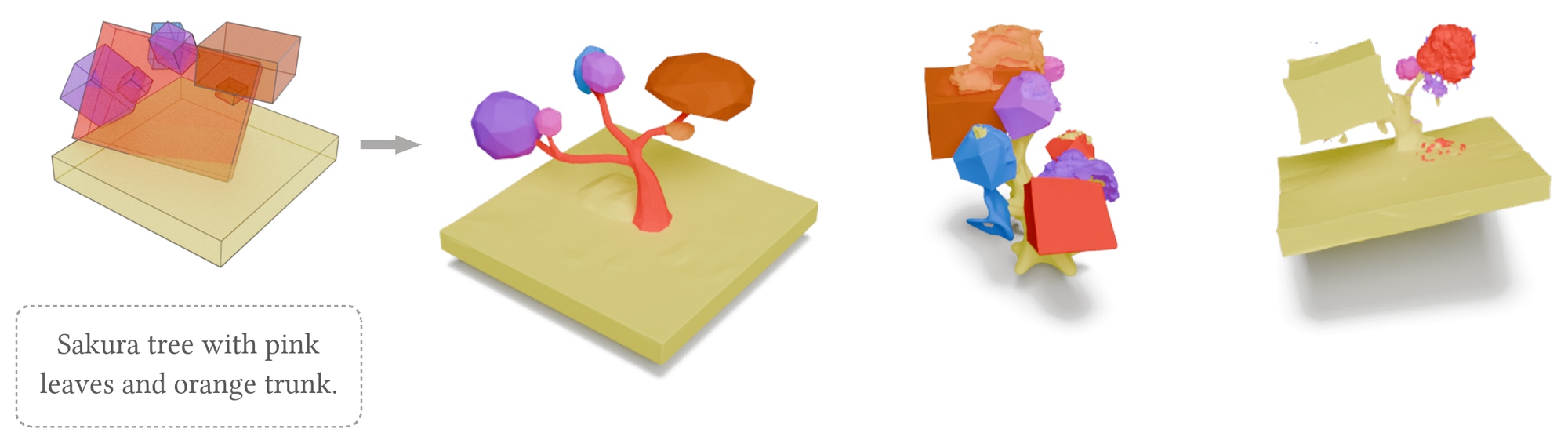}\par\vspace{0pt}}}
\vfill
{\centering{\includegraphics[width=0.98\textwidth]{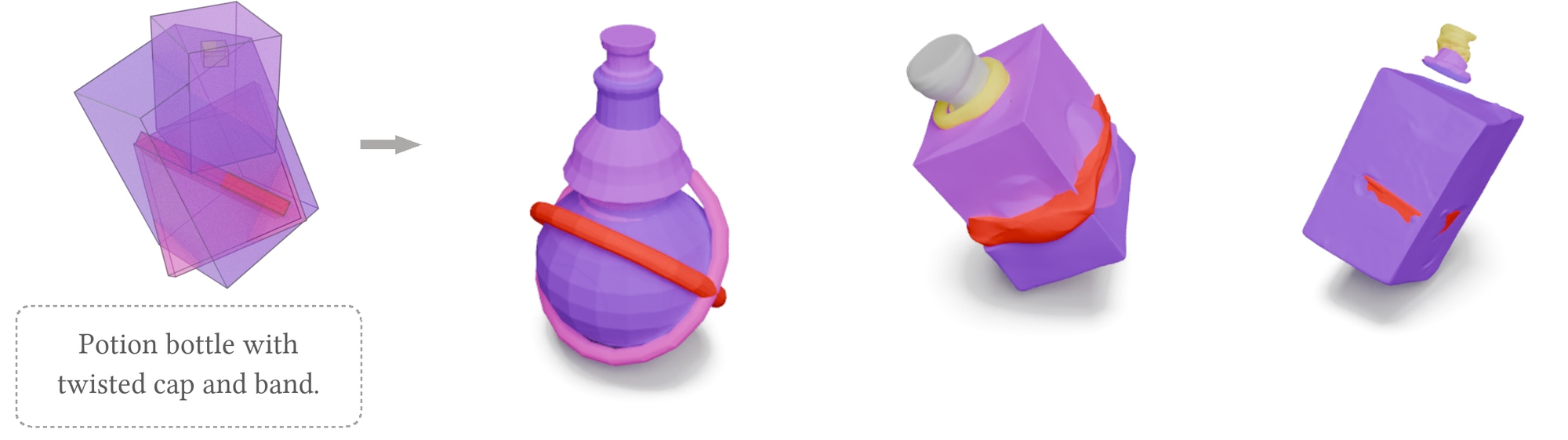}\par\vspace{0pt}}}
\vfill
{\centering{\includegraphics[width=0.98\textwidth]{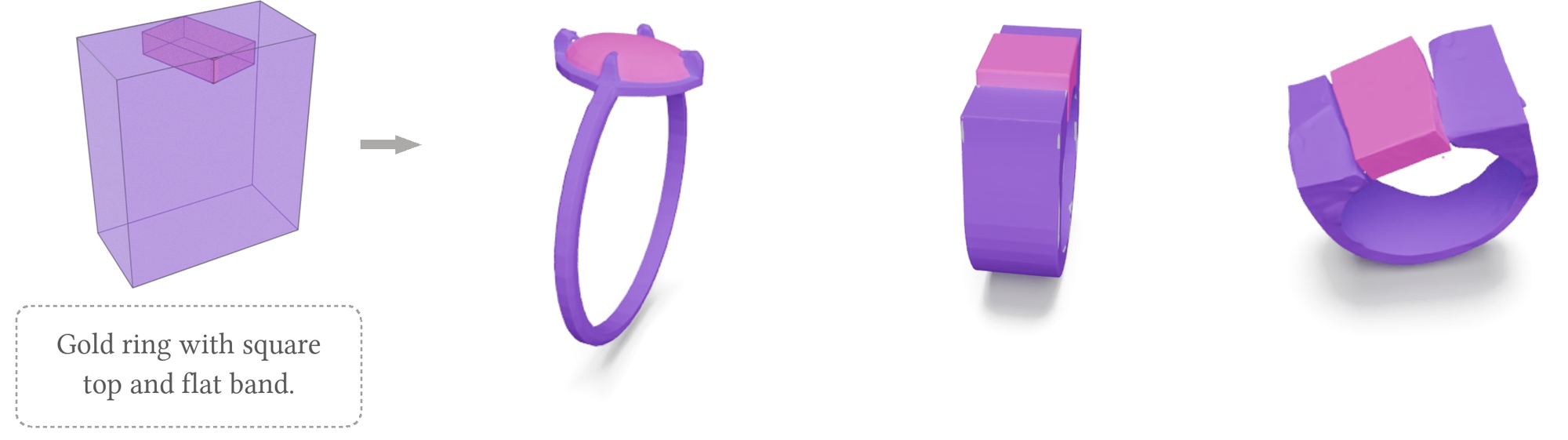}\par\vspace{0pt}}}
\vfill
{\centering{\includegraphics[width=0.98\textwidth]{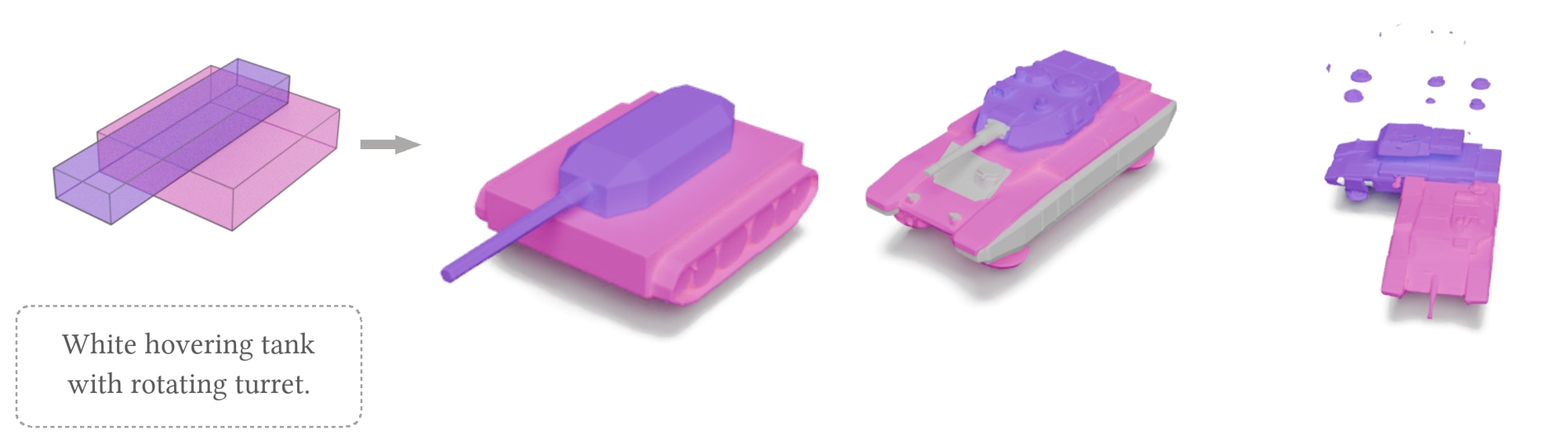}\par\vspace{0pt}}}
\vfill
{\centering{\includegraphics[width=0.98\textwidth]{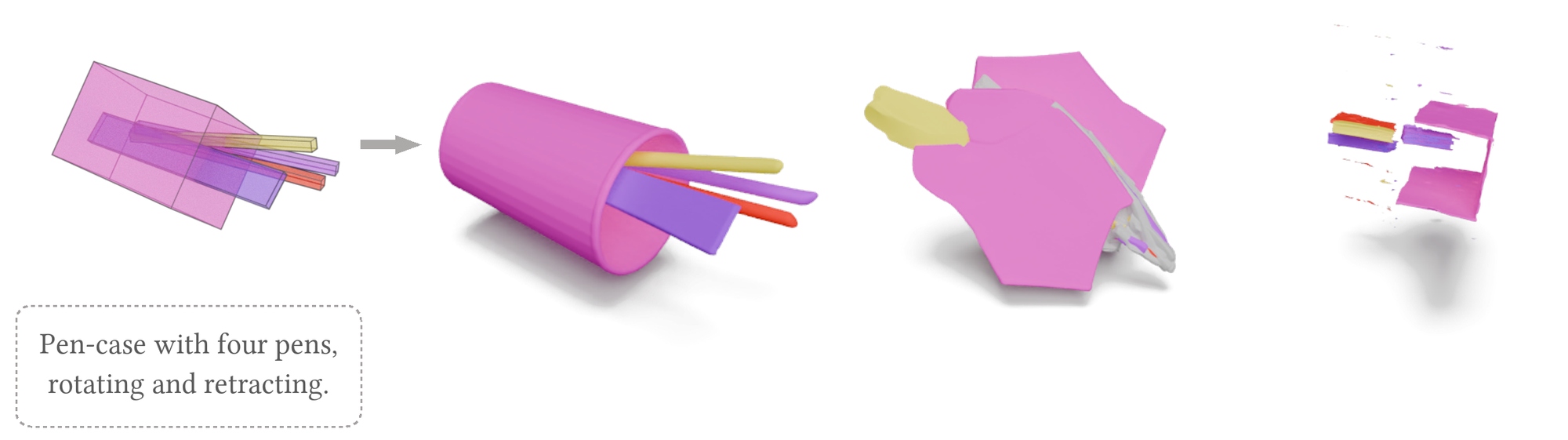}\par\vspace{0pt}}}
\clearpage
{\centering{\includegraphics[width=0.98\textwidth]{fig/rows_d4/header}\par\vspace{0pt}}}
{\centering{\includegraphics[width=0.98\textwidth]{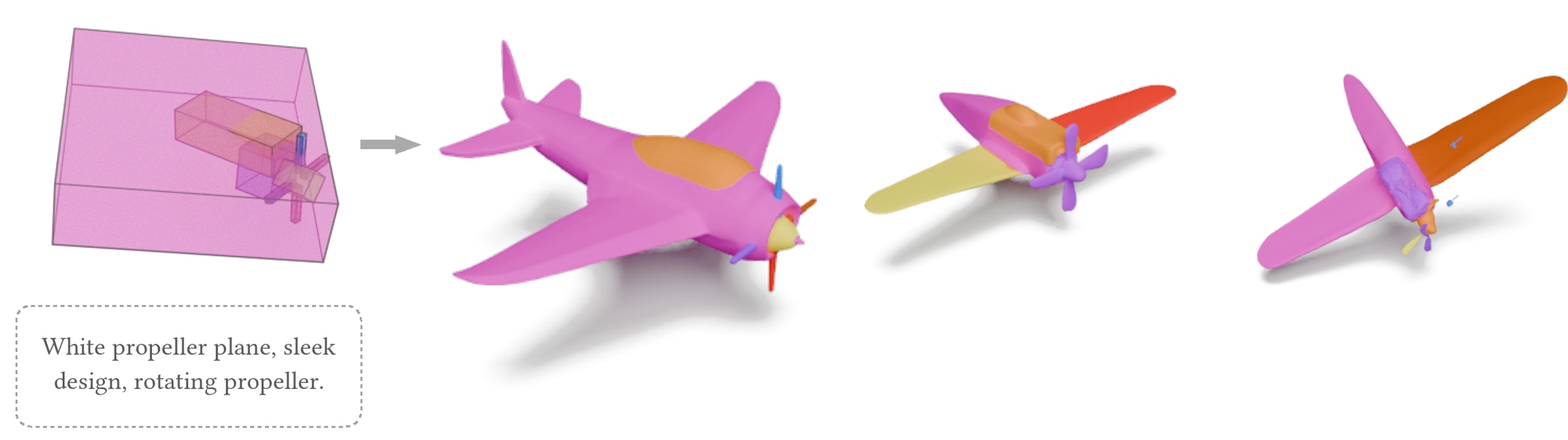}\par\vspace{0pt}}}
\vfill
{\centering{\includegraphics[width=0.98\textwidth]{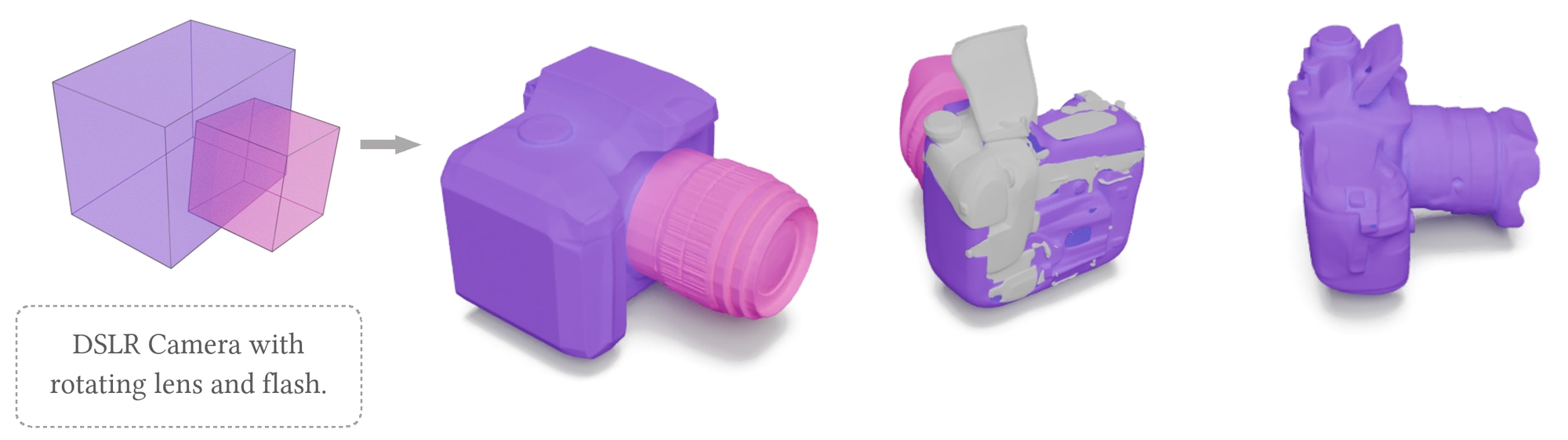}\par\vspace{0pt}}}
\vfill
{\centering{\includegraphics[width=0.98\textwidth]{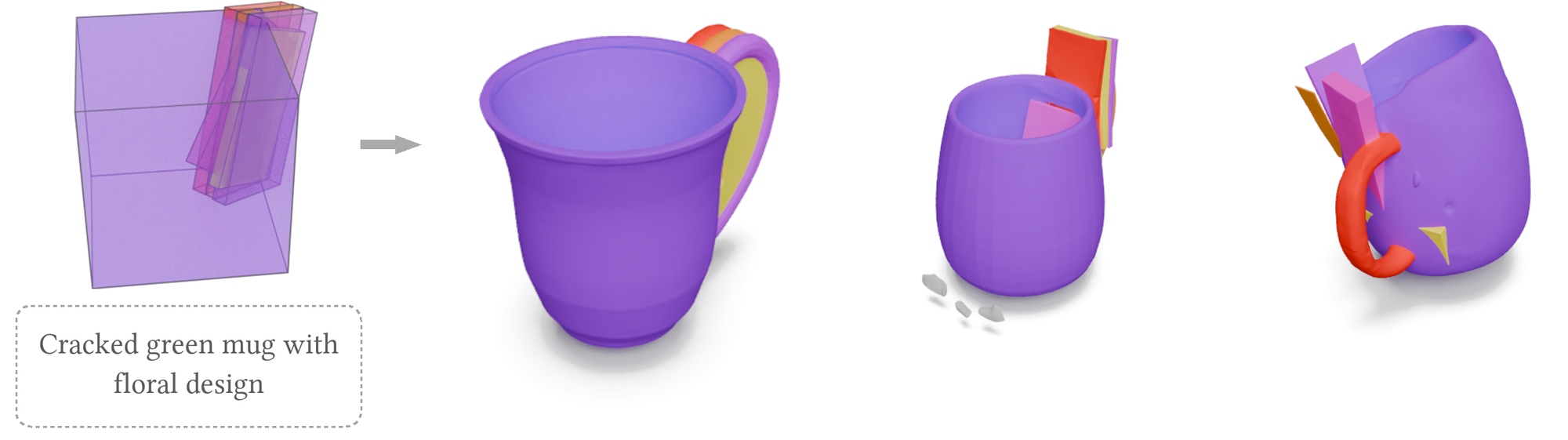}\par\vspace{0pt}}}
\vfill
{\centering{\includegraphics[width=0.98\textwidth]{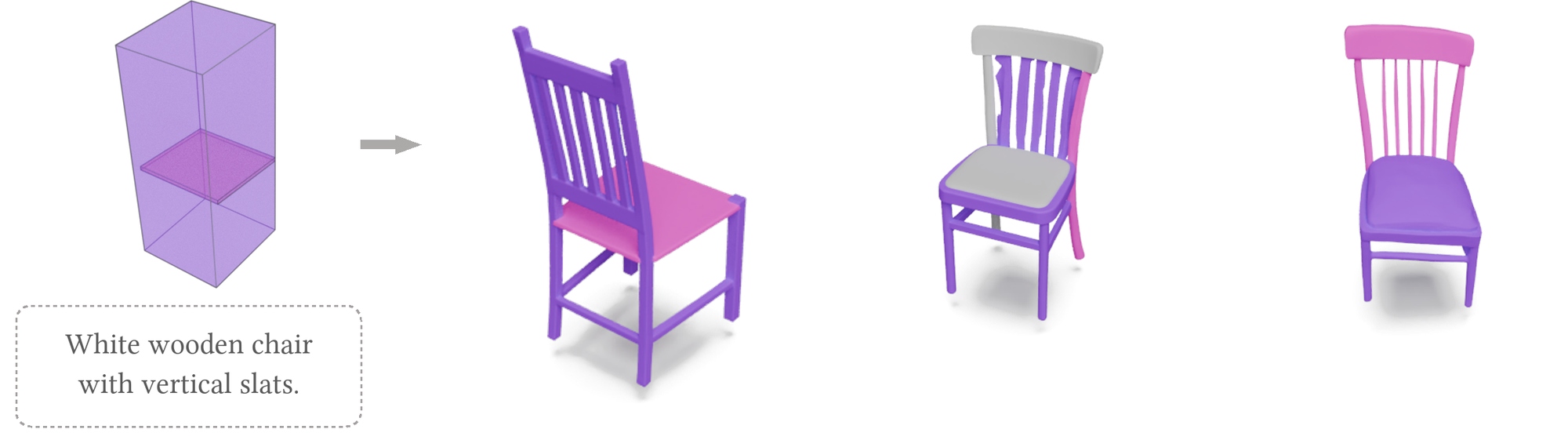}\par\vspace{0pt}}}
\vfill
{\centering{\includegraphics[width=0.98\textwidth]{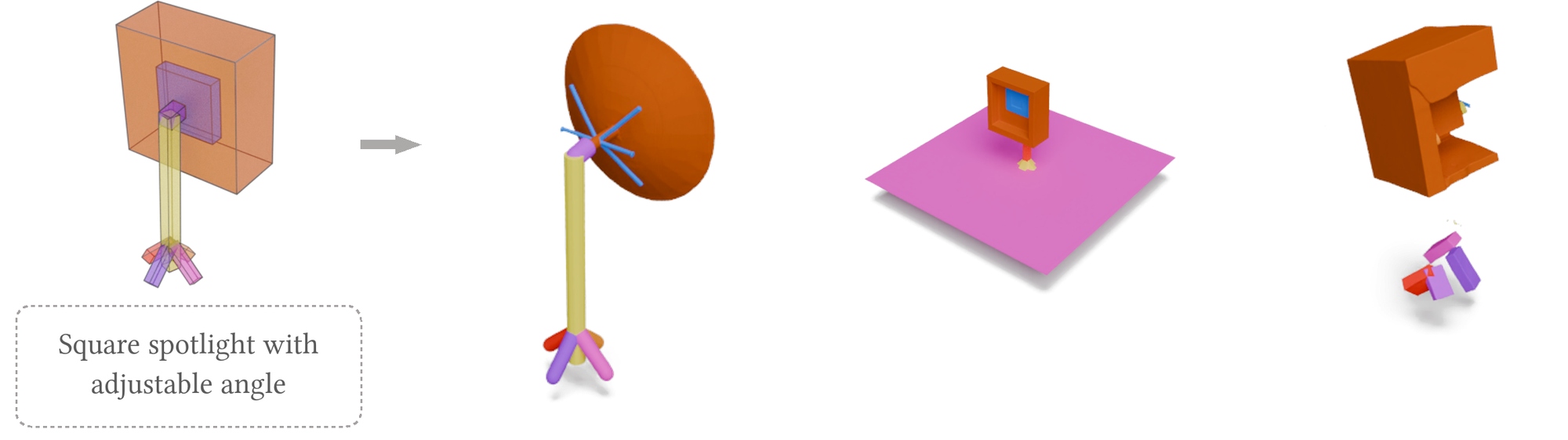}\par\vspace{0pt}}}
\clearpage
{\centering{\includegraphics[width=0.98\textwidth]{fig/rows_d4/header}\par\vspace{0pt}}}
{\centering{\includegraphics[width=0.98\textwidth]{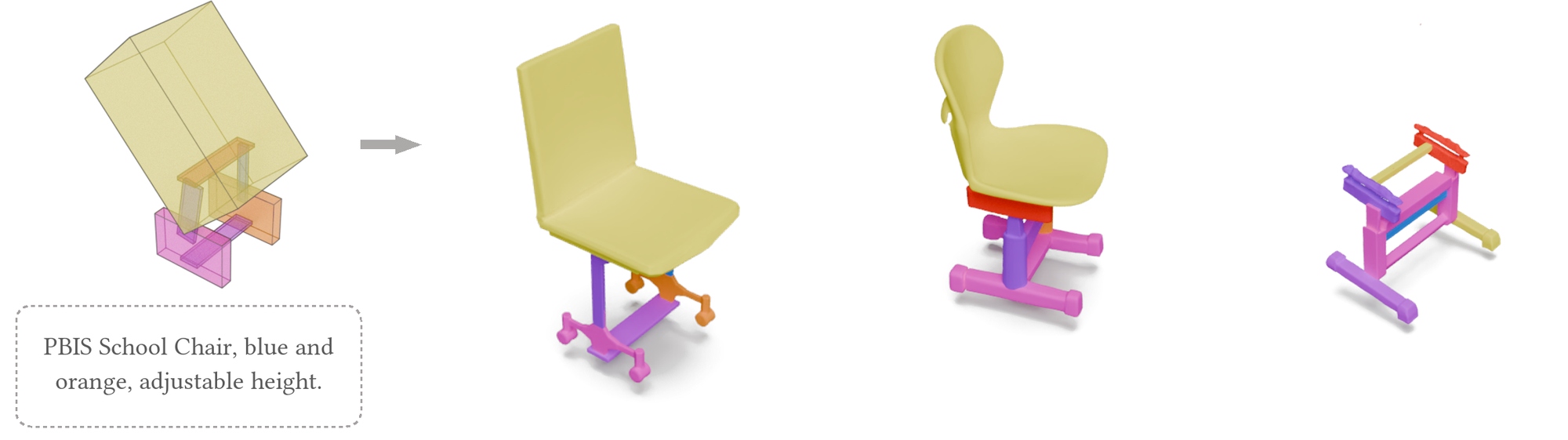}\par\vspace{0pt}}}
\vfill
{\centering{\includegraphics[width=0.98\textwidth]{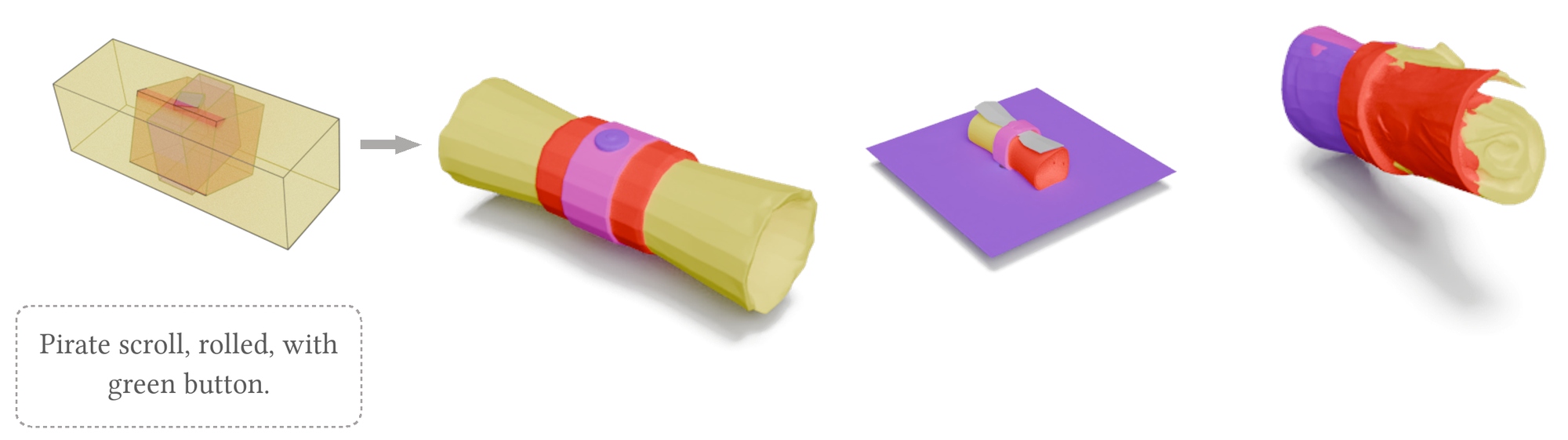}\par\vspace{0pt}}}
\vfill
{\centering{\includegraphics[width=0.98\textwidth]{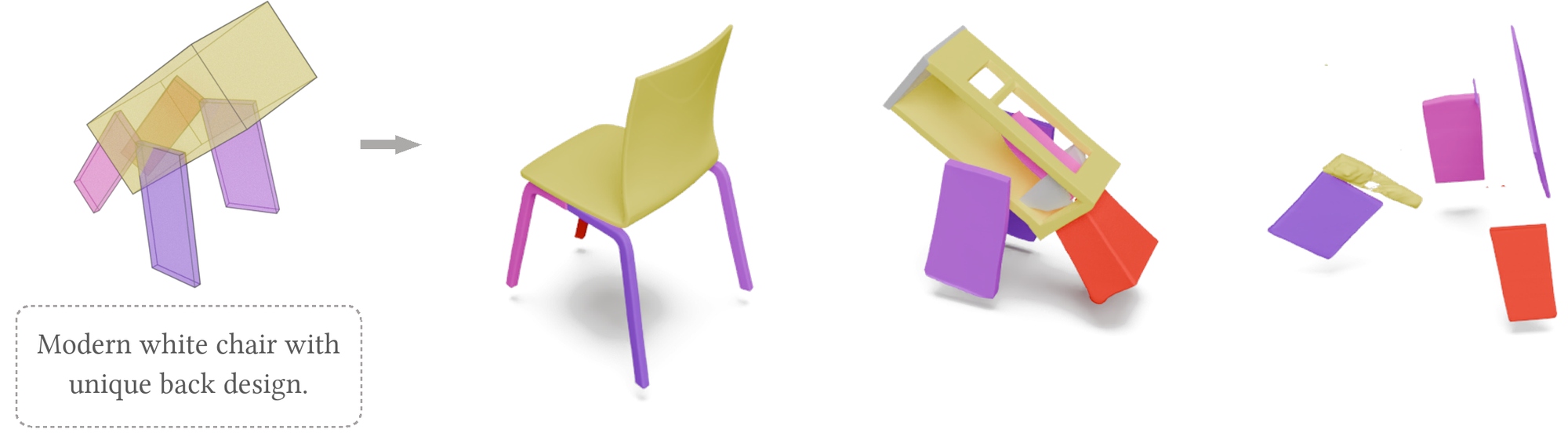}\par\vspace{0pt}}}
\vfill
{\centering{\includegraphics[width=0.98\textwidth]{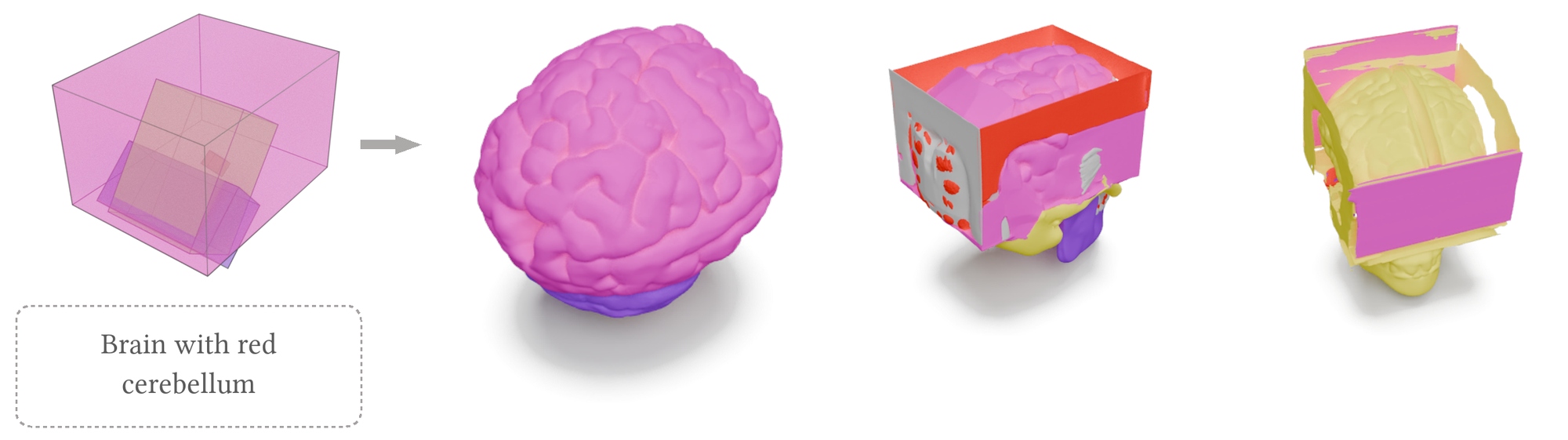}\par\vspace{0pt}}}
\vfill
{\centering{\includegraphics[width=0.98\textwidth]{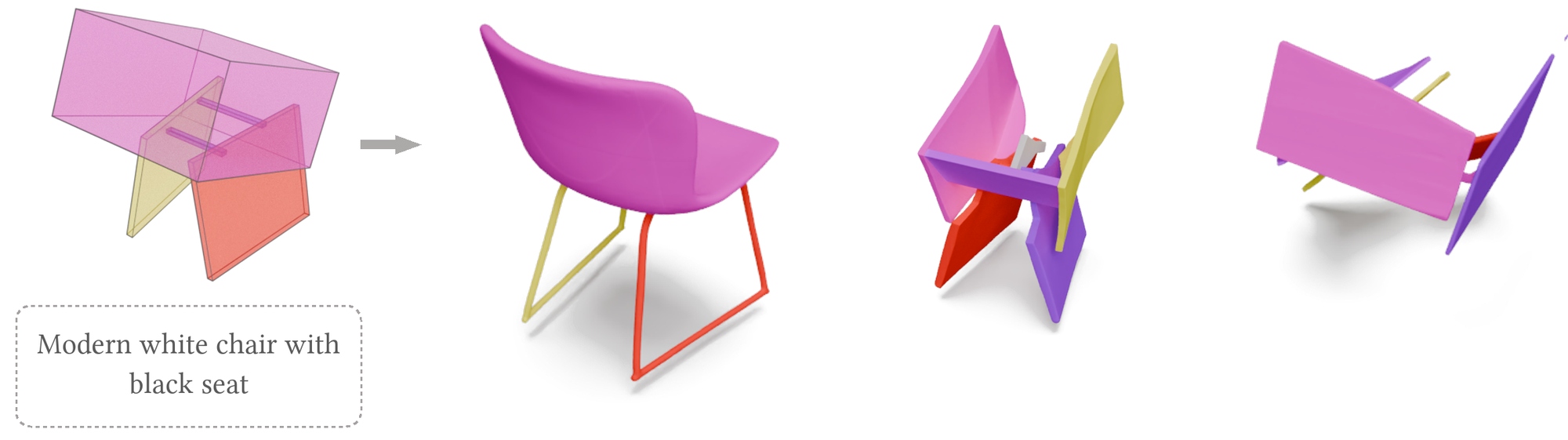}\par\vspace{0pt}}}
\clearpage

\FloatBarrier
\section{Failure Cases}
\label{sec:failure_cases}

\begin{figure*}[h]
    \centering
    \includegraphics[width=0.88\textwidth]{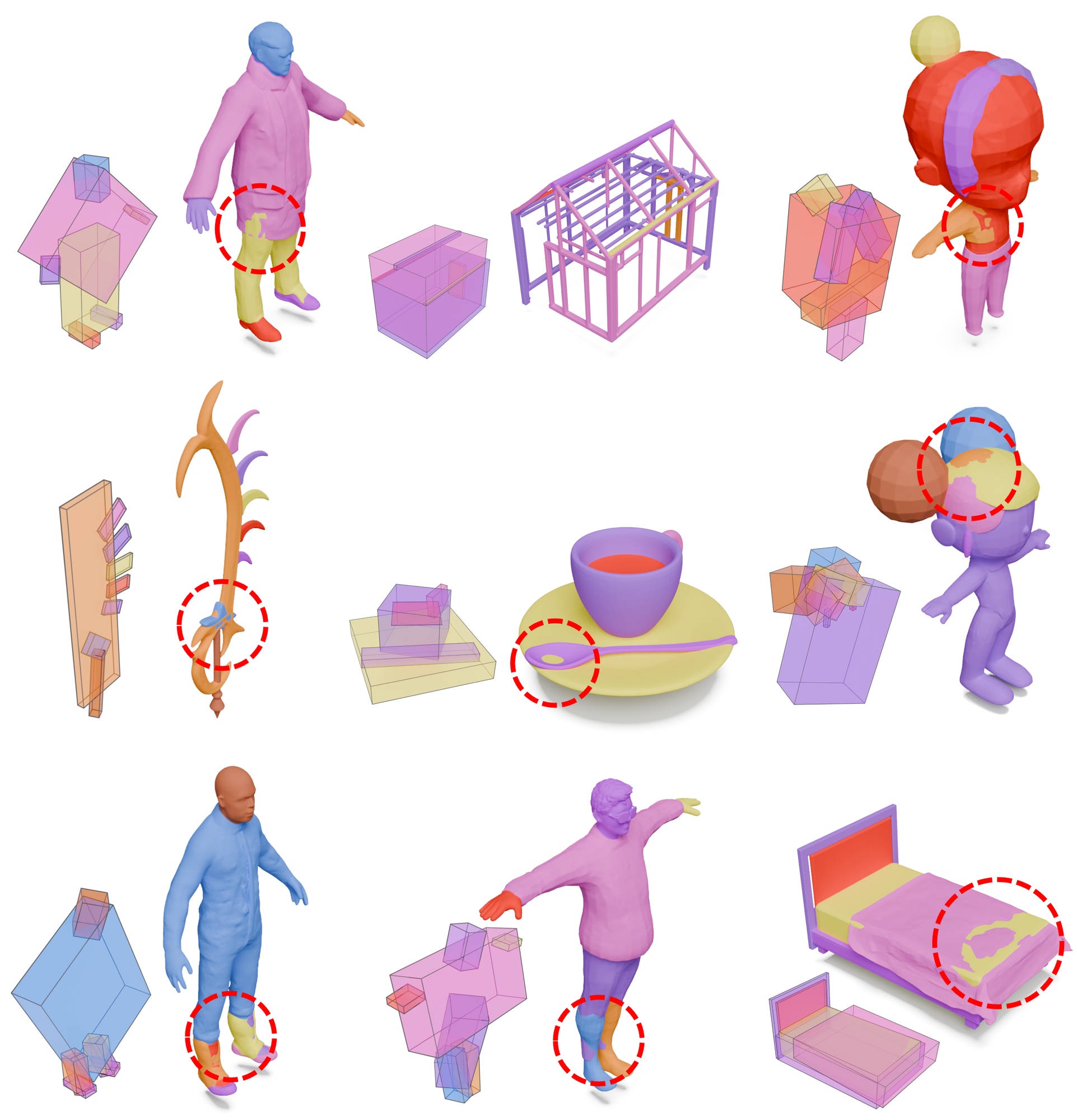}
    \vspace{1em}
    \caption{\textbf{Failure Case Examples.}
    We highlight a few failure cases of our method, which mostly include z-fighting artifacts between parts, and awkward part geometries especially prevalent in complex, highly overlapping layouts.
    }%
    \label{fig:failure_cases}
\end{figure*}

While our method achieves strong results across a wide range of inputs, we observe a few characteristic failure modes, illustrated in~\cref{fig:failure_cases}. The most common artifacts are mild z-fighting along the boundaries between adjacent parts, which arises when neighboring part surfaces are generated in close spatial proximity without explicit awareness of one another. We also occasionally observe awkward part geometries in cases with complex, highly overlapping layouts, where the model must reconcile a large number of competing spatial constraints within a small region. Importantly, these issues are localized and do not compromise the overall structural coherence or semantic alignment of the generated shapes, and we expect them to be straightforward to address in future work through improved cross-part consistency mechanisms or light post-processing.

\vfill
\vfill

\newpage
\FloatBarrier
\section{Implementation Details}
\label{sec:impl_details}

\FloatBarrier
\subsection{Architecture Details}
\label{sec:architecture_details}

\begin{figure*}[h]
    \centering
    \includegraphics[width=0.80\textwidth]{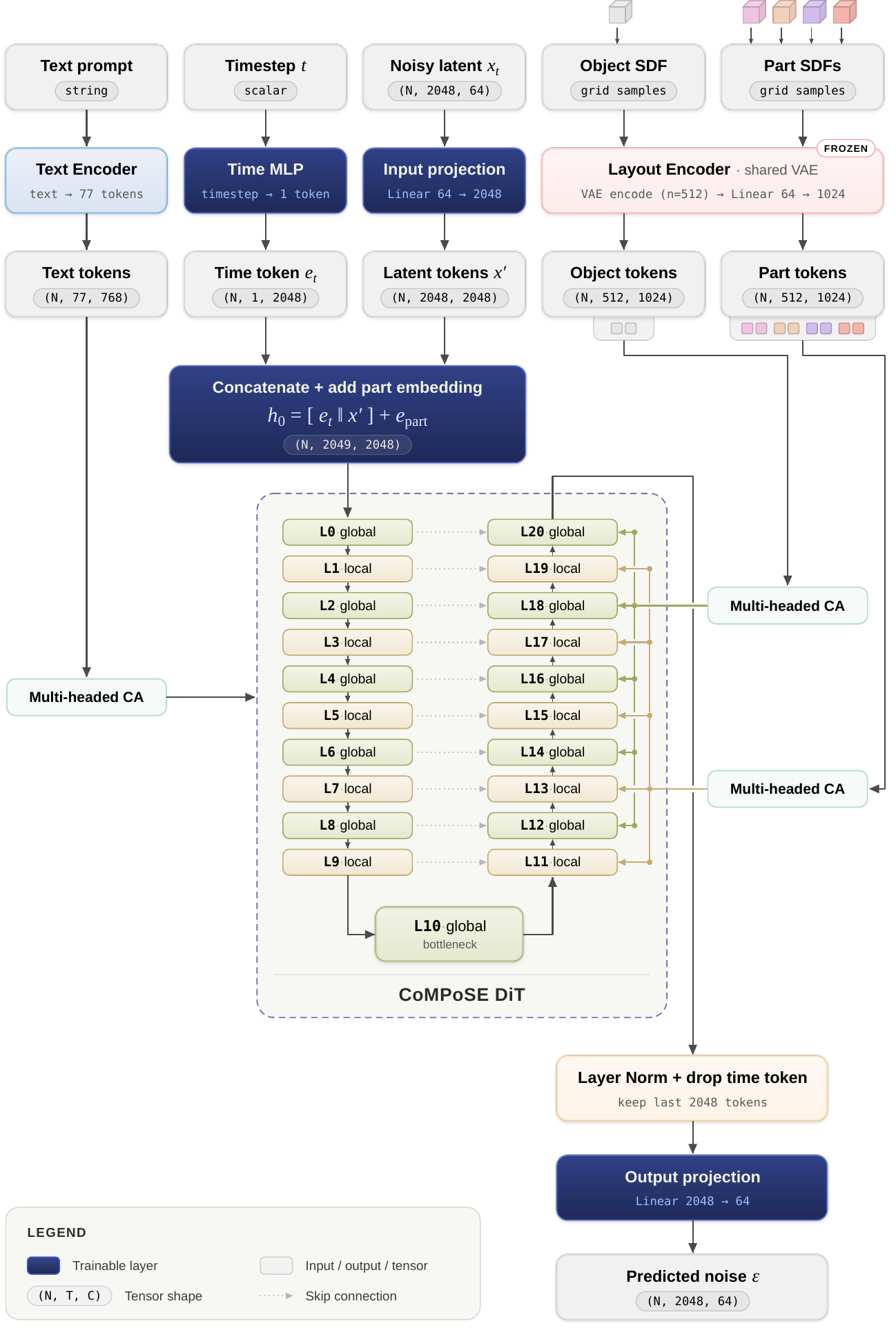}
    \vspace{1em}
    \caption{\textbf{Complete Architecture Diagram for the CompoSE Transformer.}
    We provide a complete architecture diagram for our CompoSE model, showing the part-control encoder, the layout encoder, the cross-attention fusion blocks, and the decoder.
    }%
    \label{fig:compose_arch}
\end{figure*}

We provide a complete architecture diagram of our CompoSE model in \cref{fig:compose_arch}, illustrating the flow from the part-control and layout encoders, through the cross-attention fusion blocks, to the decoder that produces the final output.

\vfill
\vfill

\newpage
\FloatBarrier
\subsection{CFG Control Annealing} 
\label{sec:cfgca}

\noindent We employ classifier-free guidance (CFG)~\cite{cfg_ho_2022} for improved sampling quality, but we find that combining it with layout conditioning requires special care: in the dual cross-attention mechanism of our layout control DiT blocks, setting the layout control strength near zero for negative samples causes excessive bounding-box conditioning, whereas a strength of one degrades output quality. Our solution is to anneal the bounding-box conditioning strength with an exponential decay across layers and throughout the denoising process, for negative CFG samples only, which substantially improves visual quality; we ablate this choice in~\cref{tab:method_ablation}. We detail the resulting schedule below.\\

\point{CFG} We use classifier-free guidance with a negative prompt to enhance the diversity and quality of the generated parts; the guidance scale and negative prompt are given in~\cref{sec:training}.\\

\begin{figure*}[h]
    \centering
    \includegraphics[width=0.95\textwidth]{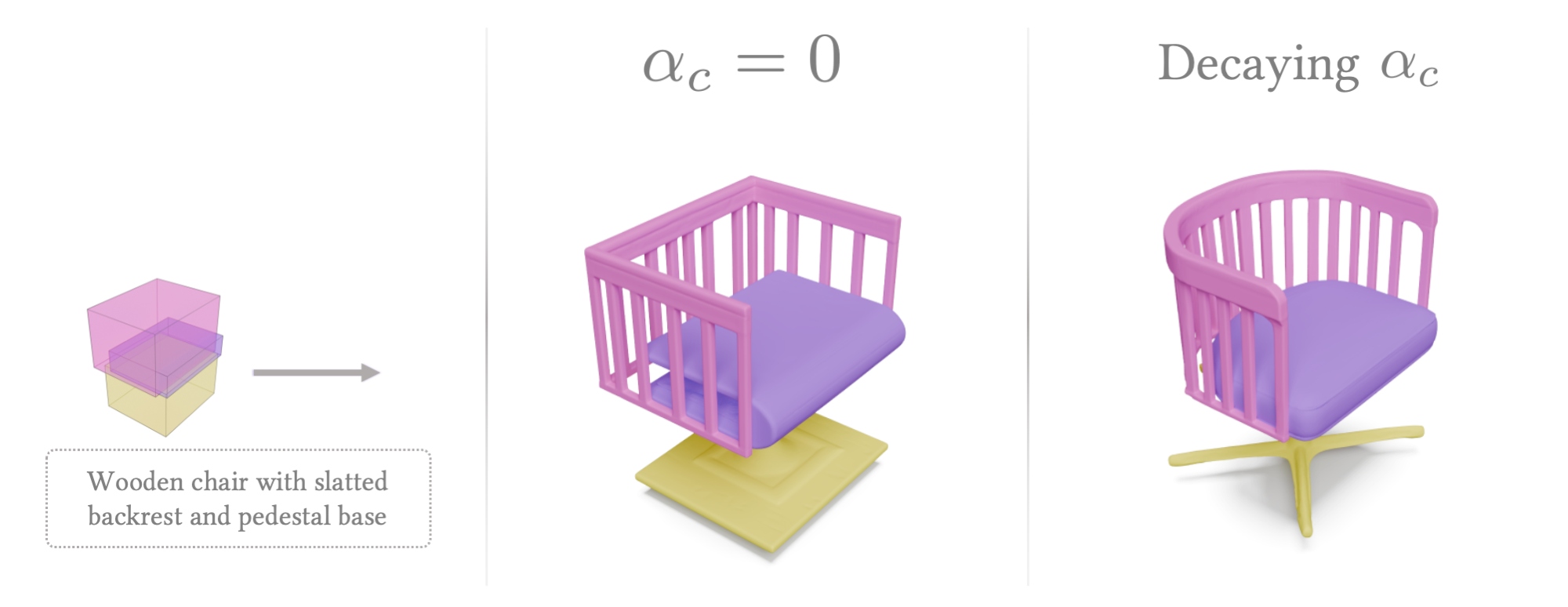}
    \caption{\textbf{Removing Control Guidance for Negative Samples.}
    We illustrate the effect of completely removing layout control for the negative CFG samples during inference (that is, setting $\alpha_c=0$ for negative samples throughout the denoising process).
    Without any layout control on negative samples (left), the synthesized shapes tend to exhibit square-like geometries.
    }%
    \label{fig:no_cfg_control}
\end{figure*}

\point{Control Annealing} Our proposed CFG Control Annealing (CFGCA) is applied as a layer-wise multiplicative decay of the layout control strength $\alpha_c$ during the denoising process, only for negative CFG samples.
For positive samples, using $\alpha_c = 0$ leads to poor layout adherence.
For negative samples, when using $\alpha_c = 0$ throughout the denoising process, synthesized shapes tend to be square-like or boxy. We illustrate this effect in \cref{fig:no_cfg_control}.
This is because negative CFG samples, being unconditioned on layout, are spatially unrestricted.
Classifier-free guidance then overcorrects by forcing the generated shape to conform tightly to the layout boundaries, resulting in box-like geometries.

Instead, by gradually reducing the layout control strength for negative samples, we push our model to strongly adhere to the layout in the later denoising steps, while allowing more freedom in the earlier steps for enhanced diversity and shape quality.\\

\begin{figure*}[h]
    \centering
    \vspace{0.5em}
    \includegraphics[width=\textwidth]{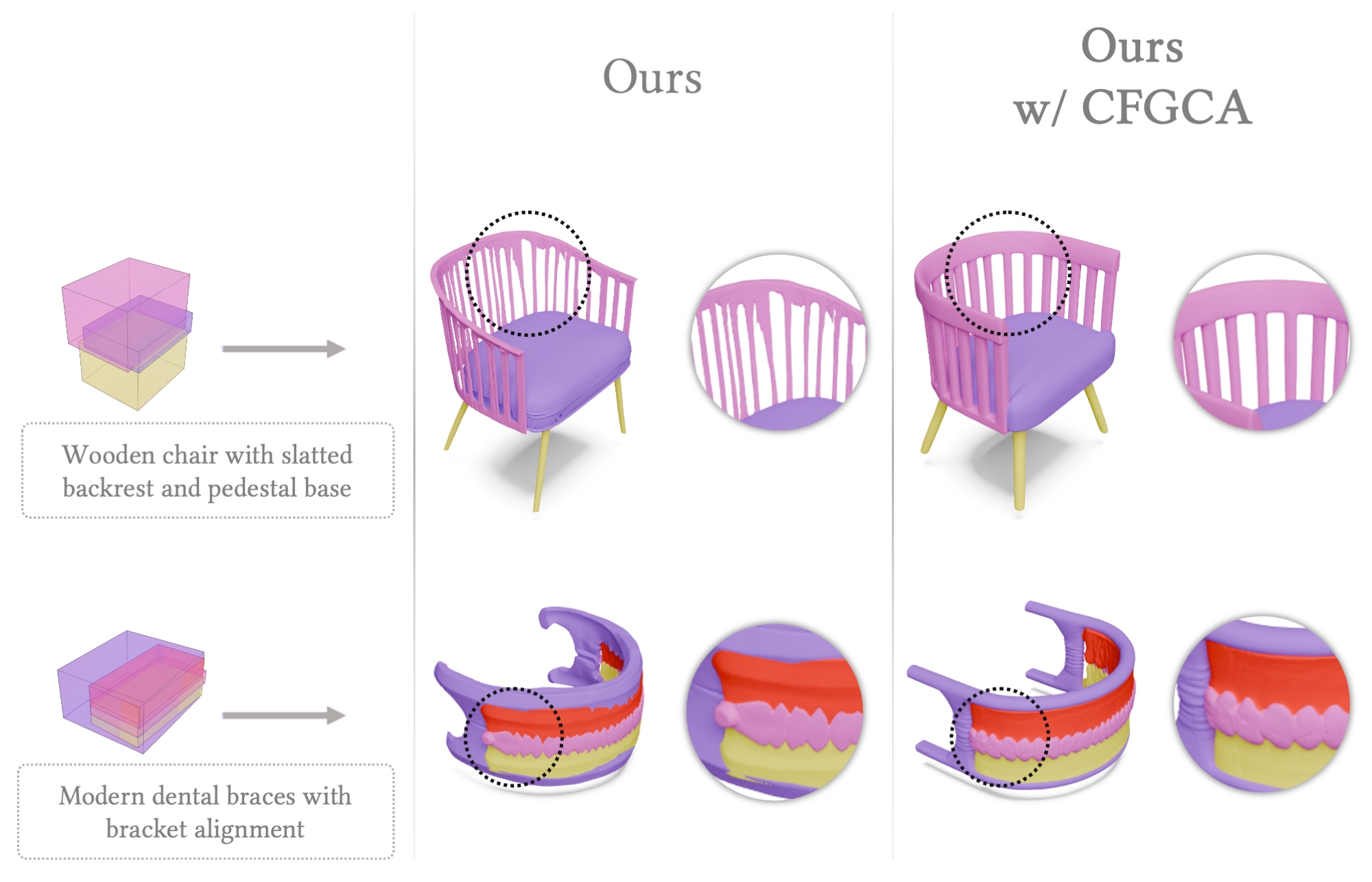}
    \vspace{0.5em}
    \caption{\textbf{Effects of CFG Control Annealing.} 
    We illustrate the effect of annealing the layout control strength during inference for the negative CFG samples.
    Without annealing, synthesized samples can exhibit undesirable artifacts (left).
    With our proposed CFG Control Annealing (right), the generated shapes display higher quality geometry more consistently.
    This is especially noticeable in samples containing regular geometric patterns (e.g., backrest slats, dental braces).
    }%
    \vspace{0.5em}
    \label{fig:cfga_ablation}
\end{figure*}

\point{Qualitative ablation} We provide additional qualitative results demonstrating the effect of CFG Control Annealing (CFGCA) during inference.
We sample shapes starting from the same layout, text prompt and initial noise latent, and show multiple generations with and without control annealing.
The results are shown in \cref{fig:cfga_ablation}.
Overall, we observe that CFGCA helps produce shapes of higher quality without geometric artifacts, while still adhering well to the input layout.
These observations confirm the quantitative results reported in~\cref{tab:method_ablation}.

\vfill
\vfill

\newpage
\vspace*{\fill}
\vfill

\clearpage
\FloatBarrier
\subsection{Layout Optimization and Artifact Filtering}
\label{sec:layout_optimization}

\begin{figure*}[h]
    \centering
    \resizebox{0.83\textwidth}{!}{%
        \includeinkscape{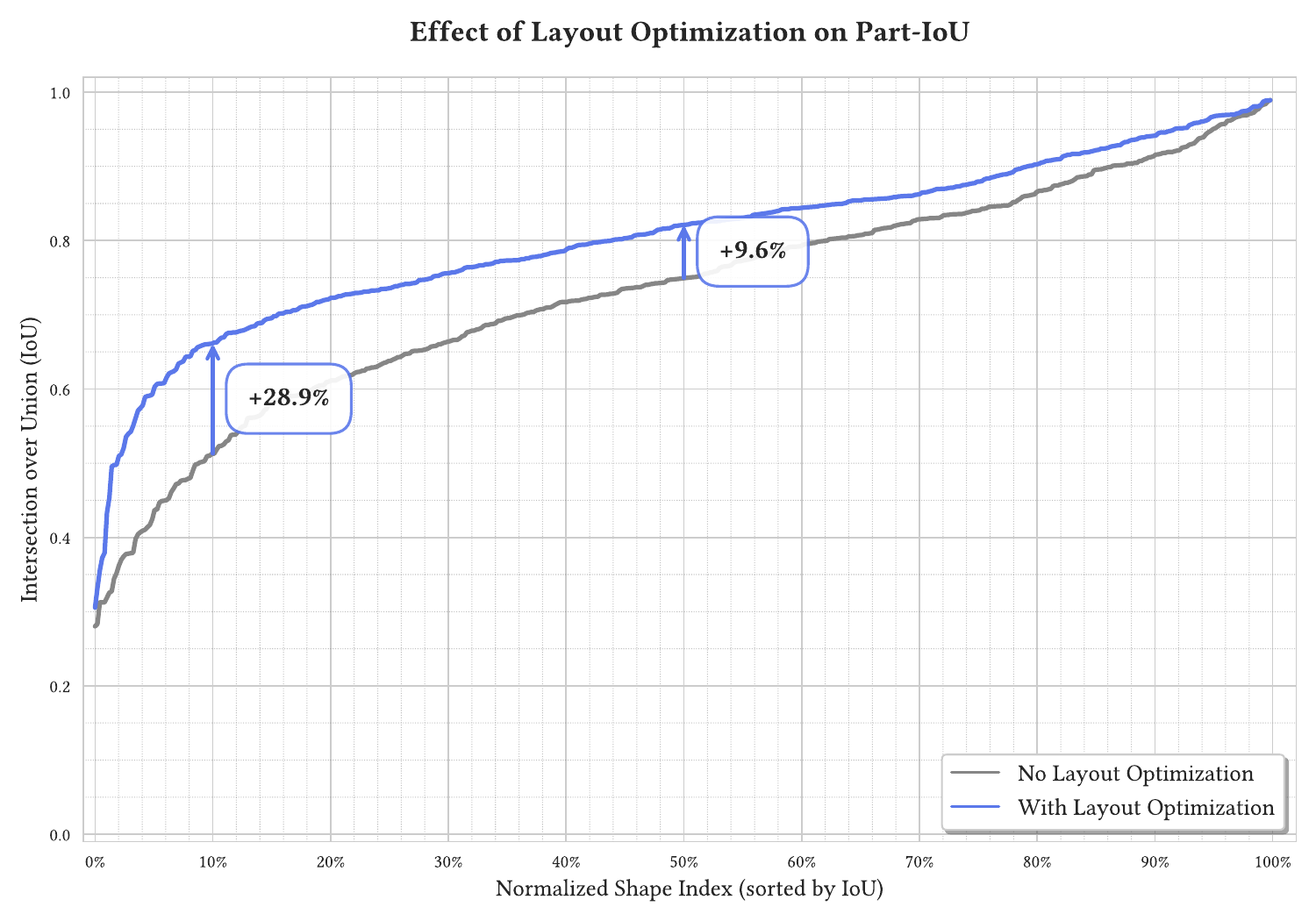}
    }
    \caption{\textbf{Effects of Layout Optimization on Part-IoU.}
    We plot the effects of our layout optimization and artifact filtering, described in this section.
    We sort shapes by increasing average part-IoU (left to right).
    We plot the average part-IoU before (gray) and after (blue) layout optimization.
    For the shapes in the bottom 10\% of most misaligned shapes, we gain an improvement of around 28\% in part-IoU after optimization.
    These gains decrease to around 9\% for the top 50\% best-aligned shapes, showing that our optimization is mostly beneficial for poorly aligned shapes.
    }%
    \label{fig:layout_opt_quant}
    \vspace{-0.15em}
\end{figure*}

\noindent Because our method operates in the latent space of a VAE, conversions to and from 3D are subject to reconstruction errors. Since we encode our geometric controls through that same VAE, these errors also distort the conditioning signal and can leave synthesized parts slightly misaligned with the input layout. Fortunately, we know the desired 3D location of each part at inference time, which lets us correct both this misalignment and the artifacts that arise from encoding parts in isolation. We detail the two procedures below.\\

\point{Misalignment} Due to our synthesis method operating in latent space, and the fact that our layout controls are also encoded into latent features, encoding/decoding inaccuracies can lead to misalignments between the generated shapes and the input part controls.
To mitigate these effects, we implement a simple beam-search-based layout optimization procedure during inference.
We optimize a shape-preserving transformation (similarity transformation) of the input layout to maximize the part-IoU between the generated parts and the transformed layout.
Note that this optimization is performed without any alteration of the synthesized geometry.\\

\point{Artifacts} Furthermore, some generated parts may exhibit artifacts such as floating geometry or disconnected fragments. This is mainly due to the distribution shift between the training data of the VAE, which is mainly composed of normalized and centered holistic shapes, and our fine-tuning data, which consists of unnormalized part segments. Previous works~\cite{PartCrafter_Lin_2025, PartPacker_Tang_2025} have also observed similar artifacts when using pre-trained VAEs for part-based reconstruction, and proposed to completely retrain a VAE on a large dataset of part segments to mitigate this issue. However, this approach is computationally prohibitive and requires access to a large dataset of parts.
In our case, having access to ground-truth part layouts and focusing on part instances allows us to implement a simple artifact filtering strategy based on part-IoU scores.
Specifically, since our model generates part instances as single connected components tightly fitting within the input bounding box, mesh components appearing outside their corresponding control boxes are highly likely to be artifacts.\\

\noindent Using this fact, we:

\begin{itemize}
    \item decompose each synthesized part mesh into its connected components, and only consider parts with more than one component as candidates for artifact removal.
    \item identify the largest connected component of each part, and compute its IoU with the input bounding box.
    \item discard all other components and retain only the largest one, if its IoU is above a fixed threshold.
\end{itemize}

\begin{figure*}[h]
    \vspace{0.5em} 
    \centering
    \resizebox{0.98\textwidth}{!}{%
        \includegraphics{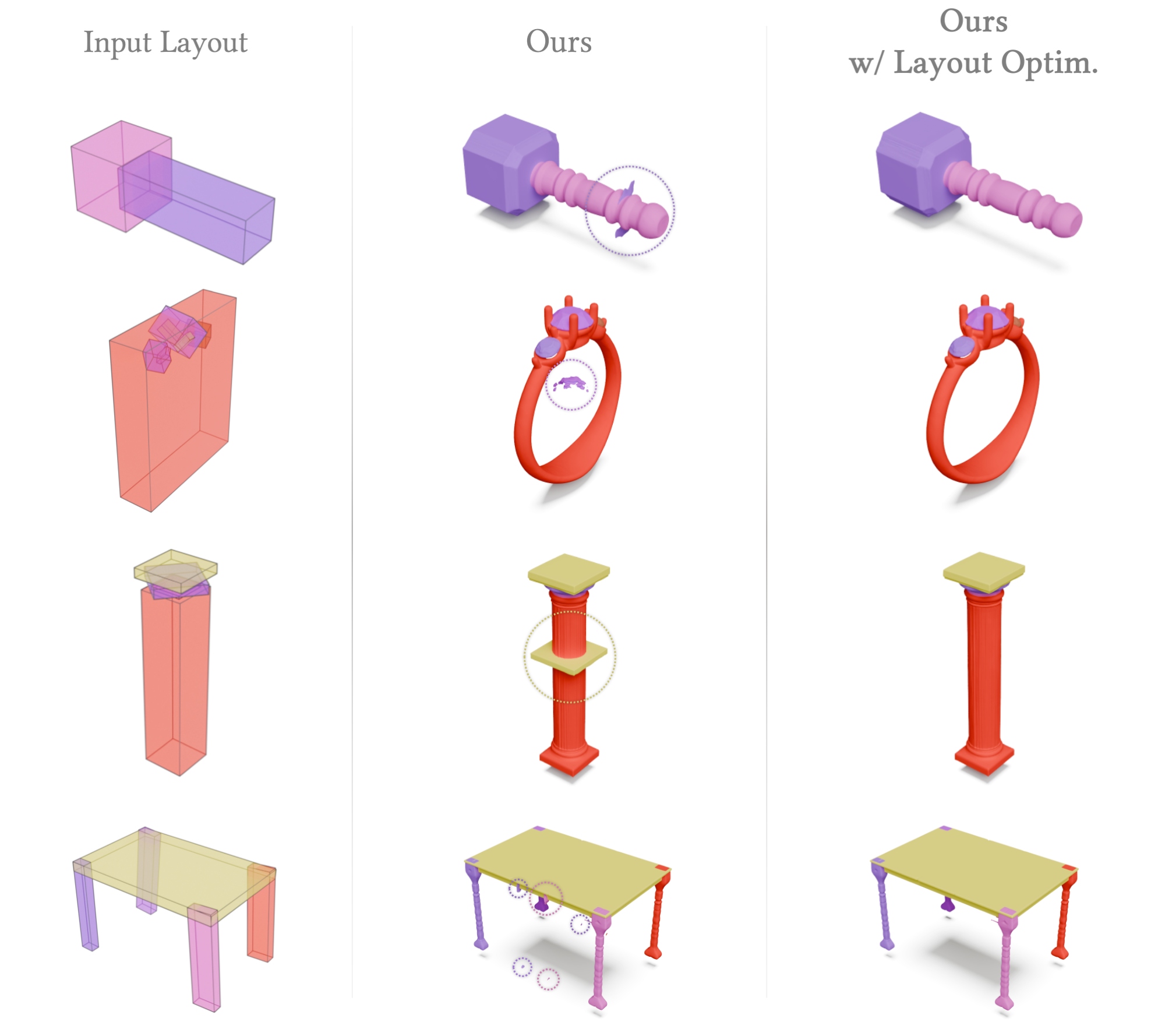}
    }
    \vspace{0.25em} 
    \caption{\textbf{Effects of Layout Optimization on Part-IoU.}
    We sample from the shapes most misaligned before optimization to illustrate the effects of our method.
    Before filtering, some parts exhibit floating geometry or disconnected fragments (middle, circled).
    These artifacts are especially prevalent in parts with flat or thin geometries (e.g., tabletops, planes).
    After optimization and filtering, the parts better fit their input bounding boxes, and artifacts are removed consistently (right).
    Importantly, this process does not affect the initial geometry.
    }%
    \label{fig:layout_opt_qual}
\end{figure*}

\vspace{0.25em}
\noindent This simple filtering strategy effectively removes floating artifacts in almost all cases, without requiring retraining of the VAE.
We illustrate qualitative results of our layout optimization and artifact filtering in \cref{fig:layout_opt_qual}.

\vfill
\vfill

\clearpage
\FloatBarrier
\subsection{Training Details}
\label{sec:training}

\point{First stage} We train our model end-to-end with a standard rectified flow loss~\cite{rectifiedflow_liu_2022}. During training, we use a mix of simple objects (consisting of few parts) and complex ones (consisting of many parts) to improve the generalization ability of our method.
Our base model is trained on 68k monolithic shapes and 23k multi-part samples from Objaverse~\cite{objaverse_deitke_2023}, for a total of \textbf{92k} objects and \textbf{170k} unique parts. Importantly, we do not use additional supervision such as part-level text prompts during training. We train this base model for 90k steps with a batch size of 128 distributed over 8$\times$A100 GPUs, which takes approximately two weeks.
We use the AdamW~\cite{adamw_loschilov_2019} optimizer with a learning rate of $5\times10^{-5}$, a weight decay of 0.01, and $\epsilon=10^{-8}$.
We clone the weights of a pre-trained TripoSG DiT~\cite{triposg_li_2025} model for initialization, and modify the DiT blocks to allow for layout conditioning as described in the main paper.
Training batches are composed of shape instances sampled randomly from our dataset, in a way that ensures that the total number of parts per batch is equal to our selected batch size.
We keep a fixed 15\% of training shapes to be monolithic shapes (i.e., with a single part) to minimize forgetting of the base capabilities.
We train using $D=512$ tokens to represent each part latent.
We further fine-tune the model with $D=2048$ tokens in a second stage, described below.

\point{Second stage} For the higher-token-resolution variant with $D=2048$ tokens, we fine-tune the model for an additional 100k steps on an expanded dataset mix combining Objaverse~\cite{objaverse_deitke_2023}, 3DCoMPaT~\cite{3dcompat_li_2022, 3DCoMPaT++_slim_2023}, and HY3DBench~\cite{hy3dbench_tencent_2026} samples, all pre-processed with the pipeline described in~\cref{fig:dataset_pipeline}, for a total of \textbf{219k} unique shapes and \textbf{1.39M} parts.
For 3DCoMPaT and HY3DBench, we use the labeled part segments as initial proposals for our merging stage as we find that they are of sufficiently high quality.
Finally, we select a further \textbf{43k} high-quality multi-part samples from the resulting dataset mix to fine-tune the model for an additional 2k steps, using heuristics based on inter-part IoU. This second training stage is performed on 8$\times$H100 GPUs.\\

\point{Third stage} To support layouts of up to 16 parts per shape, we further fine-tune the $D=2048$ model for an additional 50k steps, following the same procedure and dataset mix as in the second stage, but raising the maximum number of parts per training sample to 16.\\

\point{Inference} We use classifier-free diffusion guidance (CFG~\cite{cfg_ho_2022}) during inference, with a CFG scale of 6.5 for all of our experiments, as we find that this value provides a good balance between text and layout adherence and shape quality.
We employ the negative prompt ``Low-poly, minimal, blocky'' to discourage overly simplistic shapes.
Additionally, we use Temporal Score Rescaling~\cite{tsr_xu_2025} with a scale of $k=0.98$ to further enhance shape detail.
For all of our experiments, we use 50 denoising steps with the sampler introduced in Stable Diffusion 3~\cite{sd3_esser_2024}.
Mesh extraction is performed using Marching Cubes~\cite{marching_lorensen_1987}.
We perform inference with CFG Control Annealing (CFGCA) as described in \cref{sec:cfgca}, with a layer-wise decay factor of $\beta=0.99$.

\clearpage
\FloatBarrier
\section{Dataset}
\label{sec:dataset}
Our main first-stage dataset is derived from the Objaverse~\cite{objaverse_deitke_2023} collection through a fully automated pipeline designed to produce part-segmented shapes and their corresponding text-layout pairs without any manual effort.

\FloatBarrier
\subsection{Dataset Creation}
\label{sec:dataset_creation}

\begin{figure*}[p]
    \centering
    \includegraphics[width=\textwidth,height=0.885\textheight,keepaspectratio]{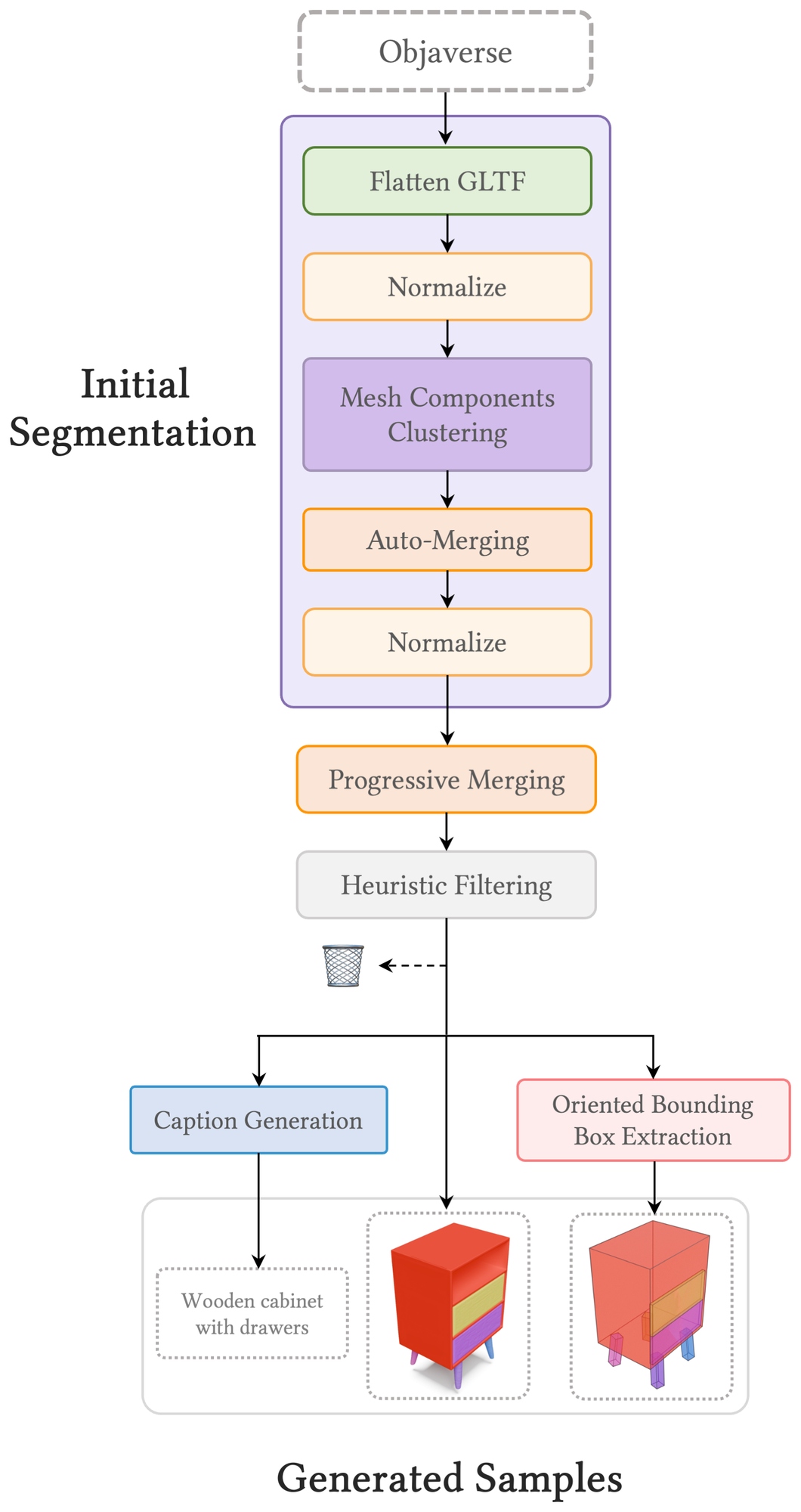}
    \vspace{1em}
    \caption{\textbf{Dataset Processing Pipeline.}
    We illustrate our fully automated dataset processing pipeline that converts raw 3D meshes from the 3D asset datasets into part-segmented shapes with corresponding text-layout pairs for training.
    Layouts are composed of oriented bounding boxes (OBBs) computed for each part segment.
    Text captions are generated using the InternVL3-8B~\cite{chen2024internvl} vision-language model and focus on geometric and part properties.
    }%
    \label{fig:dataset_pipeline}
\end{figure*}

There is no off-the-shelf dataset consisting of part-segmented shapes, bounding box layouts, and corresponding text captions that is suitable for our purposes, so we create one by applying a novel fully automated data processing pipeline to Objaverse~\cite{objaverse_deitke_2023}-sourced samples.
In our second training stage, we also process a mix of pre-segmented samples from various sources~\cite{3dcompat_li_2022, 3DCoMPaT++_slim_2023, hy3dbench_tencent_2026} using the same pipeline to increase the diversity of our training data.
We illustrate our processing pipeline in \cref{fig:dataset_pipeline}. Our pipeline consists of the following stages:\\[-1em]

\point{Initial Segmentation} We begin by decomposing each raw 3D mesh into an initial set of low-level segments, which we will merge into semantically meaningful part instances in our next stage.
We use the geometrically disconnected components of the mesh as our initial set of segments.
We note that the Objaverse GLTF files already contain an artist-defined hierarchy of segments, so we could use these segments as our initial set of segments to merge.
However, we found that the artist-defined segments in Objaverse can be spatially incoherent within an object, and inconsistent across objects, so we choose to discard them, a step we refer to as \emph{GLTF flattening}.
We then center each mesh at the object-space origin and scale it uniformly so that its longest dimension fits within a unit cube, in a process we refer to as \emph{normalization}.
Afterwards, we apply an auto-merging step to fuse segments with planar geometries and negligible volumes with their largest intersecting segment, and we re-normalize.
\\[-1em]

\point{Progressive Segment Merging} To obtain semantically meaningful parts from our initial segments, we use a progressive merging strategy guided by segment volume. In each iteration, we identify the smallest segments and, for each one, tag it to be merged with its largest connected neighbor. All tagged merges are then executed in a single pass (per iteration). This iterative procedure continues until the number of segments for the shape is within a target range.\\[-0.5em]

\point{Heuristic Object Filtering} To ensure the quality of our training data, we discard unsuitable objects using a series of heuristics. We filter out shapes with poor part segmentation quality based on three criteria: (1) the volume ratios between the largest part and the rest of the parts; (2) the average intersection-over-union (IoU) between parts; and (3) the geometric integrity of each resulting part.
We use empirically tuned thresholds for the first two criteria, and discard any shape in which a final part does not form a single connected surface (e.g., a part consisting of several disjoint triangle islands), to ensure clean, monolithic part instances.\\[-0.5em]

\point{Oriented Bounding Box Extraction} Once we have obtained semantically meaningful parts, we generate a spatial layout of coarse geometric control shapes that will guide our synthesis method. For each part, we compute its minimum oriented bounding box (OBB) using an efficient implementation~\cite{open3d_zhou_2018} of the exact minimum bounding box algorithm described by Jyl\"anki~\cite{minobb_jylanki_2015}.\\[-0.5em]

\point{Text Caption Generation} The final component of our training data is the global text prompt. To generate a high-quality caption for each processed object, we render images of the complete object, and we feed these images to the powerful InternVL3-8B~\cite{chen2024internvl} vision-language model.
We use collages of four views arranged in a 2$\times$2 grid to provide sufficient context for caption generation.
We employ a prompt that encourages captions focusing on geometry and part descriptions, avoiding color and texture which are not relevant to our synthesis task.
This process provides a clean and descriptive text prompt that is grounded to the final geometry.

\clearpage
\FloatBarrier
\subsection{Statistics}
\label{sec:dataset_stats}

\begin{figure*}[h]
    \centering
    \vspace{0.5em} 
    \resizebox{0.88\textwidth}{!}{%
        \includeinkscape{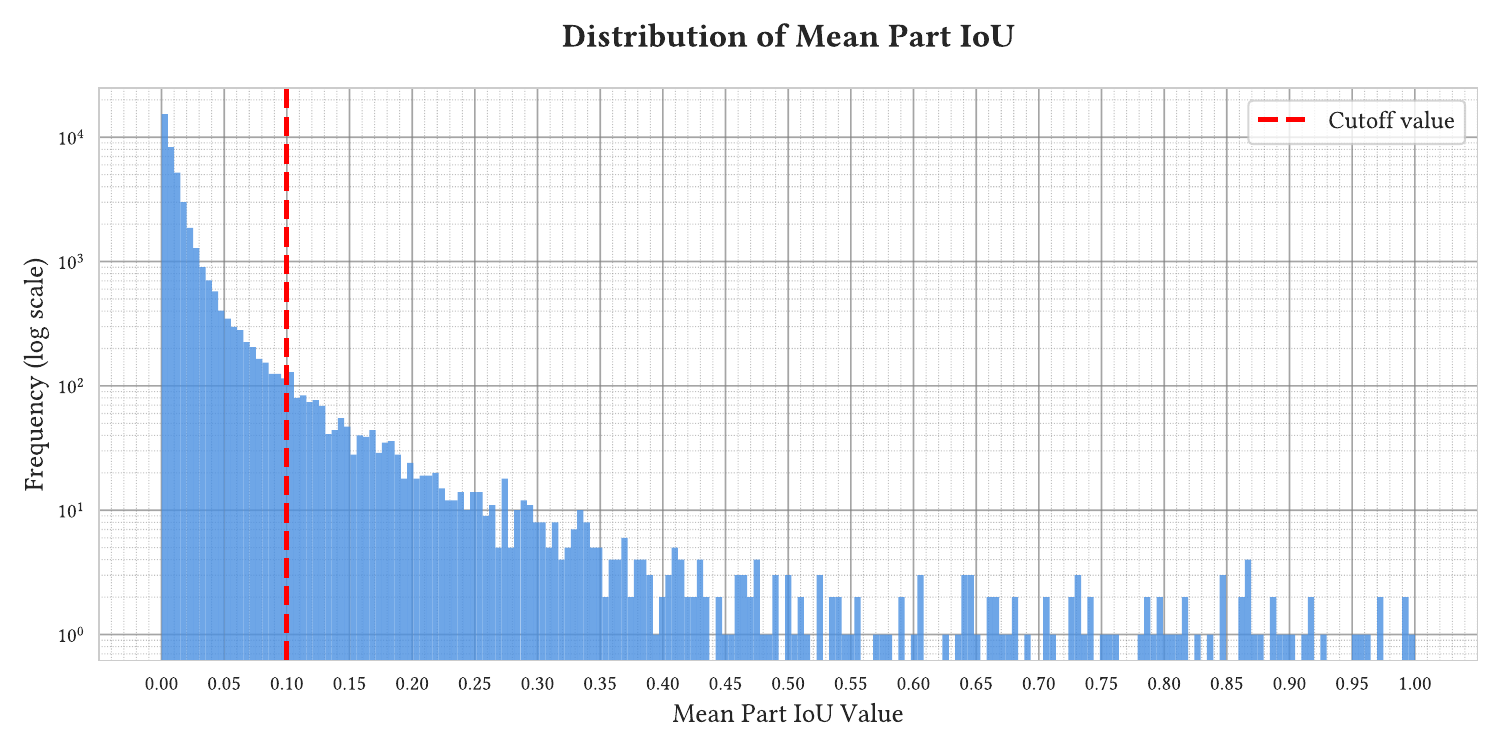}
        \vspace{0.5em}
    }
    \resizebox{0.88\textwidth}{!}{%
        \includeinkscape{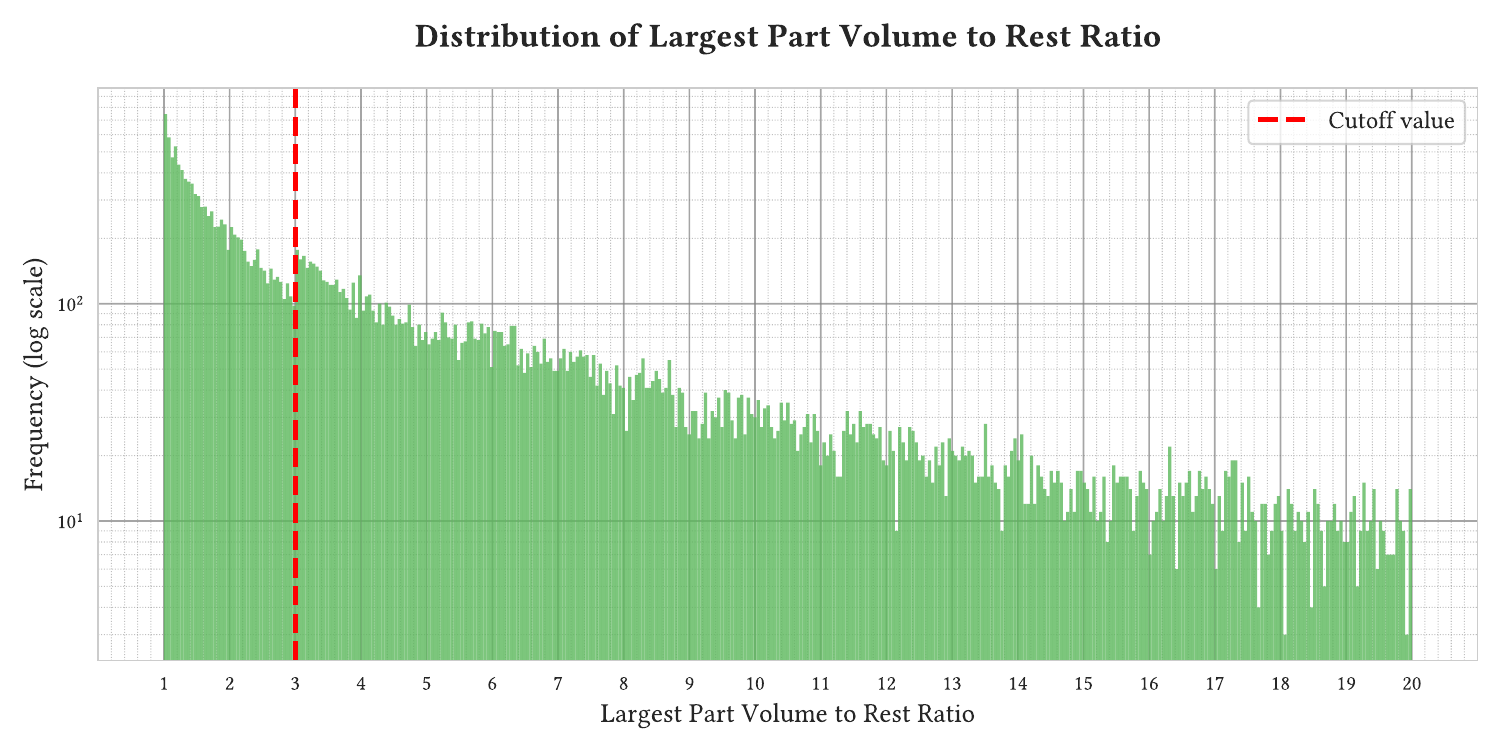}
        \vspace{0.5em}
    }
    \caption{\textbf{Dataset Statistics.}
    We plot the distribution of mean part-IoU (top) and largest-to-rest part volume ratio (bottom) in our processed dataset.
    Frequencies are shown on a logarithmic scale to better visualize the long tails of the distributions.
    We also show in red our cutoff thresholds used for heuristic filtering during dataset processing.
    }%
    \label{fig:data_statistics}
\end{figure*}

We plot the distribution of mean part-IoU and largest-to-rest part volume ratio in our processed dataset in \cref{fig:data_statistics}.
Alongside the distributions of mean part-IoU and largest-to-rest part volume ratio, we also plot the cutoff thresholds used for heuristic filtering during dataset processing in red.
These cutoffs are selected empirically to remove badly segmented shapes, and more generally poor-quality shapes.
We use a maximum mean part-IoU of 0.10 and a maximum largest-to-rest part volume ratio of 3.

Shapes with a high mean part-IoU are not beneficial to the training of our synthesis model, since their parts are highly overlapping and do not provide clear geometric cues.
Similarly, heavily imbalanced part layouts are also not useful, as they lead to degenerate training samples where a single part occupies a significant portion of the output space.

In addition to this filtering, we also discard duplicated shapes, and we filter models based on their caption using a list of keywords to maximize the quality of the geometry extracted.
After filtering, our final dataset contains \textbf{68k} monolithic shapes and \textbf{23k} multi-part shapes. Monolithic shapes are rebalanced during training to represent 15\% of training samples.

\clearpage
\FloatBarrier
\section{Additional Ablations}
\label{sec:additional_ablations}

We report additional ablations that were omitted from the main text for space: a sweep over the classifier-free guidance scale $\omega$ used by SpaceControl~\cite{spacecontrol_fedele_2026}, an evaluation of our identity-preserving part rescaling procedure against two alternatives, and a model ablation study of our design choices. We also report a comparison of inference time against the baselines.

\FloatBarrier
\subsection{SpaceControl Classifier-Free Guidance Scale}
\label{sec:cfg_scale_ablation}

In~\cref{tab:complete_guided_synthesis} of the main paper, we report SpaceControl's result at a single operating point, $\omega=3$, which we selected as it achieves the best Voxel-IoU among all guidance scales we tested, the only metric on which SpaceControl outperforms our method.
For completeness, \cref{tab:spacecontrol_cfg_ablation} reports the full sweep over $\omega \in \{3, 6, 7.5, 9\}$, as well as a variant augmented with image conditioning via ControlNet~\cite{controlnet_zhang_2023}.
Increasing $\omega$ consistently improves Object-FID at the cost of Object-IoU, as stronger guidance pushes the model toward more detailed, prompt-aligned geometry that deviates from the dense superquadric volume defined by the input proxies.
We find $\omega=7.5$ offers the best trade-off between visual fidelity and layout adherence among the scales we tested, while $\omega=3$ remains the strongest choice specifically for Voxel-IoU, motivating our choice of operating point in the main comparison.

\begin{table}[h]
\centering
\caption{\textbf{SpaceControl Classifier-Free Guidance Scale Ablation.}
    We report SpaceControl's~\cite{spacecontrol_fedele_2026} performance across classifier-free guidance scales $\omega$, as well as a variant augmented with image conditioning via ControlNet~\cite{controlnet_zhang_2023} (CN).
    }
\label{tab:spacecontrol_cfg_ablation}
\arrayrulecolor{black!25}%
\setlength{\arrayrulewidth}{0.3pt}%
\renewcommand{\arraystretch}{1.1}%
\resizebox{0.75\textwidth}{!}{%
\small
\setlength{\tabcolsep}{8pt}%
\begin{tabular}{@{}l ccc c@{}}
\arrayrulecolor{black}\toprule
\textbf{Variant} & \textbf{Object-FID}$\downarrow$ & \textbf{Object-IoU}$\uparrow$ & \textbf{Voxel-IoU}$\uparrow$ & \textbf{CLIP}$\uparrow$ \\
\arrayrulecolor{black}\midrule
$\omega=3$                     & 42.96 & 0.807 & \textbf{0.776} & 26.73 \\
$\omega=6$                     & 40.20 & 0.794 & 0.761          & 26.83 \\
$\omega=7.5$                   & 39.27 & 0.790 & 0.750          & 26.86 \\
$\omega=9$                     & \textbf{38.26} & 0.785 & 0.751          & \textbf{26.93} \\
$\omega=7.5$ + CN~\cite{controlnet_zhang_2023} & 39.55 & 0.785 & 0.754          & 26.77 \\
\arrayrulecolor{black}\bottomrule
\end{tabular}%
}
\end{table}

\FloatBarrier
\subsection{Identity-Preservation under Part Rescaling}
\label{sec:identity_preservation_ablation}

We ablate our identity-preserving rescaling procedure (\cref{sec:method-editing}), which adjusts a part's bounding box while preserving its geometric identity via KV-reinjection.
We synthesize 70 edited shapes by manually collecting part rescaling operations applied on base shapes synthesized by our method, spanning multiple object categories.
For each edit, we compute F-Score at $\tau \in \{0.01, 0.02, 0.05\}$ between the resized part and its counterpart in the unedited shape.
We compare against two alternatives sharing the same target box: \textit{Anisotropic Scaling}, which stretches the original geometry, and \textit{Part Substitution}, which regenerates the part from scratch without anchoring to the original latent.
As shown in~\cref{tab:editing_evaluation}, our method achieves the highest F-Score at every threshold (0.38, 0.58, 0.70), substantially outperforming Anisotropic Scaling which distorts fine details, and Part Substitution which is stylistically consistent but geometrically different, in line with the qualitative results provided in~\cref{fig:part-level-editing} of the main paper.

\begin{table}[H]
    \vspace{0.5em}
\centering
\caption{\textbf{Identity-Preservation under Part Rescaling.}
    We measure how faithfully each baseline preserves a part's geometry
    after rescaling its bounding box, reporting F-Score at
$\tau \in \{0.01, 0.02, 0.05\}$ ($\uparrow$) against the original part.
            }
\arrayrulecolor{black!25}%
\setlength{\arrayrulewidth}{0.3pt}%
\resizebox{0.75\textwidth}{!}{%
\small
\renewcommand{\arraystretch}{1.2}
\setlength{\tabcolsep}{12pt}%
\begin{tabular}{@{}l@{\hskip 18pt} c c c@{}}
\arrayrulecolor{black}\toprule
\textbf{Method} &
\textbf{F@0.01} $\uparrow$ &
\textbf{F@0.02} $\uparrow$ &
\textbf{F@0.05} $\uparrow$ \\
\arrayrulecolor{black}\midrule
    Anisotropic Scaling   & 0.23 & 0.38 & 0.55 \\
    Part Substitution     & 0.26 & 0.46 & 0.64 \\
\arrayrulecolor{black!25}\midrule
\textbf{Identity-Preserving} & \textbf{0.38} & \textbf{0.58} & \textbf{0.70} \\
\arrayrulecolor{black}\bottomrule
\end{tabular}%
            }
\arrayrulecolor{black}%
\label{tab:editing_evaluation}
\vspace{-0.5em}
\end{table}

\clearpage
\FloatBarrier
\subsection{Model Ablation Study}
\label{sec:model_ablation_study}

We conduct an ablation study to validate our key design choices, with results summarized in~\cref{tab:method_ablation}.\\[-0.5em]

\point{Part Condition Encoding} We compare our SDF-based encoding against (1) a \textit{parametric encoder} (box parameters via MLP) and (2) a \textit{point-based encoder} (DGCNN~\cite{dynamicgraph_wang_2019} on box surface samples, fine-tuned end-to-end). Reusing the SDF-based VAE for guidance boxes ensures latent-space alignment with shape parts --- a choice~\cref{tab:method_ablation} shows is critical for geometric fidelity and layout adherence.\\[-0.5em]

\point{Inference Strategies} CFG Control Annealing is crucial for visual quality, improving Object-FID from 19.32 to 14.44, while layout optimization --- a test-time refinement step detailed in~\cref{sec:layout_optimization} --- raises Part-IoU from 0.740 to 0.805 by correcting minor spatial drifts.\\[-0.5em]

\point{Token Resolution} Scaling per-part tokens from 512 to 2048 clearly improves all metrics, confirming our approach benefits from scale.

\begin{table}[H]
    \vspace{1em}
    \centering
    \caption{\textbf{Ablation study.} We ablate various components including choice of part-control encoders, CFG Control Annealing, layout optimization, and token resolution.}
    \vspace{0em}
    \setlength{\tabcolsep}{1pt}%
    \renewcommand{\arraystretch}{1.2}
    \maxresizebox{\textwidth}{%
        \small
        \begin{tabular}{*{7}{c}}
            \toprule
            \noalign{\vskip 4pt}
            \begin{tabular}{c} {Layout Encoding} \\[-5pt] {\small } \end{tabular}
            & \begin{tabular}{c}{CFG Control Annealing}\\[-5pt] {\small (\cref{sec:cfgca})} \end{tabular}
            & \begin{tabular}{c}{Layout Optimization}\\[-5pt] {\small (\cref{sec:layout_optimization})} \end{tabular}
            & \begin{tabular}{c}{Token Resolution}\\[-5pt] {\small } \end{tabular}
            & \begin{tabular}{c}{Object-FID}\\$\downarrow$\end{tabular} & 
            \begin{tabular}{c}{Part-FID}\\$\downarrow$\end{tabular} & 
            \begin{tabular}{c}{Part-IoU}\\$\uparrow$\end{tabular} \\ 
            \midrule
                 Parametric &   &   & 512 & 20.17 & 20.14 & 0.289 \\
                DGCNN~\cite{dynamicgraph_wang_2019} &   &   & 512 & 38.58 & 43.10 & 0.008 \\
                        SDF &   &   & 512 & 19.32 & 21.87 & 0.729 \\
                        SDF & \cmark &   & 512 & 14.44 & 19.38 & 0.740 \\
                        SDF & \cmark & \cmark & 512 & 14.40 & 19.70 & 0.805 \\
                        SDF & \cmark & \cmark & 2048 & \textbf{10.58} & \textbf{9.24} & \textbf{0.819} \\
            \bottomrule
        \end{tabular}%
    }
    \label{tab:method_ablation}
    \vspace{0em}
\end{table}

\clearpage
\subsection{Inference Time}
\label{sec:inference_time}

We evaluate inference time against the baselines on an NVIDIA H100 GPU, with results summarized in \cref{tab:inference_cost}. At resolution 512, our method is the fastest of all methods at $N_p=1$ (3.7s), and even at $N_p=16$ it stays at 20.8s, far below the other proxy-guided baselines (Fantasia3D, Coin3D, and Spice-E), which need between 146.2s and 522.6s for a single part. At resolution 2048, we start at 5.5s, ahead of SpaceControl (9.1s) and PartCrafter (7.0s), but our per-part decoding makes cost grow with the number of parts, reaching 80.3s at $N_p=16$. OmniPart is the only method whose cost is independent of $N_p$ (4.1s throughout), while PartCrafter scales at a rate comparable to our 2048 model.

\begin{table}[H]
    \vspace{1em}
    \centering
    \caption{\textbf{Inference Time.} We compare the average inference time (in seconds) of our method with varying number of parts $N_p$, measured on a single NVIDIA H100 GPU. \\
    \textcolor{gray}{$^\dagger$Fantasia3D~\cite{fantasia3d_chen_2023} being explicitly optimized for multi-GPU inference, we report its timing using 4 H100 GPUs for fairness.} \\
    \textcolor{gray}{Rows shown in gray (Image-Conditioned Part Synthesis) report timings that exclude the Nano Banana Pro~\cite{nanobananapro_2025} image-conditioning step required to produce their input; since our method requires no such step, this comparison is not entirely fair.}
        }
    \renewcommand{\arraystretch}{1}%
    \resizebox{0.8625\textwidth}{!}{%
    \small
    \begin{tabular}{lccccc}
    \arrayrulecolor{black}\toprule
    \multirow{2}{*}{Method} & \multicolumn{5}{c}{\textbf{Inference Time (sec.)} $\downarrow$} \\
    \cmidrule(lr){2-6}
    & $N_p=1$ & $N_p=2$ & $N_p=4$ & $N_p=8$ & $N_p=16$ \\
    \midrule
    \multicolumn{6}{@{}l@{}}{\textit{Proxy-Guided Synthesis}} \\
    \arrayrulecolor{black!25}\midrule
            Fantasia3D~\cite{fantasia3d_chen_2023}\textcolor{gray}{$^\dagger$} & 245.7 $\pm$ 3.0 & \iconog & \iconog & \iconog & \iconog \\
            Coin3D~\cite{dong2024coin3d} & 522.6 $\pm$ 13.8 & \iconog & \iconog & \iconog & \iconog \\
            Spice-E~\cite{spice_sella_2024} & 146.2 $\pm$ 1.7 & \iconog & \iconog & \iconog & \iconog \\
            SpaceControl~\cite{spacecontrol_fedele_2026} & 9.1 $\pm$ 4.1 & \iconog & \iconog & \iconog & \iconog \\
    \arrayrulecolor{black}\midrule
    \multicolumn{6}{@{}l@{}}{\textit{Image-Conditioned Part Synthesis}} \\
    \arrayrulecolor{black!25}\midrule
    \rowcolor{gray!15}
            \textcolor{gray!70!black}{PartCrafter~\cite{PartCrafter_Lin_2025}} & \textcolor{gray!70!black}{7.0 $\pm$ 0.6} & \textcolor{gray!70!black}{10.5 $\pm$ 0.1} & \textcolor{gray!70!black}{27.2 $\pm$ 0.6} & \textcolor{gray!70!black}{36.6 $\pm$ 0.2} & \textcolor{gray!70!black}{75.1 $\pm$ 0.1} \\
    \rowcolor{gray!15}
            \textcolor{gray!70!black}{OmniPart~\cite{OmniPart_Yang_2025}} & \multicolumn{5}{c}{\textcolor{gray!70!black}{4.1 $\pm$ 0.3}} \\
    \arrayrulecolor{black}\midrule
            \textbf{Ours}\,\textsubscript{\textsc{512}} & 3.7 $\pm$ 0.1 & 4.1 $\pm$ 0.1 & 5.8 $\pm$ 0.1 & 10.5 $\pm$ 0.1 & 20.8 $\pm$ 0.1 \\
            \textbf{Ours}\,\textsubscript{\textsc{2048}} & 5.5 $\pm$ 0.1 & 8.7 $\pm$ 0.1 & 16.0 $\pm$ 0.1 & 33.7 $\pm$ 0.1 & 80.3 $\pm$ 0.2 \\
    \bottomrule
    \end{tabular}%
    }
    \label{tab:inference_cost}
    \vspace{0em}
\end{table}

\clearpage

\twocolumn

\end{document}